 \documentclass[twocolumn]{aastex62}
 \hypersetup{linkcolor=red,citecolor=blue,filecolor=magenta,urlcolor=cyan}

\shorttitle{Cold Classical TNOs}
\shortauthors{Audrey Thirouin and Scott S. Sheppard}
\begin{document}
 
\title{Lightcurves and Rotational Properties of the Pristine Cold Classical Kuiper Belt Objects}

\correspondingauthor{Audrey Thirouin}
 \email{thirouin@lowell.edu}

\author[0000-0002-1506-4248]{Audrey Thirouin}
\affil{Lowell Observatory, 1400 W Mars Hill Rd, Flagstaff, Arizona, 86001, USA.}

\author[0000-0003-3145-8682]{Scott S. Sheppard}
\affiliation{Department of Terrestrial Magnetism (DTM), Carnegie Institution for Science, 5241 Broad Branch Rd. NW, Washington, District of Columbia, 20015, USA.}



\begin{abstract}

We present a survey on the rotational and physical properties of the dynamically low inclination Cold Classical trans-Neptunian objects. The Cold Classicals are primordial planetesimals and contain relevant information about the early phase of our Solar System and planet formation over the first 100 million years after the formation of the Sun. Our project makes use of the Magellan and the Lowell's Discovery Channel Telescopes for photometric purposes. We obtained partial/complete lightcurves for 42 Cold Classicals. We use statistical tests to derive general properties about the shape and rotational frequency distributions of the Cold Classical population, and infer that the Cold Classicals have slower rotations and are more elongated/deformed than the other trans-Neptunian objects. Based on the available full lightcurves, the mean rotational period of the Cold Classical population is 9.48$\pm$1.53~h whereas the mean period of the rest of the trans-Neptunian objects is 8.45$\pm$0.58~h. About 65$\%$ of the trans-Neptunian objects (excluding the Cold Classicals) have a lightcurve amplitude below 0.2~mag compared to the 36$\%$ of Cold Classicals with small amplitude. We present the full lightcurve of one new likely contact binary: 2004~VC$_{131}$ with a potential density of 1~g~cm$^{-3}$ for a mass ratio of 0.4. We also have hints that 2004~MU$_{8}$ and 2004~VU$_{75}$ are maybe potential contact binaries based on their sparse lightcurves but more data are needed to confirm such a find. Assuming equal-sized binaries, we find that only $\sim$10-25~\% of the Cold Classicals could be contact binaries, suggesting that there is a deficit of contact binaries in this population compared to previous estimates and compared to the abundant ($\sim$40-50$\%$) possible contact binaries in the 3:2 resonant (Plutino) population. This estimate is a lower limit and will increase if non equal-sized contact binaries are also considered. Finally, we put in context the early results of the \textit{New Horizons} flyby of (486958) 2014~MU$_{69}$.

\end{abstract}

\keywords{Kuiper Belt Objects: individual (2004~VC$_{131}$, 2004~VU$_{75}$, 2004~MU$_{8}$); Techniques: photometric}


\section{Dynamically Cold Classical Trans-Neptunian Objects} \label{sec:intro}

 In this study, we target the dynamically Cold Classical trans-Neptunian objects (TNOs), one of the trans-Neptunian sub-population. The Cold Classicals (CCs) with semi-major axes from $\sim$40 up to $\sim$48~AU, have low inclinations and low eccentricities (e$<$0.24) \citep{Gladman2008}. Generally, in the case of the inclination, the cut-off is at 4$^\circ$-5$^\circ$. However based on surface colors analysis, a limit at $\sim$12$^\circ$ seems more appropriate \citep{Peixinho2008}.  

Among the entire trans-Neptunian belt, the CCs are the least evolved TNOs \citep{Batygin2011}. Over the years, it has been argued that the CCs have likely been formed in-situ and thus have remained far from the Sun, and have never undergone any catastrophic dynamical evolution. For all of these reasons, the CCs are pristine planetesimals and therefore have important indications about the early age of our Solar System regarding composition, rotational properties, dynamics, accretion and collisional theories. 

The CC population displays several properties which make them stand out compared to the other TNO populations. Their surfaces are very-red/ultra-red, and according to a recent photometric survey, the CCs are also distinguishable in the z-band \citep{Pike2017, Benecchi2009}. Another characteristic feature of the CC population is the large amount of resolved equal size wide binaries. Thanks to extensive surveys with the \textit{Hubble Space Telescope}, \citet{Noll2008, Noll2014} inferred that all the large CCs with H$\geq$6~mag are equal size wide binaries. In the entire CC population, the fraction of resolved binaries is 22$_{-5}^{+10}$$\%$ compare to the 5.5$_{-2}^{+4}$$\%$ in the other dynamical populations \citep{Stephens2006}. Because the CCs are primordial, their rotational properties are primordial too, but care has to be taken with the binary systems as tidal effect can affect their rotations.  

The \textit{NASA's New Horizons} spacecraft flew by a small CC in January 2019 named (486958) 2014~MU$_{69}$ \citep{Stern2019}. Therefore, it is crucial to have context for the flyby results and the best way for this purpose is to study a large sample of CCs and derive as much information as possible. Also, based on a multi-chords stellar occultation (and confirmed by the flyby), 2014~MU$_{69}$ is a contact binary and so it is useful to find more of them for comparison and understand their formation/evolution as well as constrain their fraction across the trans-Neptunian belt  \citep{Moore2018}.  

We present here a survey dedicated to the rotational features of the CC population. Following, we will present our survey (Section 2), as well as our sparse/complete lightcurves of 42 CCs (Section 3). In Section 4, we will derive information about the shape and rotational frequency distribution of the CC population. We will also compare the contact binary fraction of the CCs and the Plutinos, as well as the main rotational properties of the CCs and the rest of the TNOs (Section 5). Also, we will provide some context for the second flyby of the \textit{NASA New Horizons} mission (Section 6). Finally, we will summarize our findings.  
 
%%%%%%%%%%%%%%%%%%%%%%%%%%%%%%%%%%%%%%%%%%%%
%%%% Observations
%%%%%%%%%%%%%%%%%%%%%%%%%%%%%%%%%%%%%%%%%%%%

\section{Survey Design, Observational Facilities and Data}
 
\begin{figure*}
\includegraphics[width=20cm, angle=0]{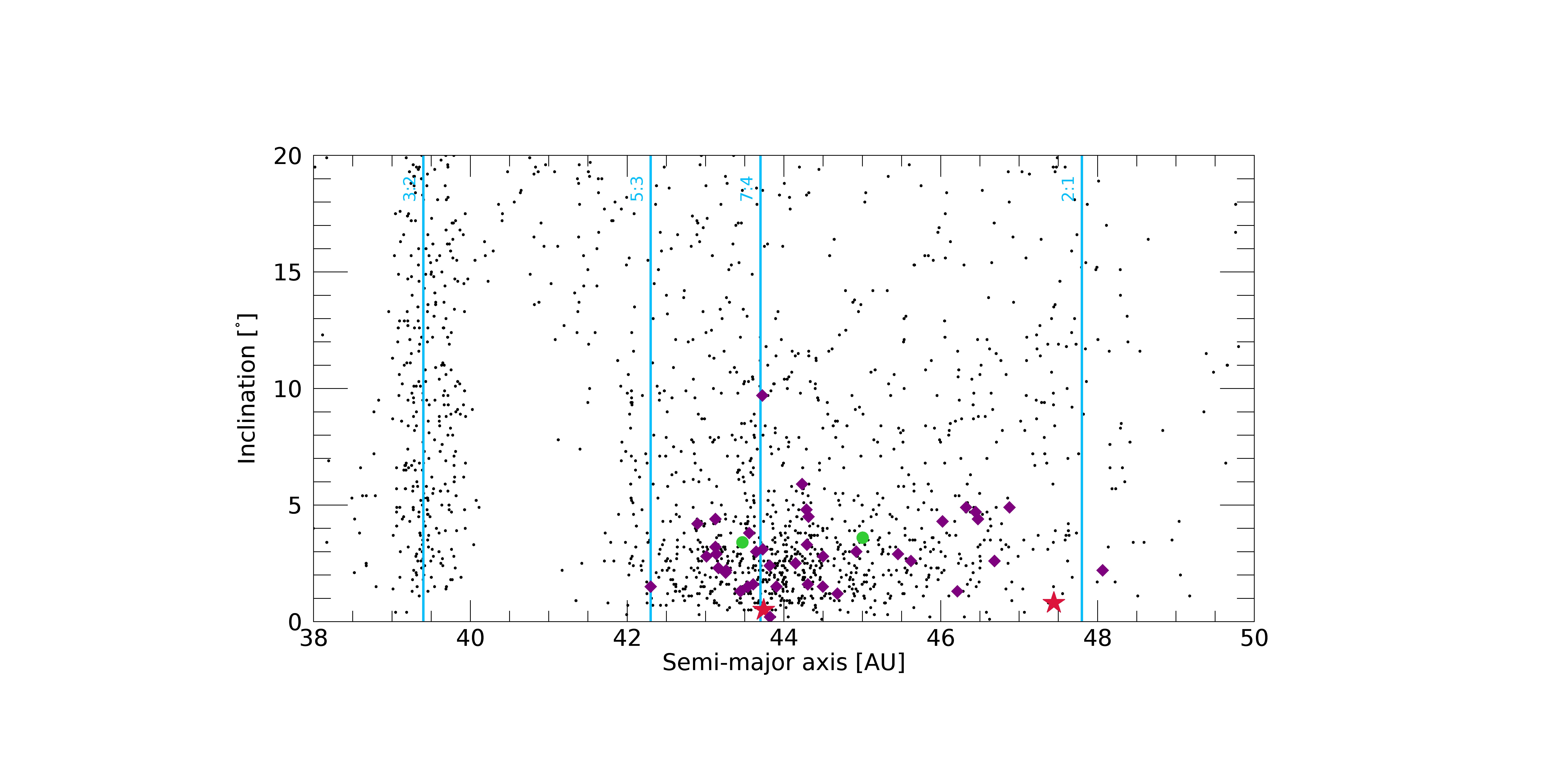}
\includegraphics[width=20cm, angle=0]{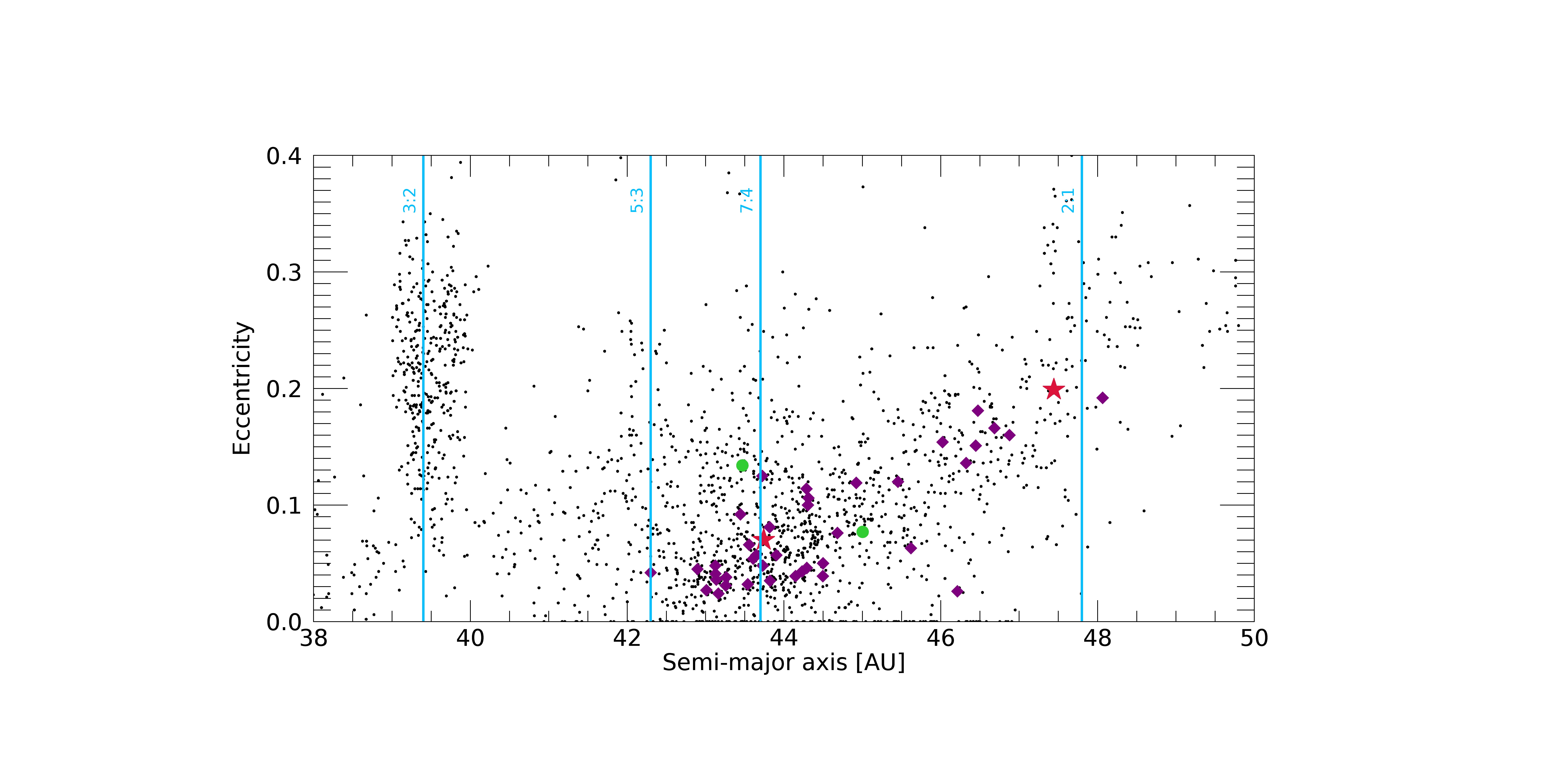}
 \caption{Black dots are all the TNOs discovered with semi-major axes from 38 to 50~AU. Purple diamonds are the CCs with partial/complete lightcurves reported in this work, red stars are the two likely contact binaries with full lightcurves: 2002~CC$_{249}$ \citep{ThirouinSheppard2017}, and 2004~VC$_{131}$ (this work), green circles indicates two candidates for likely contact binaries: 2004~MU$_{8}$, and 2004~VU$_{75}$ (this work). Blue continuous lines are the 3:2, 5:3, 7:4 and 2:1 Neptune's resonances. Orbital elements available at the Minor Planet Center (MPC, November 2018).  }
\label{fig:MPC}
\end{figure*}

We carried out a survey dedicated to the CCs with the 6.5~m Magellan-Baade Telescope and the Lowell Observatory's 4.3~m Discovery Channel Telescope (DCT). So far, we compiled a sample of 42 partial/complete rotational lightcurves for CCs with absolute magnitudes from 5 to 7.2~mag (Tables~\ref{ObsLog} and \ref{Summary_photo}). In Figure~\ref{fig:MPC}, all known TNOs with a semi-major axis from 38 to 50~AU, an inclination up to 20$^\circ$, and an eccentricity up to 0.4 are plotted. Objects reported in this work are highlighted with different colors and symbols (see Section~\ref{sec:analysis} for more details). Our targets have an inclination i$\leq$5$^{\circ}$ (cut-off generally used to distinguish Hot from Cold Classicals), except 2004~OQ$_{15}$ with an inclination of 9.7$^{\circ}$ and 2014~GZ$_{53}$ with i=5.9$^{\circ}$. 

We aim to identify the primordial characteristic(s) of this population compared to the rest of the TNOs. We also want to test different size regimes, pushing towards smaller objects as they are underrepresented in the literature, as well as the discovery of potential contact binaries also lacking in the literature. In majority, the resolved binary CCs have been studied in the past and thus we are focusing on the single objects and non-resolved binaries (i.e. contact/close binaries).  

Our survey is designed to test the short- and long-term variability of the CC population. In fact, based on \citet{Sheppard2008, Duffard2009, Thirouin2010, Benecchi2013, Thirouin2014}, TNOs have a mean periodicity of about 8~h but, binary/multiple systems have the tendency to rotate slower than the single objects. Therefore, it is crucial to test a potential variability over a couple of hours (i.e. the short-term variability), and variability over several days (i.e. the long-term variability). Thus, when we observe a TNO for sparse lightcurve, we ideally want to observe it over a night and re-observe it a couple of days up to month after the first run. With such a strategy, we can evaluate the likelihood of a short and/or long rotational period. This strategy highly depends of the observing schedule and the weather conditions, and thus in some occasions, we can only test the short-term variability.    

Our survey strategy is to observe a large sample of CCs for partial lightcurves which allow us to constrain the rotational period and the variability. These lightcurves are crucial to identify interesting targets with a large amplitude, typically $>$0.4~mag, but also are crucial once we have to calculate the contact binary fraction, the fraction of spherical or elongated objects, and thus have a distribution of shapes for the CC population. If an object is showing a large amplitude (larger than 0.4~mag) over a few hours or days, we will attempt to get its full lightcurve. But, if an object is not showing a significant variability ($<$0.2~mag), we will not obtain its full lightcurve. For objects with a moderate amplitude (around 0.3~mag), we will also schedule them for full lightcurve, but they have a lower priority than the large amplitude objects, and thus may not be re-observed based on the amount of observing time available to us. Our cut-off at 0.4~mag allows us to favor objects with an elongated shape and potential contact binaries which are our highest priorities. With such a strategy, we may miss some very slow rotators with a large amplitude, and to avoid this we try to re-observe our targets a couple of days/months after the first observations (if the observing schedule and weather allow us to do so). Unfortunately, slow rotators (with large or small amplitude) are difficult to detect for the ground mainly because of the large amount of observing time needed as well as the 24~h aliasing effect. Therefore our survey and all other ground-based surveys are biased against slow rotators. So far, the rotational periods of the likely/confirmed contact binaries are between $\sim$6 and $\sim$16~h with the exception of one with a period of $\sim$35~h \citep{Thirouin2018}. Because we are trying to have at least an observing block of about 5~h per object, we are able to cover a reasonable amount of the object's rotation and thus identify any object of interest. In the case of 2014~JL$_{80}$ with a period of about 35~h, two nights were needed to confirm the large amplitude. The fact that our team was able to identify 2014~JL$_{80}$ as an object of interest demonstrates that our strategy is adequate for our purpose \citep{Thirouin2018}.     

Our two main facilities are the Magellan-Baade and the Lowell's Discovery Channel telescopes. At Las Campanas Observatory (Chile), the 6.5~m Magellan-Baade telescope is equipped with IMACS (Inamori-Magellan Areal Camera \& Spectrograph). This instrument is a wide-field imager with a 27.4$\arcmin$ diameter field (8 CCDs), and a pixel scale of 0.20$\arcsec$/pixel. The short camera mode was used for all our runs. The Lowell's DCT (Happy Jack, Arizona) is equipped with the Large Monolithic Imager (LMI), a 6144$\times$6160 pixels CCD \citep{Levine2012}. The field of view is 12.5$\arcmin$$\times$12.5$\arcmin$, and 0.12$\arcsec$/pixel is the pixel scale. 

We use a range of exposure times between 250 and 900~seconds, depending on the telescope, the weather conditions, and the filter. Generally, broadband filters\footnote{Transmission curve are at \url{http://www2.lowell.edu/rsch/LMI/specs.html}, and \url{http://www.lco.cl/telescopes-information/magellan/operations-homepage/instruments/IMACS/imacs-filters/imacs-filters-1}} are selected to maximize the signal-to-noise ratio of the TNO (VR filter at DCT and WB4800-7800 filter at Magellan). Both filters cover the 500-800~nm range. Observing details are in Table~\ref{ObsLog}. We applied our usual data calibration, reduction and analysis \citep{Thirouin2018, Thirouin2010}. The main steps are: i) use the bias and dome or twilight flats obtained every night for calibration, ii) select the optimal aperture radius with the growth curve technique \citep{Howell1989}, iii) perform the aperture photometry using the DAOPHOT routines with the optimal aperture \citep{Stetson1987}, and iv) search for periodicity using the Lomb periodogram technique and double-check the result with the Phase Dispersion Minimization \citep{Lomb1976, Stellingwerf1978}.

%%%%%%%%%%%%%%%%%%%%%%%%%%%%%%%%%%%%%%%%%%%%
%%%% Photometric results
%%%%%%%%%%%%%%%%%%%%%%%%%%%%%%%%%%%%%%%%%%%%

\section{Photometric results}
\label{sec:lightcurve}  

Following, we present partial/complete lightcurves for 42 CCs. Partial lightcurves are plotted in the Appendix~A whereas our photometry is compiled in the Appendix~B. We divide our sample as follows: i) lightcurves showing a large amplitude, ii) objects with moderate amplitude up to 0.4~mag, iii) lightcurves of wide binaries, and iv) low lightcurve amplitudes with a variability lower than 0.2~mag.

\subsection{Large amplitude Cold Classicals}

Following, we will present three objects with a large lightcurve amplitude, suggesting that they are likely contact binaries.

\paragraph{2004~VC$_{131}$}

\begin{figure*}
 \includegraphics[width=7.5cm, angle=0]{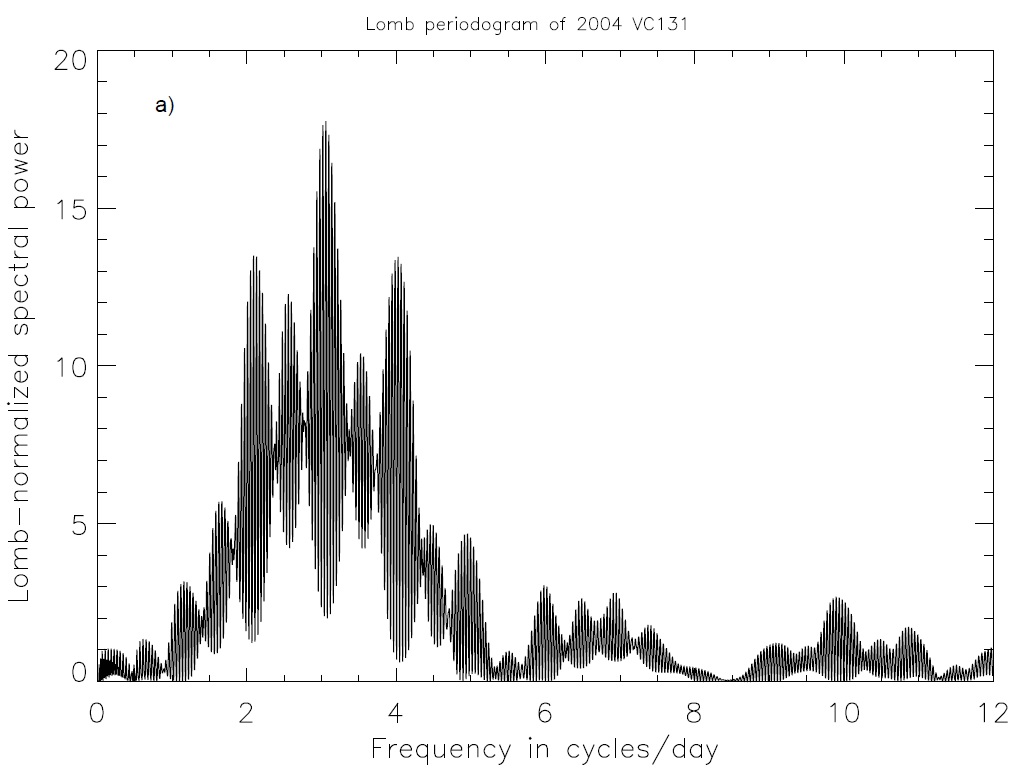}
 \includegraphics[width=8cm, angle=0]{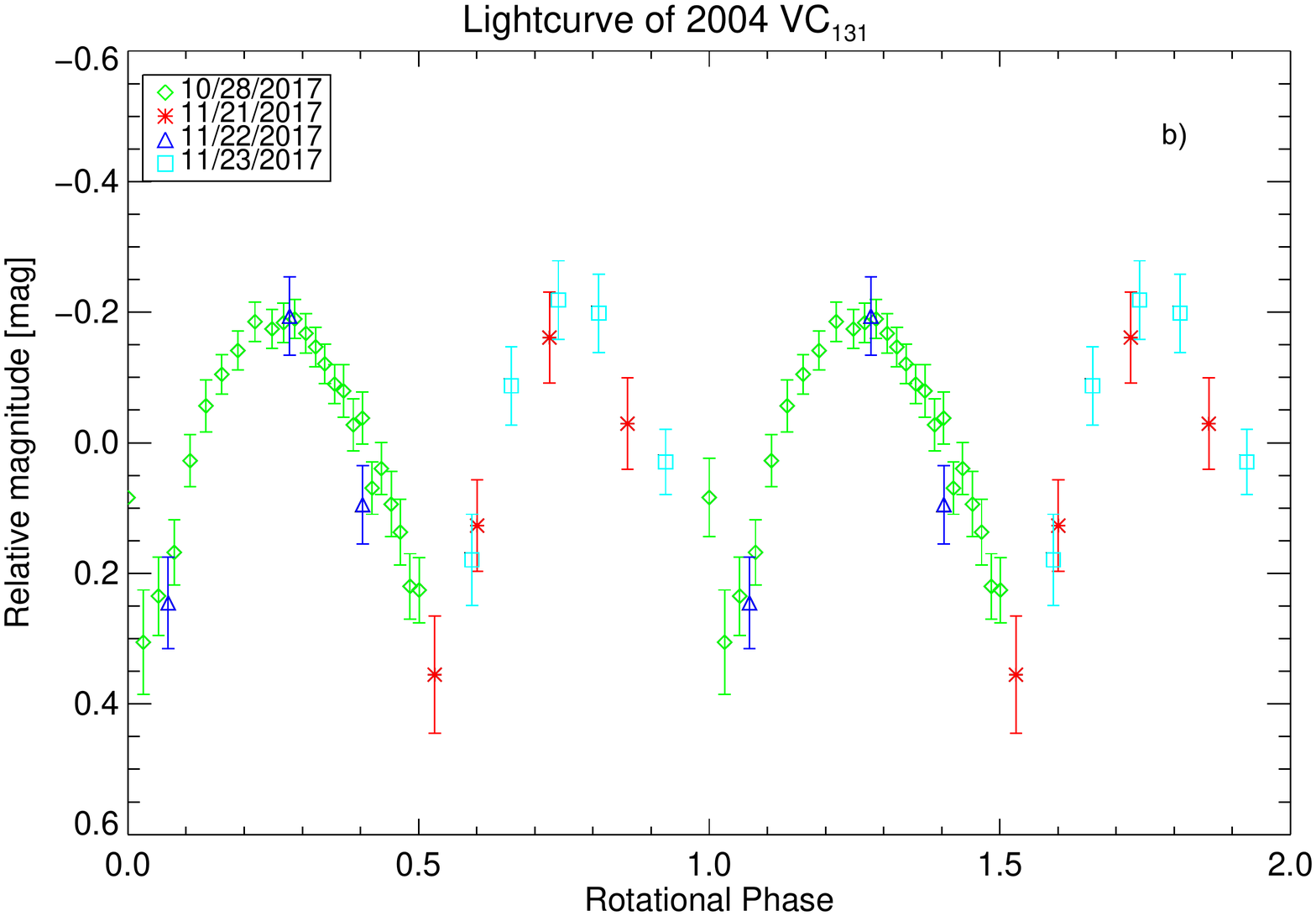}
 \includegraphics[width=8cm, angle=0]{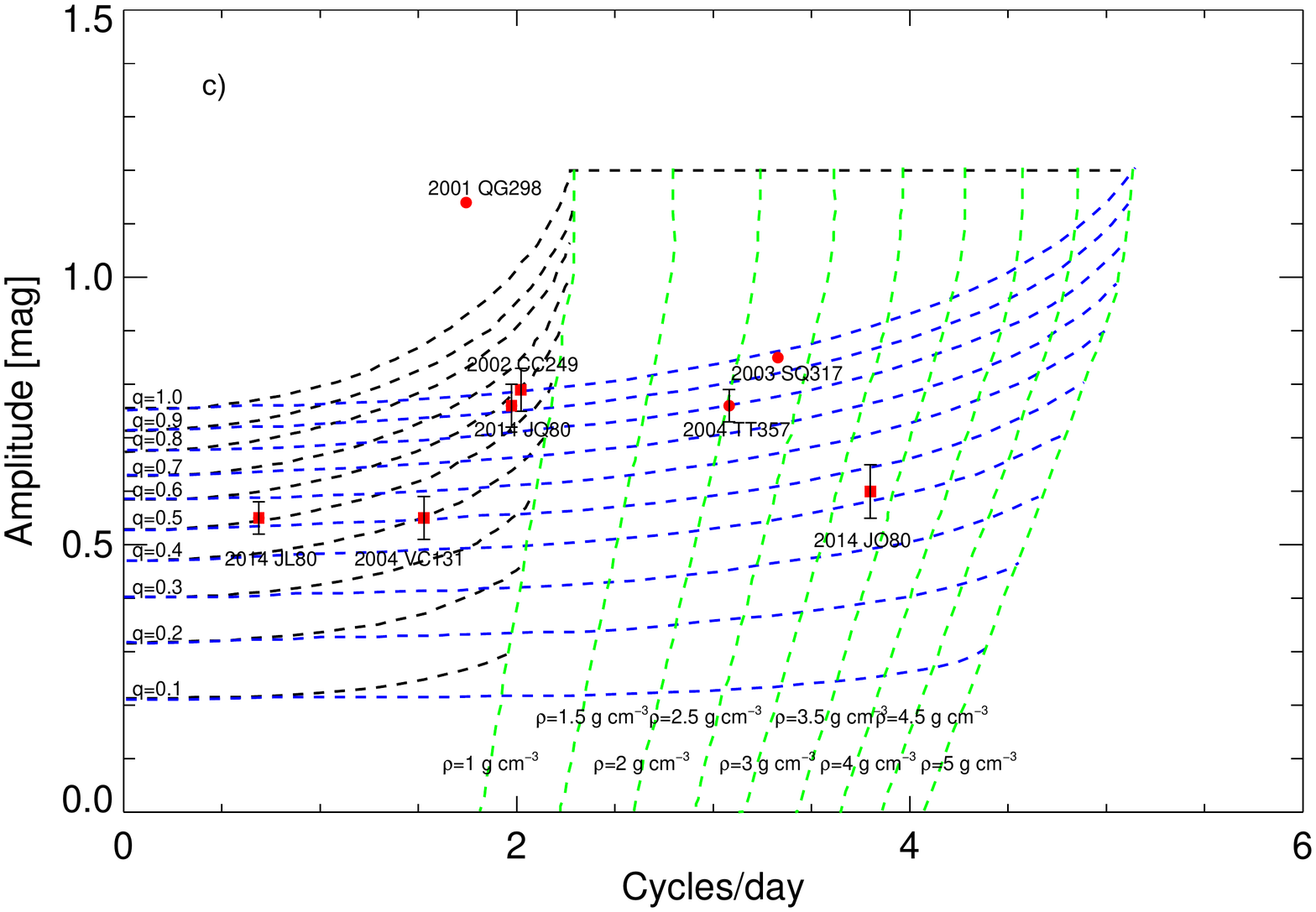}
 \includegraphics[width=8cm, angle=0]{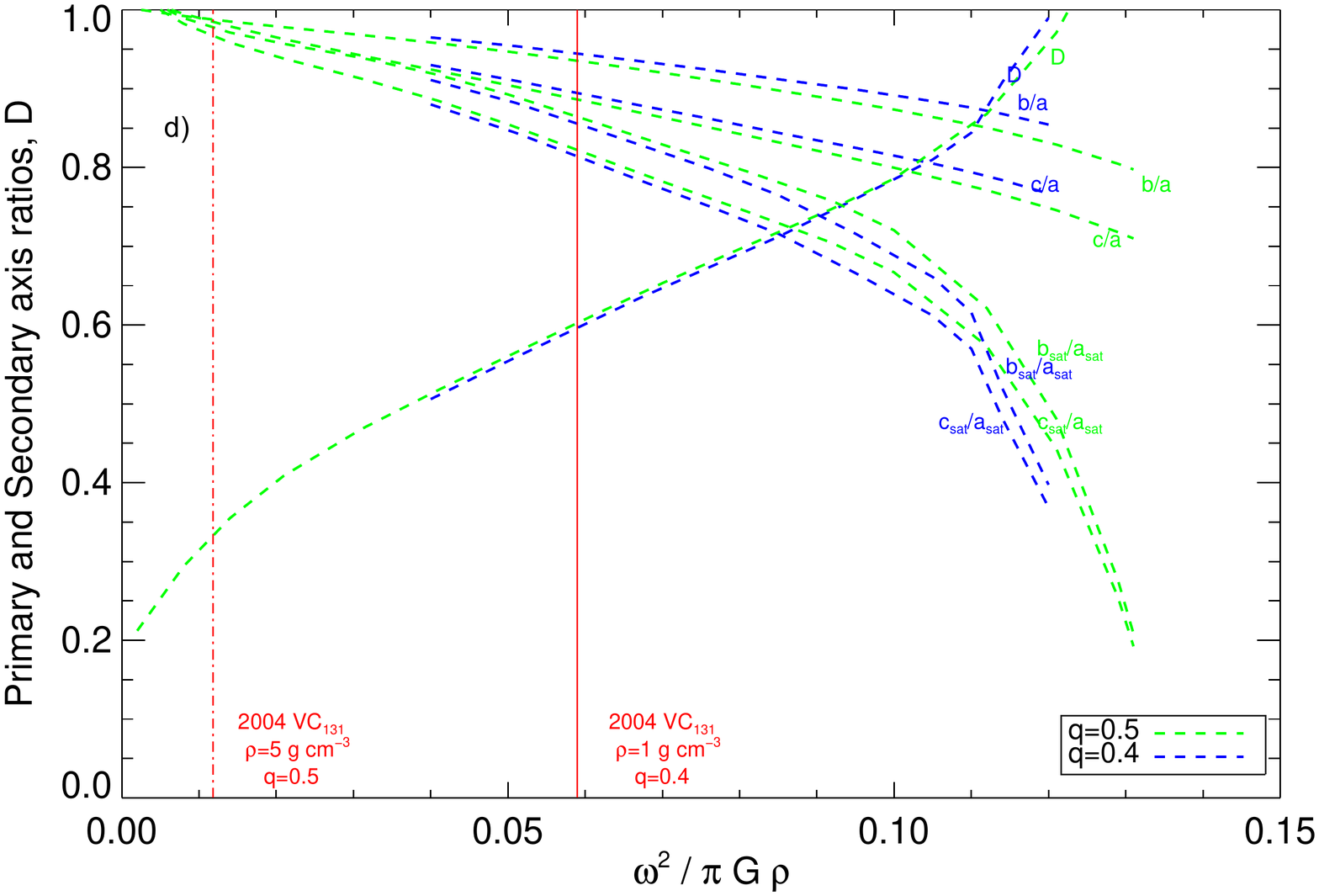}
\caption{Study of 2004~VC$_{131}$: The Lomb periodogram (plot a)) has one main peak suggesting a rotational period of 2$\times$7.85~h (plot b)). The rotational phase of the lightcurve is between 0 and 2, and so two rotations are plotted. With plots c) and d), we calculated the potential mass ratios, size and shape of the components, density, and separation (D) for a contact binary configuration.   }
\label{Fig:VC131}
\end{figure*}

We observed 2004~VC$_{131}$ over about one month in 2017. We report one isolated night in October obtained with the DCT and three consecutive nights in November with the Magellan-Baade telescope. The main peak of the Lomb periodogram is at 3.06~cycles/day or 7.85~h (Figure~\ref{Fig:VC131}). Based on the considerations reported in \citet{Thirouin2017}, we favor a double-peaked lightcurve with a 15.7~h period (Figure~\ref{Fig:VC131}). The peak-to-peak lightcurve amplitude is 0.55$\pm$0.04~mag. This large amplitude can be attributed to a contact/close binary or a triaxial ellipsoid. Following the approach described in \citet{Thirouin2017}, we derive some physical parameters about 2004~VC$_{131}$ considering a Jacobi ellipsoid and a Roche system. 

Assuming a close/contact binary system, we obtain two extreme solutions: i) a system with a mass ratio of q$_{min}$=0.4, density of $\rho_{min}$=1~g~cm$^{−3}$ or ii) a system with a mass ratio of q$_{max}$=0.5, a density of $\rho_{max}$=5~g~cm$^{−3}$. The mass ratio uncertainty is $\pm$0.05. If 2004~VC$_{131}$ has a mass ratio of 0.4, we derive for the primary: b$_{p}$/a$_{p}$=0.94, c$_{p}$/a$_{p}$=0.89 (a$_{p}$=130/58~km, b$_{p}$=122/55~km, and c$_{p}$=116/52~km with an albedo of 0.04/0.2), the secondary axis ratios: b$_{s}$/a$_{s}$=0.86, c$_{s}$/a$_{s}$=0.81 (a$_{s}$=102/46~km, b$_{s}$=88/40~km, and c$_{s}$=83/37~km, albedo of 0.04/0.2). With D=0.6, the separation between the components is 387/174~km for an albedo of 0.04/0.2. This study is summarized in Figure~\ref{Fig:VC131}. As a density of 5~g~cm$^{−3}$ is not likely in the Kuiper Belt, we do not consider this option here, but basic parameters derived assuming such a density are available in Figure~\ref{Fig:VC131} \citep{ThirouinSheppard2017}.

In the case of a Jacobi ellipsoid and considering an equatorial view, the object's elongation is a/b=1.66, and c/a=0.44 \citep{Chandrasekhar1987}. We find: a=353~km (a=158~km), b=213~km (b=95~km), and c=155~km (c=70~km) for an albedo of 0.04 (0.20). The viewing angle of 2004~VC$_{131}$ has to be larger than 62.5$^\circ$ to avoid an axis ratio a/b$>$2.31 \citep{Jeans1919}. Considering an equatorial view, the density is $\rho$$\geq$0.17~g~cm$^{−3}$ \citep{Chandrasekhar1987}.

\paragraph{2004~VU$_{75}$}
 
\begin{figure*}
   \includegraphics[width=8cm, angle=0]{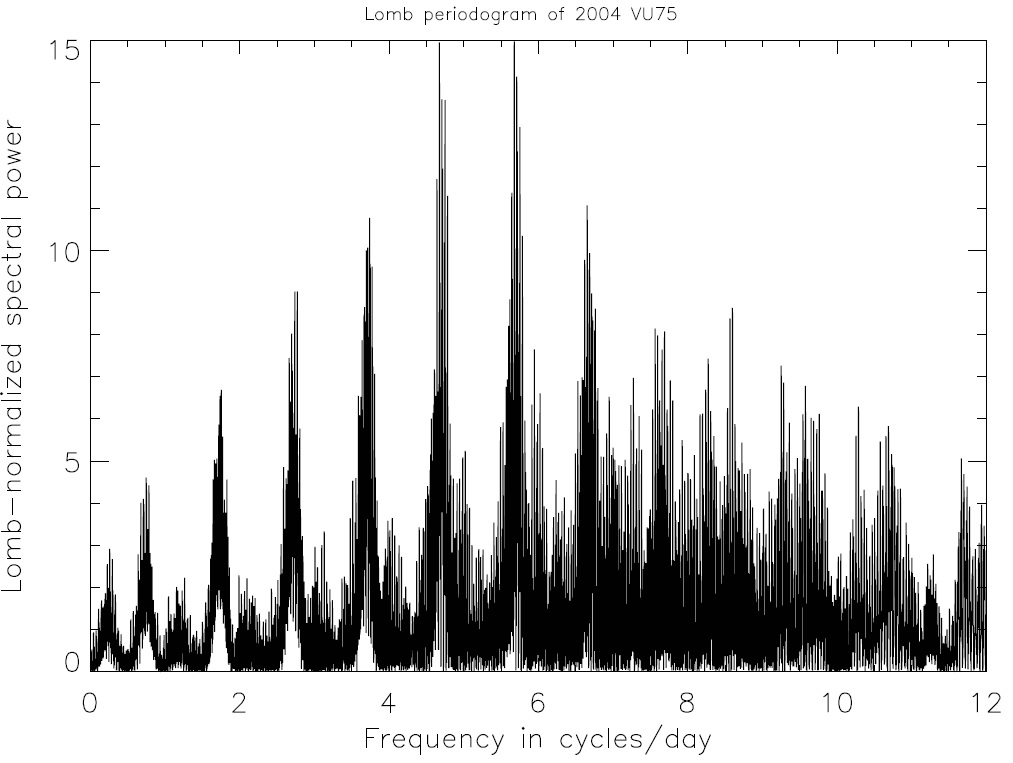} 
      \includegraphics[width=9cm, angle=0]{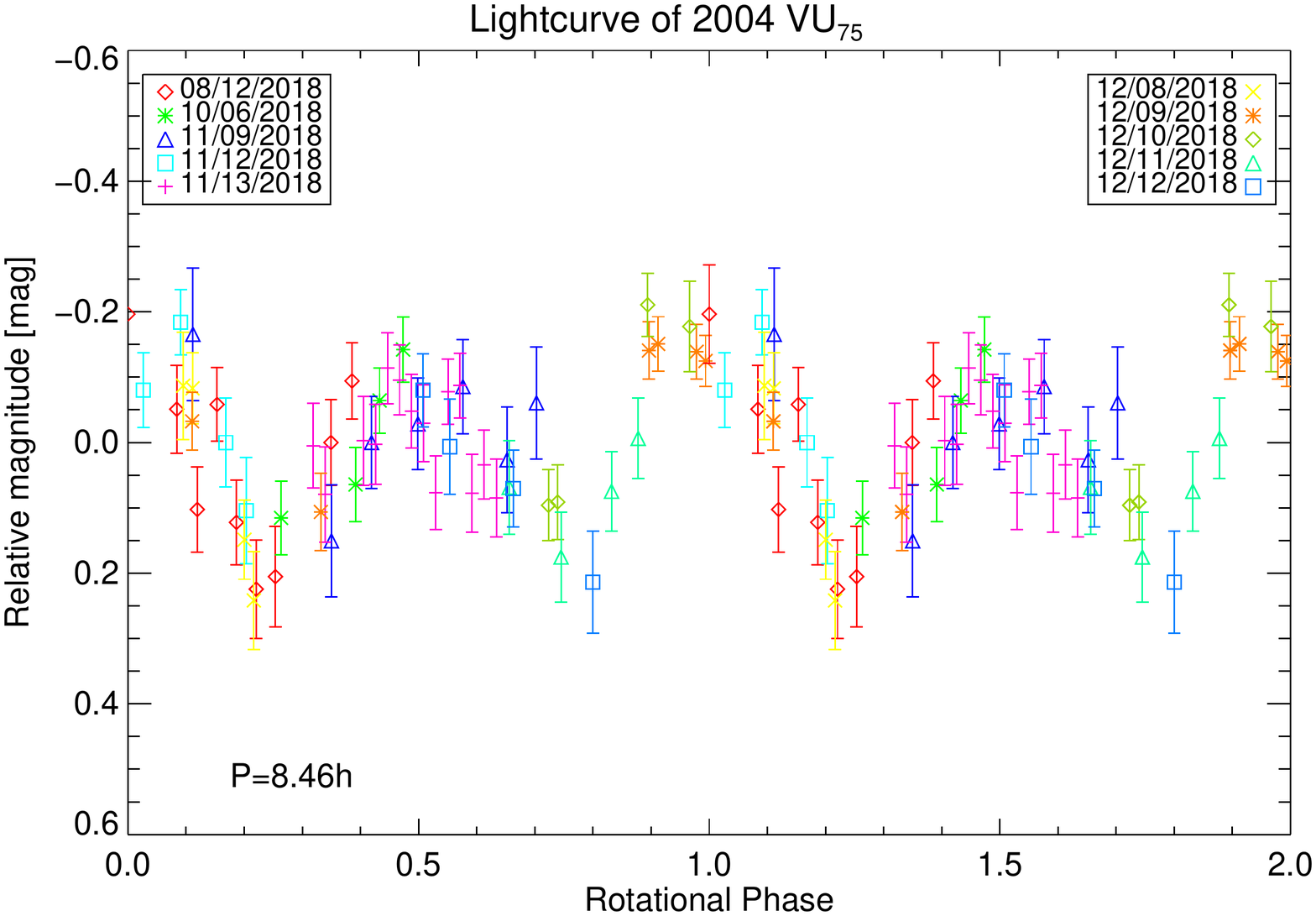}              	      \includegraphics[width=9cm, angle=0]{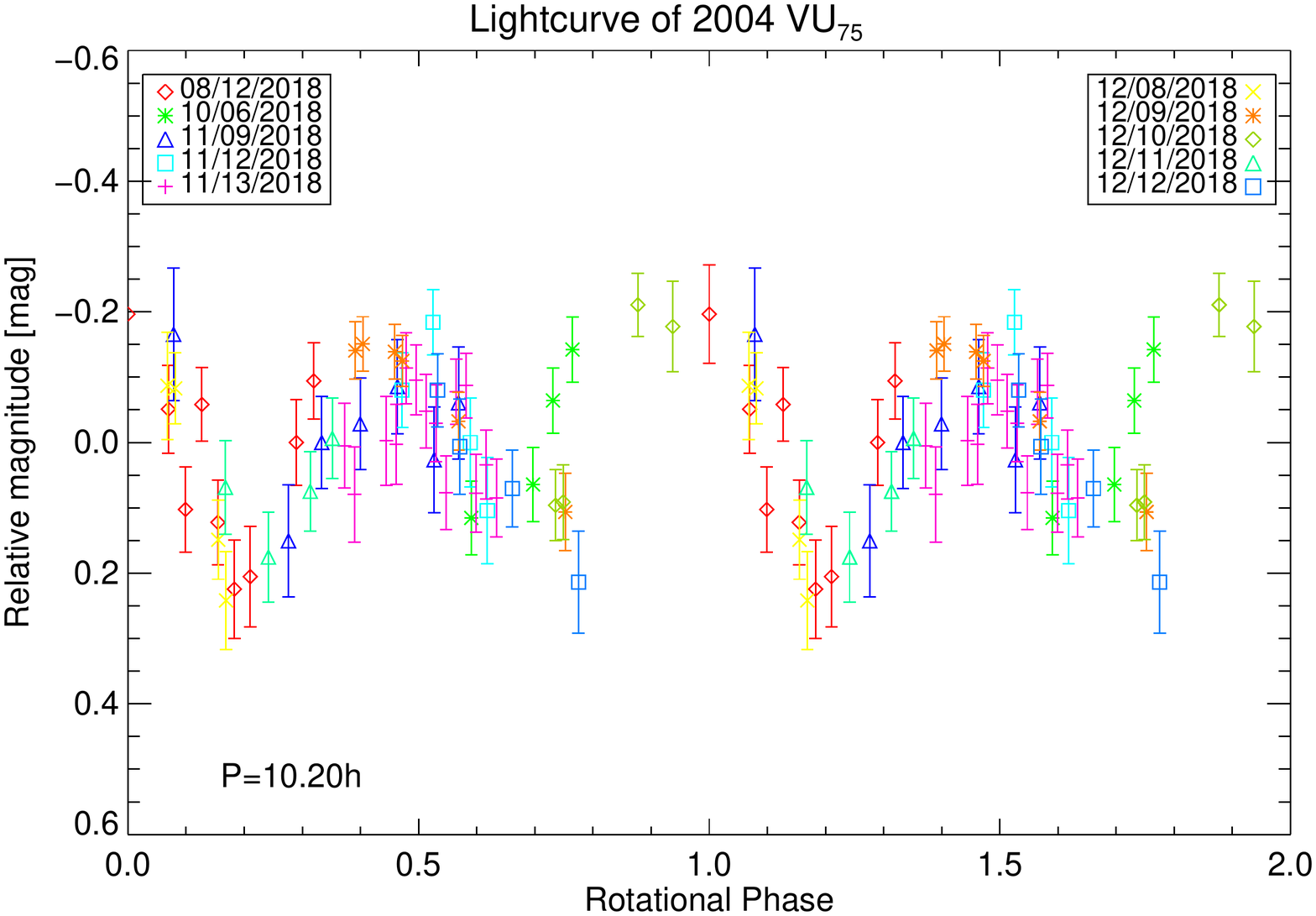}     
  \includegraphics[width=9cm, angle=0]{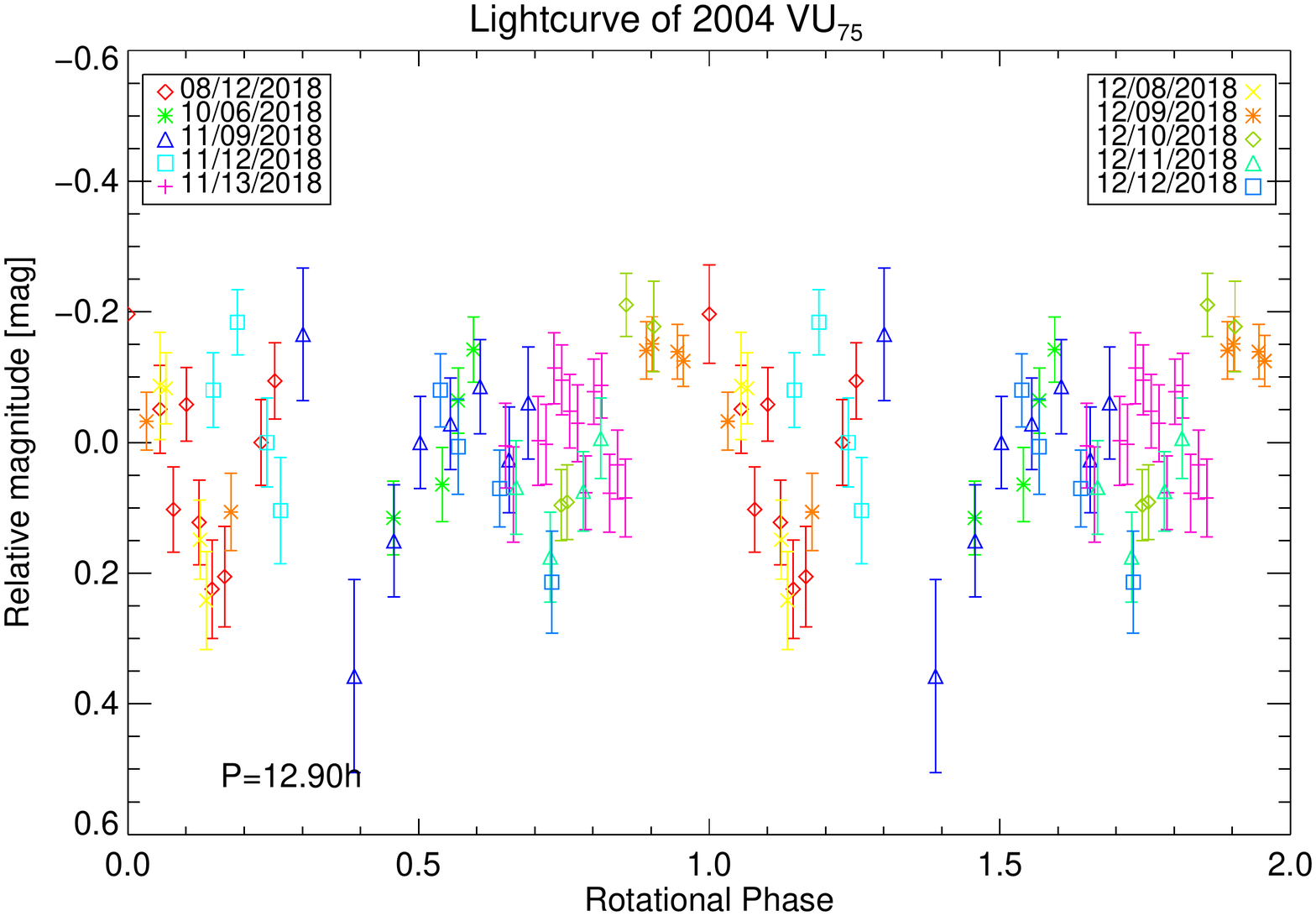}     
\caption{Lomb periodogram using all our data favors a single-peaked lightcurve rotational period of 4.23~h or 5.13~h for 2004~VU$_{75}$, which would be a double-peaked rotational period of 8.46~h or 10.2~h. Using only our Magellan data, Lomb periodogram favors a single-peaked lightcurve with P=6.45~h (double-peaked of 12.9~h). The best fit is obtained assuming a rotation of 8.46~h (upper right plot). }
\label{fig:VU75}
\end{figure*}

We observed 2004~VU$_{75}$ on several occasions with the DCT. Our first partial lightcurve was obtained in August 2018 over approximately 3~h for a variability of 0.42~mag. Our August data corresponds to the minimum of the curve which seems to be a sharp minimum with a V-shape. We re-observed this object in November over 3 non-consecutive nights. Unfortunately due to bad weather, we can only report fragmentary datasets (Figure~\ref{fig:VU75}). In December 2018, we re-observed 2004~VU$_{75}$ over 5 consecutive nights with the Magellan-Baade telescope. We performed a search for rotational period using all our data or only the high quality data and found three potential double-peaked periodicities of 12.9~h, 10.2~h and 8.4~h (Figure~\ref{fig:VU75}). We favor the double-peaked based on the large amplitude. With a range of 8 to 13~h, this object seems to have a moderate rotational period. Based on the large variability and the sharp minimum of our first lightcurve, we have some hints that 2004~VU$_{75}$ is maybe a contact binary. Therefore, more observations to confirm the nature of this object are highly desirable.  
 
\paragraph{2004~MU$_{8}$}

\begin{figure}
   \includegraphics[width=9cm, angle=0]{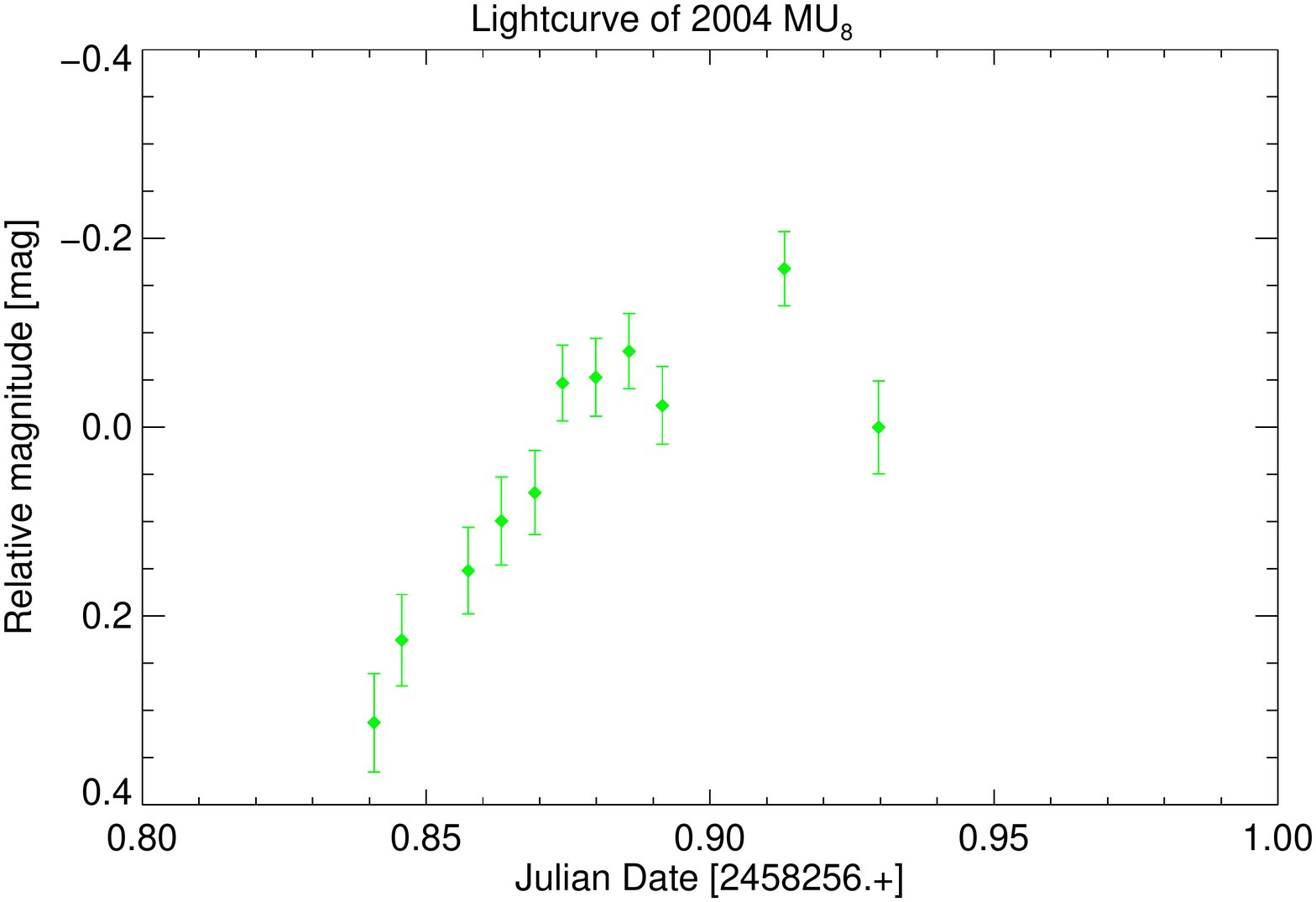}
      \includegraphics[width=9cm, angle=0]{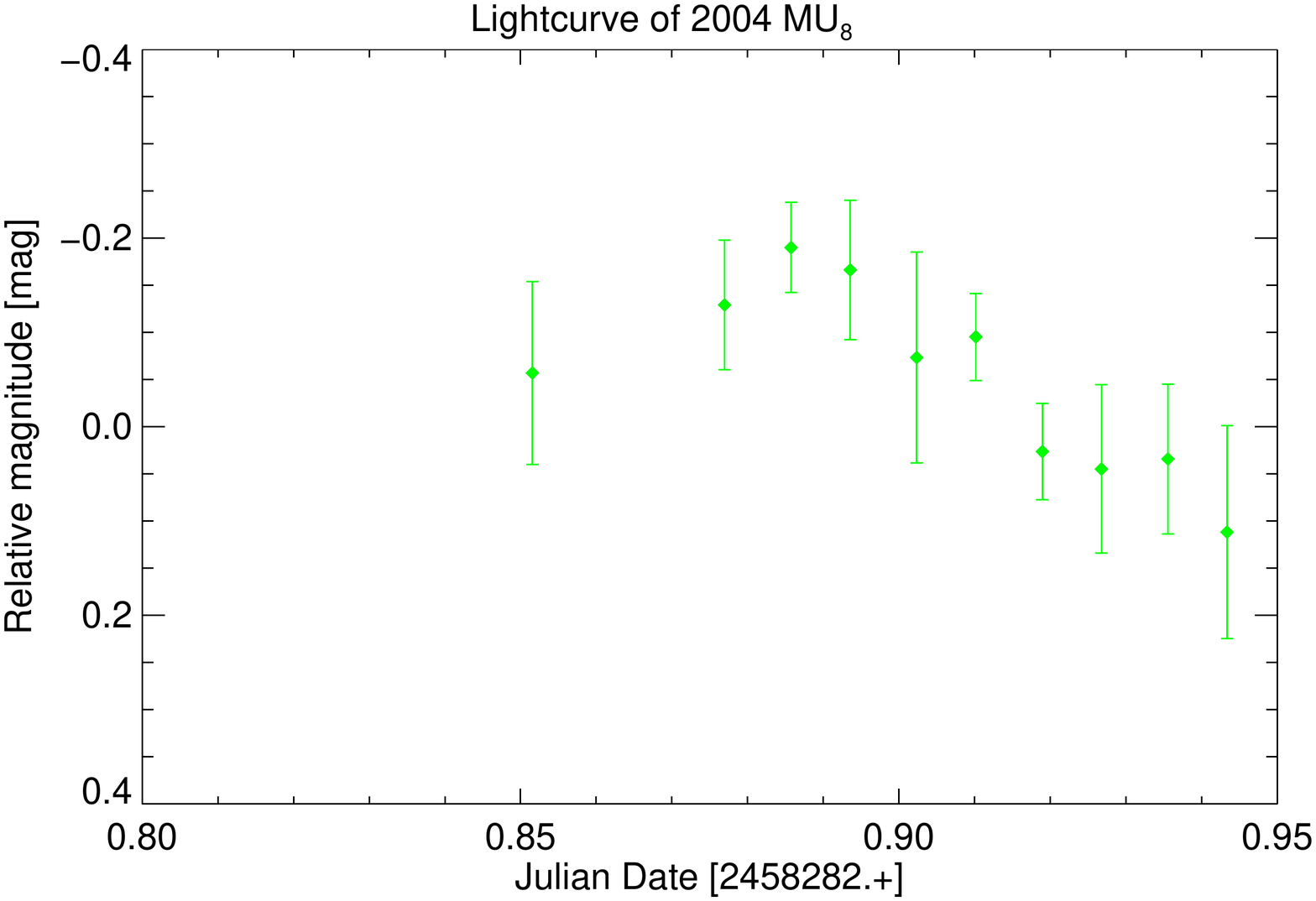}
\caption{ Partial lightcurves obtained with the Magellan Telescope (upper plot), and with the DCT (lower plot). In both cases, 2004~MU$_{8}$ displays a large amplitude in a few hours and thus it is a good candidate for likely contact binary. More data are required to obtain the full lightcurve and infer the nature of the object/system.  }
\label{fig:MU8}
\end{figure}

We observed 2004~MU$_{8}$ on two occasions with the Magellan Telescope in May 2018 and with the DCT in June 2018 Figure~\ref{fig:MU8}. Over about 2.5~h at DCT and $\sim$2h at Magellan, 2004~MU$_{8}$ displayed a variability of 0.33~mag and 0.48~mag, respectively. Unfortunately, we do not have enough data to cover the full rotation of this object and so derive its periodicity. We can only infer that the period is larger than 2.5~h and the amplitude is larger than 0.48~mag. Based on the large amplitude over a short amount of time, 2004~MU$_{8}$ is a good candidate to the likely contact binary category. There is no information in the literature about a search for satellites orbiting 2004~MU$_{8}$. \\
 
 \subsection{Moderate amplitude Cold Classicals}
 
This subsection is dedicated to the CCs displaying a moderate lightcurve amplitude. Objects with a variability above 0.3~mag are potentially interesting targets for follow-up observations at different epochs to look for changes. 

\paragraph{2014~LS$_{28}$}

\begin{figure}
  \includegraphics[width=9cm, angle=0]{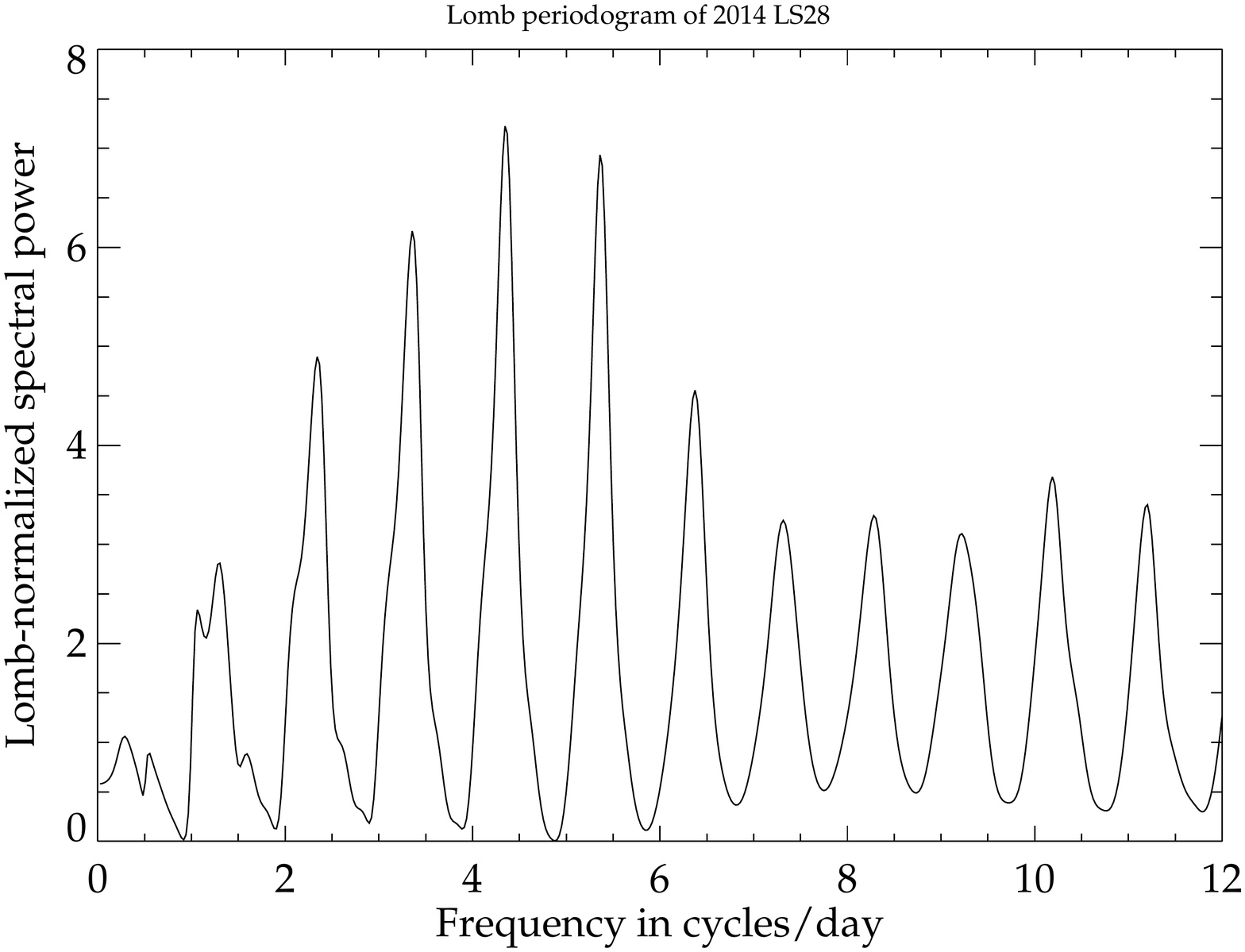}
  \includegraphics[width=9cm, angle=0]{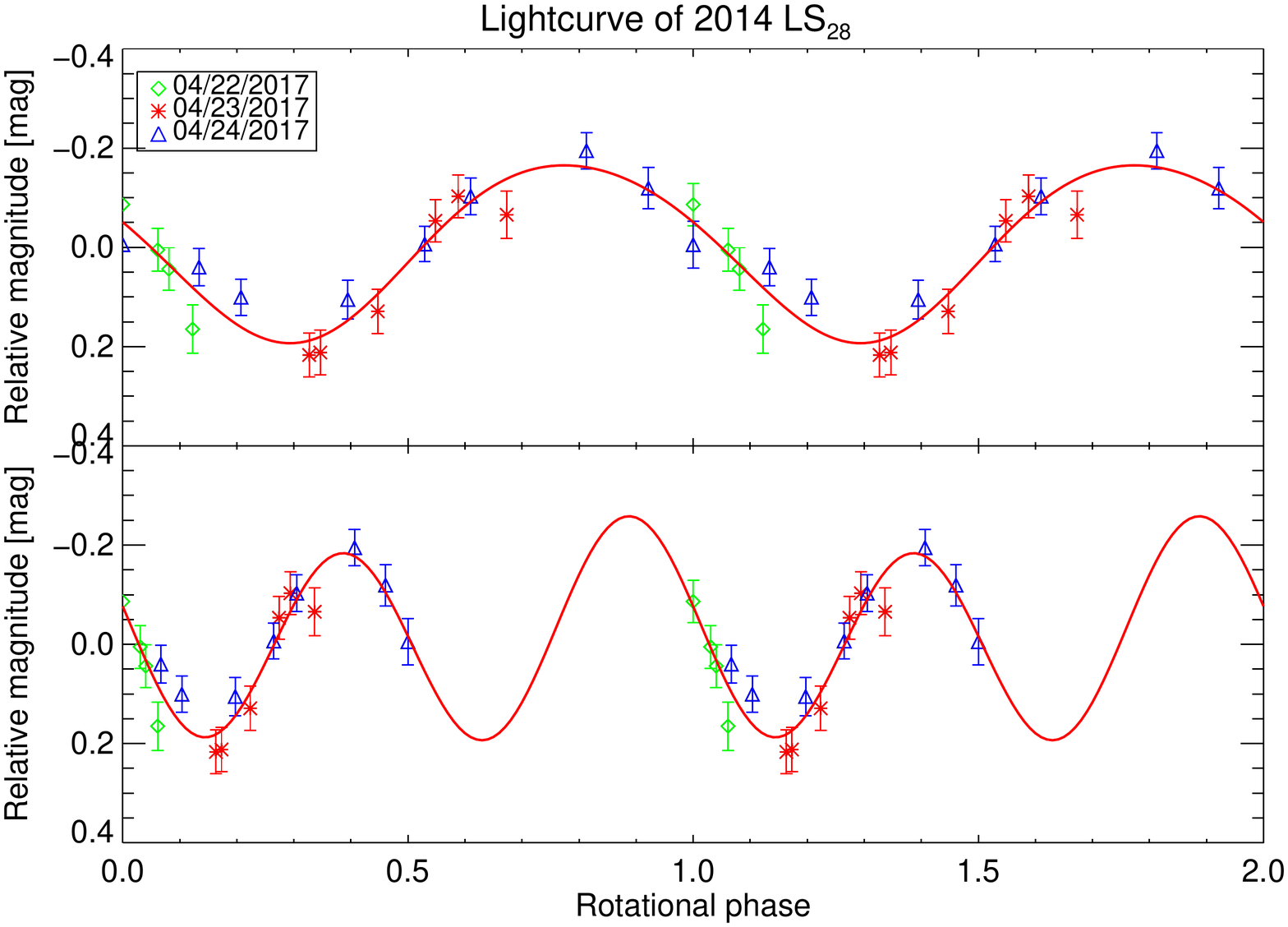}
\caption{The Lomb periodogram for 2014~LS$_{28}$ has one main peak at 4.37~cycles/day and several aliases. The lower plot are the single- and double-peaked lightcurves assuming the main peak as rotational period. }
\label{fig:LS28}
\end{figure}

From three consecutive nights of observations with the Magellan-Baade telescope in April 2017, we estimate that the periodicity of 2014~LS$_{28}$ is 11.04~h,, and the amplitude is 0.35$\pm$0.03~mag. The single-/double-peaked lightcurves with rotational periods of 5.52~h/11.04~h, respectively, are plotted in Figure~\ref{fig:LS28}. The Lomb periodogram presents several aliases of the main peak (Figure~\ref{fig:LS28}). To our knowledge, there was no satellite search for 2014~LS$_{28}$. 

Assuming that 2014~LS$_{28}$ is a triaxial object with a$>$b$>$c and an equatorial view, we derive: b/a=0.72, c/a=0.49. The density is $\rho$$\geq$0.33~g~cm$^{-3}$. Detailed procedure regarding the estimate of these values is available in \citet{Thirouin2010, Thirouin2012}.

\paragraph{(444025) 2004~HJ$_{79}$}

Based on three consecutive nights, 2004~HJ$_{79}$ shows a possible steep slope suggesting an amplitude larger than 0.20~mag. However, the rest of the curve is mostly flat. We re-observed this object in February and March 2019. The lightcurve displays similar behavior. As the amplitude is about 0.20~mag over consecutive nights, we use this value in this study. We infer that the periodicity is larger than 7.5~h. No resolved binary search is available for this object. 

\paragraph{2010~TF$_{192}$}

Over $\sim$2.5~h of observations, 2010~TF$_{192}$ has a variability of about 0.3~mag. Our partial lightcurve displays a minimum and a maximum suggesting that the amplitude of the full lightcurve is likely not much greater than 0.3~mag. There is no information in the literature about a satellite search for this object.  

\paragraph{2010~TL$_{182}$}

In about 5.5~h, 2010~TL$_{182}$ displays a variability of about 0.25~mag. To our knowledge, no search for companion was performed for this object. 

\paragraph{2014~LR$_{28}$}

In a little less than 4~h, 2014~LR$_{28}$ has a variability of about 0.25~mag. 2014~LR$_{28}$ is the largest object in our sample, and one of the largest objects in the CC population. As we will discussed in the Section~\ref{sec:analysis}, all the large TNOs are resolved wide binaries \citep{Noll2014}. However, no search for satellite has been done for this object. Therefore, it would be interesting to confirm the presence of a companion and infer if the amplitude is due to an asynchronous state.  

\subsection{Wide binary Cold Classicals}

Resolved wide binaries are not the main topic of our survey, but we observed Logos-Zoe to confirm a potential large variability and discovered a new wide binary, 2014~LQ$_{28}$. 

\paragraph{(58534) 1997~CQ$_{29}$, Logos-Zoe}

Logos was discovered in 1997 from Maunea Kea and its satellite, Zoe was identified in \textit{Hubble Space Telescope (HST)} images by \citet{Noll2002}. Zoe's orbit is well characterized with an orbital period of 309.9~days, a semi-major axis of 8220~km, an eccentricity of 0.55, and an inclination of 95.4$^\circ$ \citep{Grundy2011, Noll2004}. Logos has an estimated diameter of 80~km and Zoe is a 66~km diameter object \citep{Noll2004}. \citet{Noll2008} reported a high variability for Logos based on \textit{HST} data, however due to the sparse sampling there is no constraint for the rotational period. In March 2017, we observed Logos-Zoe for about 1~h and confirm a large variability with $\Delta m$$>$0.5~mag. 

\paragraph{2014~LQ$_{28}$}

We observed 2014~LQ$_{28}$ during 5.5~h with the Magellan-Baade telescope in September 2016. During our observing run, the seeing was 0.4$\arcsec$ allowing the discovery of a companion orbiting this object. The magnitude difference between the two components is about 0.4~mag and their separation was 0.86$\arcsec$ \citep{Sheppard2018}. The system was re-observed in October 2017 to confirm the binarity and this time the separation was 0.36$\arcsec$. The components variability are about the same of $\sim$0.1~mag over 5.5~h.   

\subsection{Low amplitude Cold Classicals}

Finally, we report objects with a low lightcurve amplitude. If an object shows signs of variability ($\Delta m$$>$0.15~mag), we will use the duration of our observing as lower limit for the rotational period and the variability as lower limit of the lightcurve amplitude. If the ligthcurves are too noisy and/or flat ($\Delta m$$<$0.15~mag), we will report an approximate amplitude. Such low amplitude lightcurves can be attributed to a nearly spherical object with limited albedo variations on its surface, a pole-on oriented object, or a very slow rotator with a variability (and periodicity) undetectable over the duration of our observing block \citep{Sheppard2008, Thirouin2014}. Therefore, the duration of our observing block cannot constrain the object's periodicity as we do not know the reason of the flat lightcurve. Only if the object is a slow rotator, our observing block duration can be a lower limit.   

\paragraph{2000~CL$_{104}$}

In March 2016, we observed 2000~CL$_{104}$ during two consecutive nights with the Magellan-Baade telescope. The duration of the observing blocks are about 5.5~h and $\sim$3.2~h, respectively. Both nights show a similar variability of about 0.2~mag. We can infer that the rotation is probably longer than 5.5~h. To our knowledge there is no other published lightcurve for 2000~CL$_{104}$. No companion was found orbiting this object \citep{Noll2008b}. 

\paragraph{(138537) 2000~OK$_{67}$}

Between July and September 2015, four nights were dedicated to the observations of 2000~OK$_{67}$ with the DCT. This object has a low variability of about 0.15~mag, consistent over all observing nights. Our longest observing block is $\sim$6~h. \citet{Noll2008b} reported no companion for 2000~OK$_{67}$.  

\paragraph{2000~OU$_{69}$}

In August 2015, we observed 2000~OU$_{69}$ for about 2.5~h. The lightcurve presents a low variability of about 0.15~mag. To our knowledge, we report here the first lightcurve for this object. According to \citet{Noll2008b}, 2000~OU$_{69}$ has no satellite.

\paragraph{2001~QS$_{322}$}

With an observing block of about 6~h, we report a partial lightcurve with a variability of  about 0.15~mag for 2001~QS$_{322}$. No companion search was performed for this object.  
 
\paragraph{(363330) 2002~PQ$_{145}$}

We dedicated three observing nights to 2002~PQ$_{145}$ between August and September 2015. We report a very low variability lightcurve with $\Delta m$$\sim$0.1~mag. Our longest observing run is about 5.5~h. \citet{Noll2008b} found no evidence for a companion orbiting 2002~PQ$_{145}$. 

\paragraph{(149348) 2002~VS$_{130}$}

With three observing nights in December 2015 with the DCT, we are not able to derive the periodicity. Based on our longest observing run of 3.5~h, the lightcurve amplitude is $\sim$0.1~mag. To our knowledge, no companion search was performed for 2002~VS$_{130}$. 

\paragraph{2003~QE$_{112}$}

With one night of data from the Magellan-Baade telescope, we report a low variability of about 0.1~mag over the duration of our observing block which is $\sim$5~h. To our knowledge, no observation to search for satellite has been performed for this object. 

\paragraph{2003~QJ$_{91}$}

We present 7 images of 2003~QJ$_{91}$ obtained over approximately 6~h. This object shows a variability of about 0.2~mag. We report here the first variability measurements for 2003~QJ$_{91}$. No search for satellite has been published. 

\paragraph{2003~QY$_{111}$}

\citet{SantosSanz2009} observed 2003~QY$_{111}$ with the Very Large Telescope in November 2003 for a \textit{BVRI} color study. They report a potential short-term variability of 0.72~mag over 156~min (see Table~3 in \citet{SantosSanz2009}). To confirm such a large variability, we re-observed this object with the DCT in October 2017. After 5.5~h of observations, we report a variability of about 0.2~mag. Therefore, the large amplitude noticed by \citet{SantosSanz2009} is unconfirmed. Assuming that this amplitude noticed in 2003 is correct, one can argue that the spin axis orientation of 2003~QY$_{111}$ changed significantly to see a different lightcurve amplitude in 2017. Similarly, if 2003~QY$_{111}$ is a close binary, the system's configuration would have changed significantly in about 14~years \citep{Lacerda2011}. However, \citet{SantosSanz2009} focused on colors and thus their images were not the best suited for lightcurve analysis. To our knowledge, 2003~QY$_{111}$ was not observed for a resolved companion.        

\paragraph{2003~SN$_{317}$}

2003~SN$_{317}$ was observed for less than 1~h with the DCT in August 2015. The partial lightcurve has an amplitude of about 0.1~mag. No search for satellite or other lightcurve have been reported in the literature for this object.   

\paragraph{2003~YU$_{179}$}

Based on 5.5~h of data obtained in February 2016, we cannot derive a secure rotational period for 2003~YU$_{179}$. The partial lightcurve displays an amplitude of about 0.2~mag. There is no report in the literature about a search for resolved companion orbiting this object. 

\paragraph{2004~EU$_{95}$}

Images of 2004~EU$_{95}$ were obtained over two consecutive nights with the Magellan telescope. With 8~h of data, the amplitude is about 0.1~mag. No literature available about a search for resolved binary. 

\paragraph{2004~HD$_{79}$}

We report two consecutive nights with the Magellan-Baade telescope in April 2017. Our longest observing block is about 7~h. As there is no clear repetition in the photometry, the period is likely larger than 8~h. The amplitude is about 0.15~mag. To our knowledge, 2004~HD$_{79}$ has not been inspected for a resolved companion.  

\paragraph{(469610) 2004~HF$_{79}$}

Based on two consecutive observing nights, we cannot derive a secure period for this object. We constrain the periodicity to be larger than 7.5~h, and the amplitude is about 0.15~mag. To our knowledge, 2004~HF$_{79}$ has not been observed for companion search. 

\paragraph{2004~HP$_{79}$}

We obtained 5 usable images of 2004~HP$_{79}$ with the DCT suggesting a lightcurve variability of $\sim$0.15~mag and a periodicity longer than 3~h. This object has not been observed for companion detection. 

\paragraph{2004~MT$_{8}$}

We observed this object over three consecutive nights during approximately 2~h every night. Each block is showing a low variability, but it seems that 2004~MT$_{8}$ has a variability of about 0.2~mag over the three nights. Based on our short observing blocks, we cannot exclude a short-term variability and thus we can only constrain the periodicity to be longer than 2~h. No satellite search was performed for this object based on the literature available.

\paragraph{2004~OQ$_{15}$}

We observed 2004~OQ$_{15}$ for about 2.5~h with the DCT. This object displays a variability of about 0.1~mag. No search for companion has been performed for this object. 
 
\paragraph{2004~PV$_{117}$}

The variability of 2004~PV$_{117}$ is low, about 0.1~mag over 3~h. No information about a satellite search is reported in the literature. 

\paragraph{2004~PX$_{107}$}

\citet{Noll2008b} searched for a companion orbiting 2004~PX$_{107}$. They report no satellite based on their data. We observed this object in July 2017 with the DCT. Our observing block is about 1.5~h. There is no clear variability based on our data (amplitude $\sim$0.1~mag).  

\paragraph{2004~PY$_{107}$}
 
 In August 2018, we observed 2004~PY$_{107}$ during approximately 2~h, and we report a low amplitude lightcurve of 0.1~mag. This object was not inspected for binarity.

\paragraph{2005~EX$_{297}$}

We observed 2005~EX$_{297}$ during one night in March 2016. Based on our data obtained over $\sim$6~h, the lightcurve variability is only about 0.1~mag. There is no published information about a satellite search for this object.  

\paragraph{2005~JP$_{179}$}

With only 2 images of 2005~JP$_{179}$ obtained in 2018, we report that the variability is larger than 0.08~mag. This amplitude is consistent with our 2019 data. There is no indication in the literature for a binary search. 

\paragraph{2005~PL$_{21}$}

In about 4~h of data obtained with the Magellan-Baade telescope in 2016, we cannot secure a rotational period for 2005~PL$_{21}$. The partial lightcurve reported here presents a low amplitude of about 0.15~mag. This object has not been the topic of a search for companion.

\paragraph{2011~BV$_{163}$}

We report $\sim$2.5~h of data for 2011~BV$_{163}$ obtained with the DCT in February 2017. The partial lightcurve has a variability of about 0.15~mag. We present here the first lightcurve for this object. No search for satellite has been done for 2011~BV$_{163}$.

\paragraph{2012~DA$_{99}$}

In May 2018, we obtained images of 2012~DA$_{99}$ with the Magellan-Baade telescope over two nights. This object shows a low variability ($\sim$0.1~mag) in about 2.2~h. Data obtained in 2019 confirm the low variability. 2012~DA$_{99}$ has not been imaged for resolved companion.   

\paragraph{2012~DZ$_{98}$}

Over about 5~h, 2012~DZ$_{98}$ displays a variability of $\sim$0.2~mag. There is no indication in the literature about a resolved binary search.  

\paragraph{2013~AQ$_{183}$}

From two nights of observations in February and March 2017, we cannot derive a secure rotational period for 2013~AQ$_{183}$. Our longest observational run is about 5~h and thus, we infer that the periodicity is longer than 5~h. Both nights present the same variability, about 0.15~mag. No satellite search has been done for 2013~AQ$_{183}$.

\paragraph{2013~EM$_{149}$}

In more than 7~h of observations, 2013~EM$_{149}$ has a very low variability of only 0.1~mag. No companion search has been performed for this object. 

\paragraph{2013~FA$_{28}$}

Based on our $\sim$2~h of observations for 2013~FA$_{28}$ obtained in February 2017, we report a low variability of about 0.1~mag. Similar variability is noticed in both of our dataset obtained in February-March 2019. To our knowledge, no search for satellite orbiting 2013~FA$_{28}$ was performed.

\paragraph{2014~GZ$_{53}$}

In $\sim$8~h, 2014~GZ$_{53}$ shows a very low variability of about 0.1~mag. This object has not been imaged for satellite search. 
 
\paragraph{2014~OA$_{394}$}

With only four usable images of 2014~OA$_{394}$ obtained over about 3~h, we are not able to estimate a periodicity. The partial lightcurve displays an amplitude of about 0.15~mag. There is no information available regarding a search for companion. 

\paragraph{2014~OM$_{394}$}

We observed 2014~OM$_{394}$ with the Magellan-Baade telescope over approximately 5.5~h in September 2016. The partial lightcurve reported here presents a very low variability, about 0.1~mag. There is no derived rotational period based on our data. To our knowledge, this object has never been the topic of a search for satellite.

%%%%%%%%%%%%%%%%%%%%%%%%%%%%%%%%%%%%%%%%%%%%
%%%% Discussion
%%%%%%%%%%%%%%%%%%%%%%%%%%%%%%%%%%%%%%%%%%%%

\section{Rotational characteristics of the Cold Classicals}
\label{sec:analysis}

\subsection{Lightcurves: Our sample + Published results}

Table~\ref{Summary_CCs} summarizes all the published lightcurves of the CC population. About 900 CCs are known, but only 43 have been observed for rotational variability (without taking into account our survey), and 10 of them (24$\%$ of the sample) are in fact known wide binaries. The lightcurve of a binary system can be resolved or unresolved. In other words, the two system's components can be separated and thus there is one lightcurve for each component or the components are unresolved and the reported photometry in the photometry of the pair. Resolved ground-based lightcurves are challenging and require excellent weather conditions to separate the components, as well as systems with large separation, and large aperture facilities. For those reasons, most of the binary system lightcurves are unresolved \citep{Thirouin2014}. Only two attempts of resolved lightcurves with the 6.5~m Magellan Telescope have been published for Teharonhiawako-Sawiskera ((88611) 2001~QT$_{297}$), and 2003~QY$_{90}$ \citep{Osip2003, Kern2006}.

Over the past years the main effort to study the rotational properties of the CCs has been focused on the binaries. However, binaries undergo tidal effects affecting the rotations of the components, and thus their rotational properties are not primordial and therefore not representative of the CC population \citep{Thirouin2014}. Several objects, we and other teams imaged have not been observed for companion search and thus we do not know if they are wide binaries (e.g., 2002~GV$_{31}$\footnote{With a long rotational period of about 29~h and a relatively large size of H=6.4~mag, 2002~GV$_{31}$ is potentially a binary system \citep{Thirouin2014, Pal2015}}, 2003~QY$_{111}$). 

The main focus on lightcurves of wide binary CCs created a bias in our understanding of the rotational properties of the CC population. Therefore, for the purpose of this work, we focus on the single CCs. It is important to mention that some of the reported CCs have not been observed for satellites with \textit{HST} and thus some of them could be wide binary systems (see Table~\ref{Summary_photo} for a complete review). Only the known binary Logos-Zoe was selected because we wanted to confirm the large amplitude reported in \citet{Noll2008}. In the case of 2014~LQ$_{28}$, we discovered during the Magellan observations that this object is an equal-sized wide binary (see Section~\ref{sec:lightcurve}). 

\citet{Noll2014} reported that 100$\%$ of the bright CCs (H$\lesssim$6~mag) are binary systems. Mainly, we select CCs with an absolute magnitude H$>$6~mag, but as we do not want to bias our sample toward "small" size objects, we also select a couple of larger CCs. Therefore, several large CCs in our sample are maybe wide binaries. We estimate that up to five CCs with H$<$6~mag (2004~HD$_{79}$, 2004~HF$_{79}$, 2014~LR$_{28}$, 2014~LS$_{28}$, and 2014~OM$_{394}$) could be wide binaries based on \citet{Noll2014} criteria. The cut-off at H$<$6~mag is approximate and so objects with an absolute magnitude around the cut-off could be wide binaries (e.g. 2013~FA$_{28}$). Because of their recent discoveries, the \citet{Alexandersen2018} targets have never been search for resolved binaries, but based on their small sizes, we do not expect them to be resolved binary systems \citep{Noll2008, Noll2014, Penteado2016}. Ultimately, only a search for resolved companion with the \textit{Hubble Space Telescope} and/or the \textit{James Webb Space Telescope} will confirm the nature of theses objects/systems.   

Also, the brightest and thus the largest CCs (typically, H$<$6~mag) have been studied for lightcurves as they are the easiest ones to observe. Therefore, our survey is focused on smaller objects with an absolute magnitude up to 7.2~mag. Recently, \citet{Alexandersen2018} observed objects up to 9.2~mag taking advantage of the large aperture of the Subaru telescope. \citet{Alexandersen2018} observed TNOs discovered by the Outer Solar System Origins Survey (OSSOS) in all dynamical groups, and only 25 objects in their sample belong to the CC population (see Table~2 of \citet{Alexandersen2018}). The maximum amplitude variation reported by \citet{Alexandersen2018} is the difference between the brightest and the faintest data point. In some cases, the lightcurves present a high dispersion and some points are outliers. Therefore, we re-estimated such amplitudes with conservative values by taking into account potential outliers and removing them for the estimates (Table~\ref{Summary_CCs}).  

The smallest CCs observed for lightcurve variability with the \textit{HST} are: 2003~BF$_{91}$ with H=11.7~mag, 2003~BG$_{91}$ with H=10.7~mag and 2003~BH$_{91}$ with H=11.9~mag \citep{Trilling2006}. These objects are in the same size range as 2014~MU$_{69}$ with H=11.1~mag. In the case of 2003~BH$_{91}$, no reasonable rotational period was derived and so this object will not be considered in our work. For 2003~BF$_{91}$, a single-peaked lightcurve with P=9.1~h and a $\Delta m$=1.09~mag was preferred, but 7.3~h was also a possibility. By assuming a single-peaked lightcurve, it is considered that 2003~BF$_{91}$ is a spheroidal object with albedo variation of its surface. However, a variability of 1.09~mag suggest very strong albedo variegation(s) on the object’s surface, which is doubtful. Therefore, a more appropriate option is to consider an ellipsoidal object with a double-peaked lightcurve. Based on the photometry and the two potential rotational periods reported by \citep{Trilling2006}, the best lightcurve is found using a periodicity of 2$\times$7.3~h (value used for the rest of our study) with 1.01~mag as amplitude. Based on the large variability, 2003~BF$_{91}$ is a likely contact binary, but one has to keep in mind that the lightcurve is very noisy \citep{Trilling2006}. For 2003~BG$_{91}$, \citet{Trilling2006} selected a single-peaked period of 4.2~h. Based on the supposed fast rotation, it is likely that 2003~BG$_{91}$ is highly elongated, however the amplitude reported is only 0.18~mag. Therefore, it is more likely that the double-peaked lightcurve with a period of 2$\times$4.2~h is more appropriate (solution used in this work). 

In conclusion, the published literature and our survey are focused on a wide range of sizes (5.0mag$\geq$H$\geq$11.9mag) and thus by merging these results, we can infer the rotational properties of the entire CC population, as well as probe the properties of different size regimes (Table~\ref{Summary_CCs}). 

\begin{figure}
\includegraphics[width=10cm, angle=0]{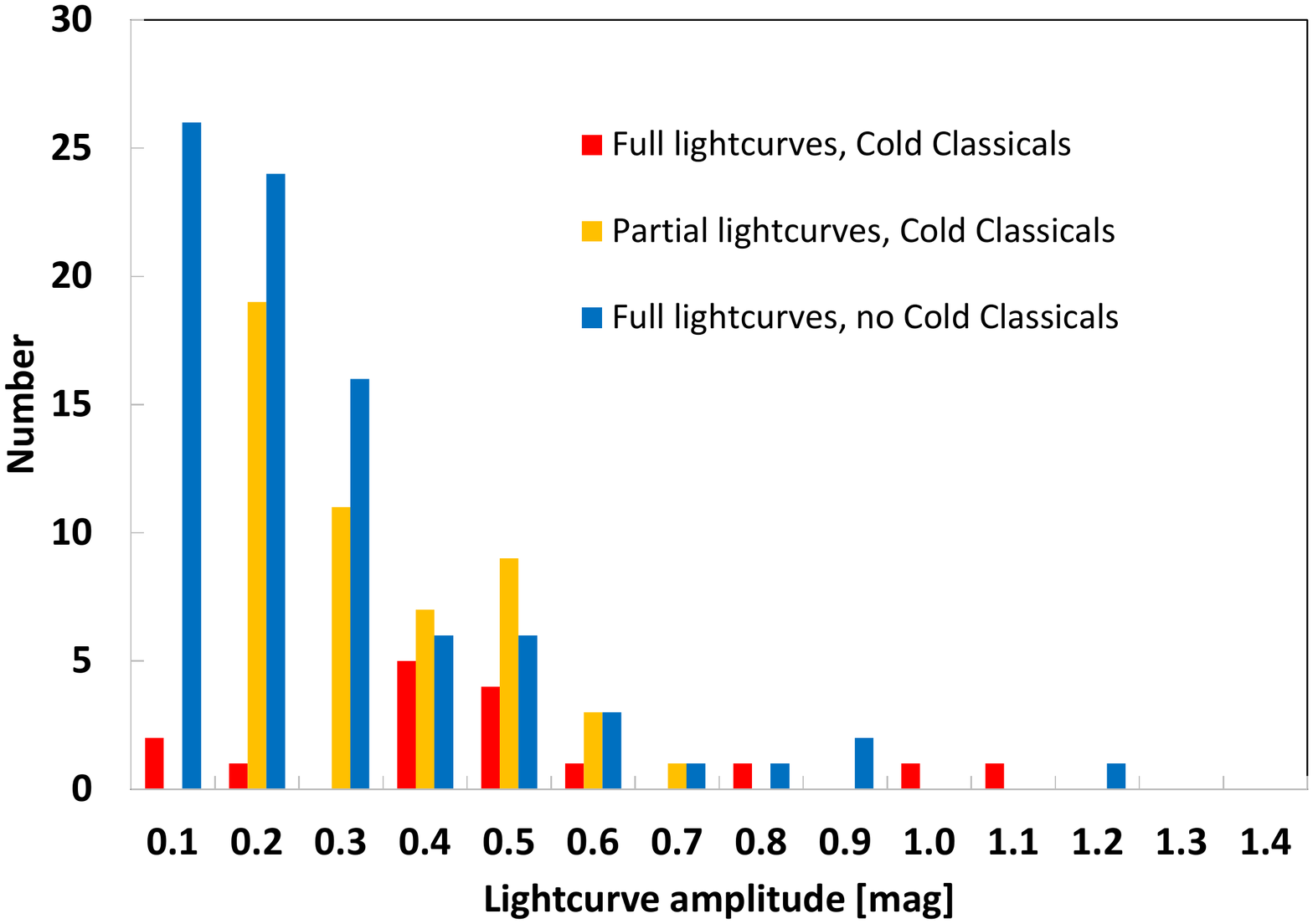}
\includegraphics[width=10cm, angle=0]{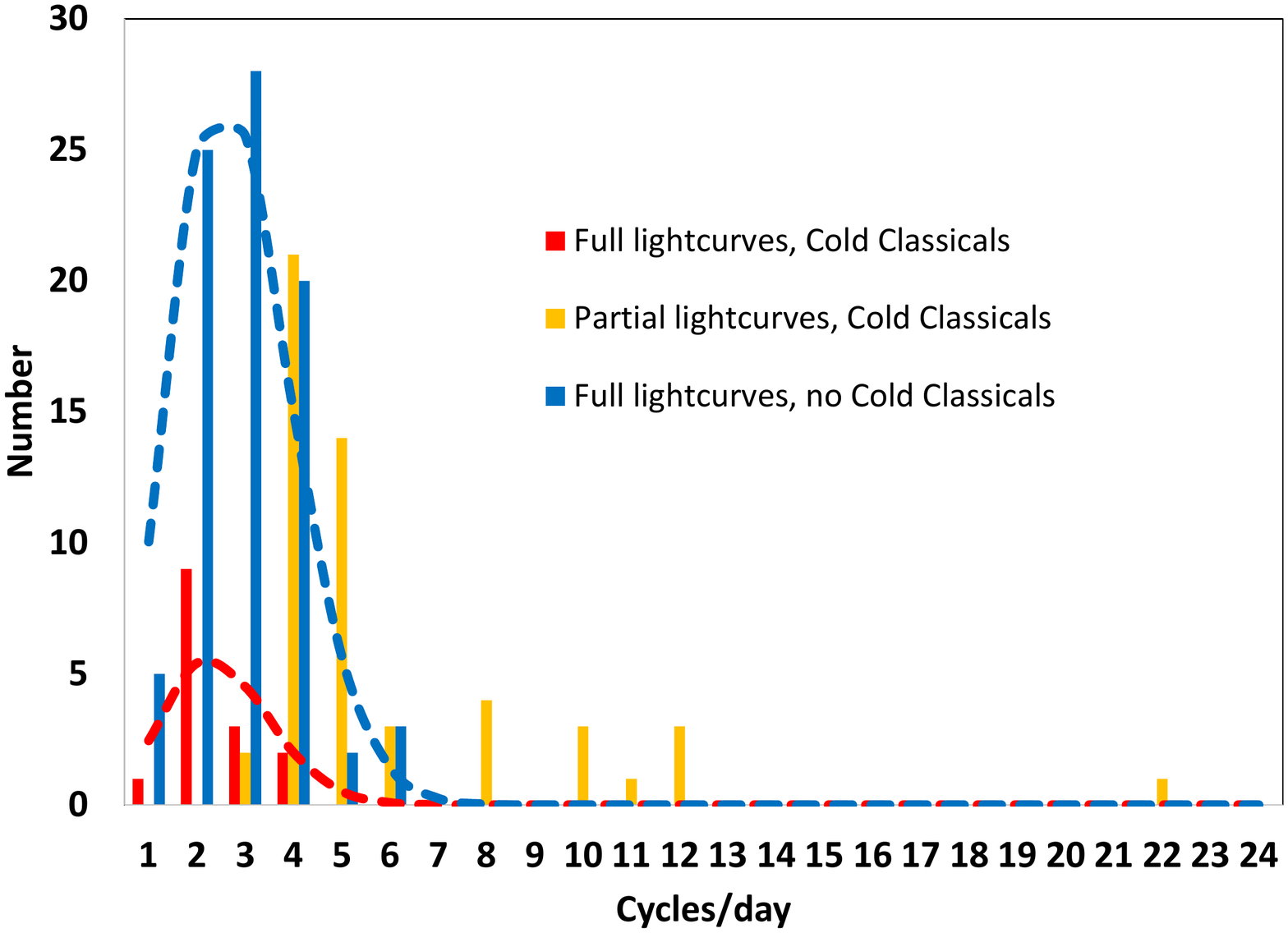}
 \caption{Histograms using the partial and full lightcurves reported in this work and the literature. The CCs tend to have more amplitude than the rest of the TNOs. The blue discontinuous line is a Maxwellian fit using only the CC full lightcurves suggesting a mean rotational period of 2.53~cycles/day (9.48~h), whereas the red discontinuous line is for the other TNOs with a mean period of 8.45~h. }
\label{fig:Histo}
\end{figure}

\subsection{Lightcurve amplitude and rotational period distributions}

In Figure~\ref{fig:Histo}, we summarize the lightcurve studies of the CC population by taking into account the full and partial lightcurves. For the partial lightcurves, we only have access to the lower limits for the rotational period and the lightcurve amplitude, whereas the full lightcurve provides us with an exact estimate for both parameters. Flat lightcurves from this work are not plotted. 

Based on the 16 full lightcurves available in the literature and this work, we report that the mean lightcurve amplitude is about 0.39~mag whereas the partial lightcurves have a mean amplitude of 0.29~mag (0.20~mag with the flat lightcurves). Based on TNO lightcurves from all dynamical groups, \cite{Duffard2009} reported that 70$\%$ of them have a low variability, $\leq$0.2~mag. Based on an updated and larger sample, we estimate that $\sim$65$\%$ of the other TNOs have an amplitude $<$0.2~mag. Only 36$\%$ of the CCs have an amplitude lower than 0.2~mag (including the flat lightcurves). Therefore, the CCs tend to have more amplitude suggesting that they are more elongated or present a higher deformation that the rest of the TNOs. As the sample of full lighcurves is dominated by resolved binaries, the larger amplitude can be attributed to the formation of these systems \citep{Thirouin2014}. But, based on our sample of partial lightcurves likely dominated by single objects and unresolved binaries, such a tendency remains. Therefore, the larger amplitude can potentially be a primordial characteristic of the CC population. 

The period distributions with the full and partial lightcurves from our survey and the literature are also plotted in Figure~\ref{fig:Histo} (flat lightcurves are not included). Considering only the full lightcurves, the Maxwellian distribution fit infers a mean period of 9.48$\pm$1.53~h. \citet{Duffard2009, Thirouin2010, Benecchi2013} calculated a mean rotational period of the entire TNO population of 7-8~h. Using an updated sample, the mean rotational period of the other TNOs is 8.45$\pm$0.58~h (Figure~\ref{fig:Histo}). Therefore, the CC population seems to rotate slower, but one has to keep in mind the large error bar in the mean rotational period due to the still limited sample. Based on Figure~\ref{fig:Histo}, Tables~\ref{Summary_photo} and \ref{Summary_CCs}, one can appreciate that most of the partial lighcurves have been obtained over 5-6~h. As the typical TNO mean rotational period is about 8~h, most of these partial lightcurves should have covered almost the full object's rotation and thus a rough period estimate should have been estimated.       

\begin{figure*}
\includegraphics[width=20cm, angle=0]{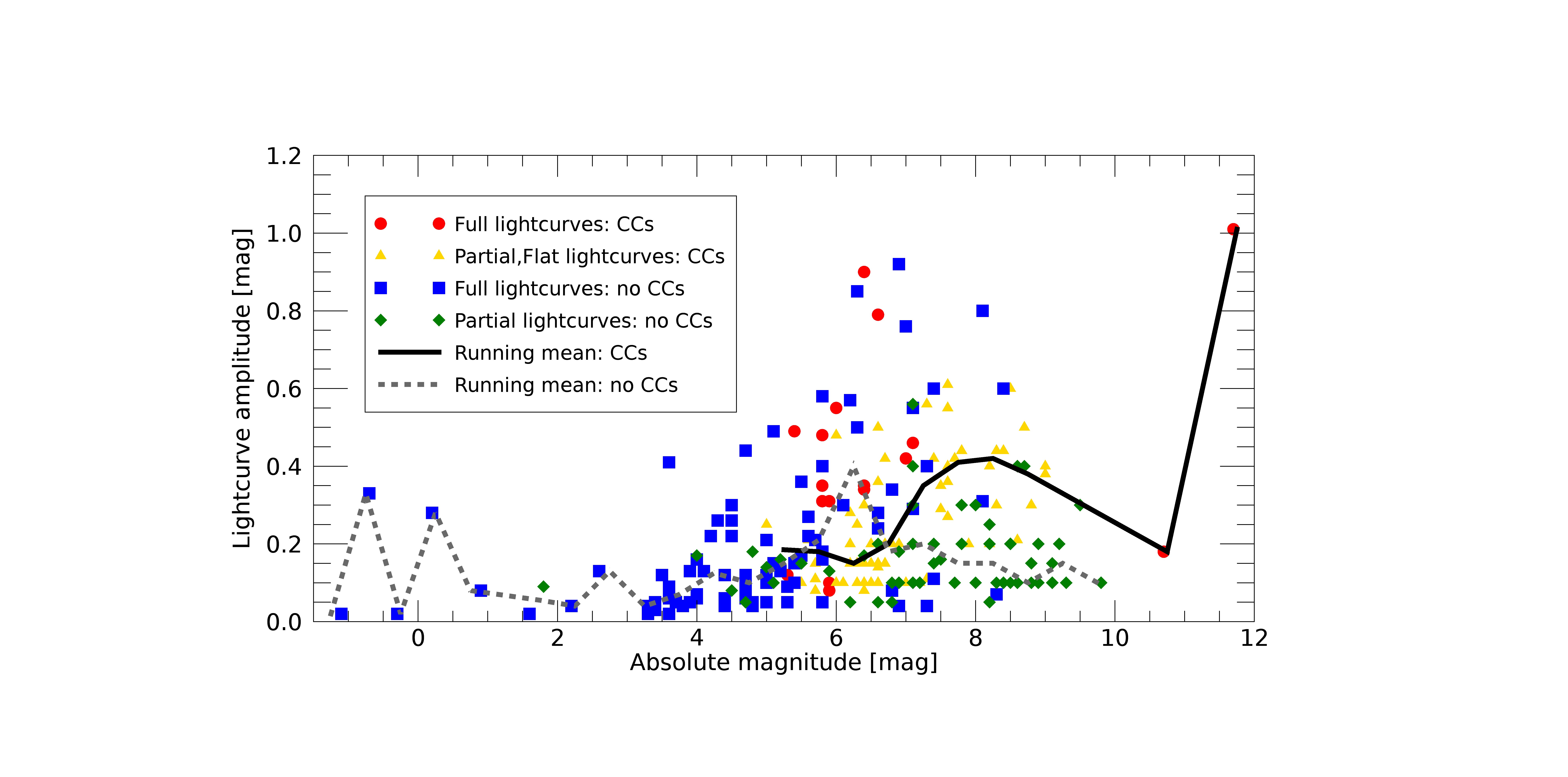} 
 \caption{We used our results and the literature to plot the lightcurve amplitude versus absolute magnitude distributions of the CC population and of the other TNOs. Different symbols and colors are used to separate the two populations and the partial/full lightcurve sample. Two running means using the partial and the full lightcurves are over-plotted, one for the CC population and one for the other TNOs. The other TNOs have a roughly flat distribution across the different size regimes, whereas the CC population is showing an increase of amplitude at small sizes. Some bins only have one or zero object.}
\label{fig:RM}
\end{figure*}

An important parameter to take into account is the size range of the CCs and the other TNOs. The CCs observed for lightcurve studies have an absolute magnitude between 5 and 11.7~mag, whereas the other TNOs belong to the size range of -1.1 up to 9.8~mag. Therefore, both samples have overlap, but one has to keep in mind that the other TNOs sample has dwarf planets and large/medium size TNOs that are not present in the CC population. Also, there are only 2 very small CCs with H$>$10~mag observed for lightcurves. Therefore, the two samples are mostly overlapping in the range of 5 to 9~mag. In Figure~\ref{fig:RM}, we plotted all the TNOs from our sample and the literature with a partial, flat or a full lightcurve. Based on the running means in Figure~\ref{fig:RM}, the CC population tends to have more amplitude at small sizes whereas the other TNOs have roughly a flat distribution across size ranges. In the case of the CC population, there is a constant increase of amplitude starting at H$\sim$6~mag. The last bins (H$>$9~mag) only have 0 or 1 object per bin and so the running mean is not adequate. In conclusion, it seems that the CC population is showing more lightcurve variability than the other TNOs. We also check for any trend between rotational period and size, but did not find any relation.
The CC population may also have slower rotations. As these properties are not noticed in the other TNO sample, we infer that they are primordial characteristics. More complete lightcurves of CCs and other TNOs, and especially small objects would help to confirm these results. Also, the slower rotation of the CCs can be due to the loss of wide binaries satellites through the conservation of angular momentum assuming that most of the CCs were born as binaries.   

\subsection{Anti-correlation/Correlation}

One can investigate trends between rotational properties and orbital elements for the CC population with the Spearman rank correlation \citep{Spearman1904}. We calculated the Spearman coefficient ($\rho$) and the significance level (SL). A correlation is strong if $\mid\rho\mid>0.6$, weak if $\mid\rho\mid>0.3$, and non-existent if $\mid\rho\mid<0.3$. The significance level is very strong if $>$99\%, strong if $>$97.5\%, and reasonably strong if $>$95\%.  

In a first step, we considered only the full lightcurves from the literature and our sample. Sila-Nunam is likely tidally locked and will not be considered in our search for correlations with rotational period, but will be for lightcurve amplitude \citep{Rabinowitz2014}. Our results are summarized in Table~\ref{Corr}, and we emphasize that the sample of objects with a full lightcurve is limited to 16 CCs (with Sila-Nunam). There is a correlation between lightcurve amplitude and rotational period suggesting that the slow rotators tend to have larger lightcurve amplitudes (i.e. objects more deformed or irregular shape). Such a tendency is confirmed by taking into account only the known resolved binaries and the potential contact binaries. As the samples are still limited, it is unclear if we are dealing with an observational bias or not. If true, and because the sample is dominated by binaries, this correlation may give us some clues about binary system formation \citep{Thirouin2014}. Also, such a tendency is not observed in the rest of the trans-Neptunian belt.  

The correlation search can also be performed using lower/upper limits as implemented in the astronomy survival analysis package named ASURV\footnote{\url{http://astrostatistics.psu.edu/statcodes/asurv}} \citep{Spearman1904, Feigelson1985, Isobe1986, Isobe1990, LaValley1990, LaValley1992}. Therefore, our second step was for statistical tests in our merged sample of full and partial lightcurves (Table~\ref{Corr}). Flat lightcurves reported here were not used for the correlation search. We noticed a weak correlation between lightcurve amplitude and absolute magnitude (i.e. smaller objects have larger amplitude). Such a tendency has been already reported in several dynamical sub-populations as well as in the entire TNO population, and is in agreement with the TNO collisional evolution \citep{Davis1997, Sheppard2008, Duffard2009, Thirouin2013, Benecchi2013, Alexandersen2018}.

In a third step, we divided our sample according to absolute magnitude: i) CCs with H$\leq$6~mag (i.e. ``large'' objects, sample dominated by resolved binaries) , ii) CCs with 6$<$H$\leq$8~mag (i.e. ``medium size'' objects, sample presumably dominated by single objects and unresolved binaries), and iii) CCs with 8$<$H$\leq$12~mag (i.e. ``small'' objects, sample likely dominated by single objects). The large objects sample shows a potential anti-correlation between period and eccentricity with a low significance level. The medium size sample shows a weak anti-correlation between period and absolute magnitude suggesting that the large objects rotates slower. A possible explanation is that the binaries dominate at large sizes and they undergo tidal effects able to slow down their rotations  \citep{Thirouin2014}. Also, it is important to point out that the potential contact binaries are in this size range and that the cut-off at H=6~mag to infer if an object is maybe a resolved binary is only approximate \citep{Noll2014}. There is a weak correlation between rotational period and absolute magnitude with only a significance level of 81~$\%$. Such a tendency if true is interesting as the medium size objects display the opposite relation. The sample composed of the smallest objects presents a reasonably strong anti-correlation between rotational period and semi-major axis. Also, there are several trends with significance levels below our threshold of confidence. For example, there is a weak anti-correlation between amplitude and eccentricity indicating that large amplitude CCs have low eccentricities. It is unclear if such tendencies are an observational bias or not as the sample at small size is still limited.

\begin{figure*}
\includegraphics[width=20cm, angle=0]{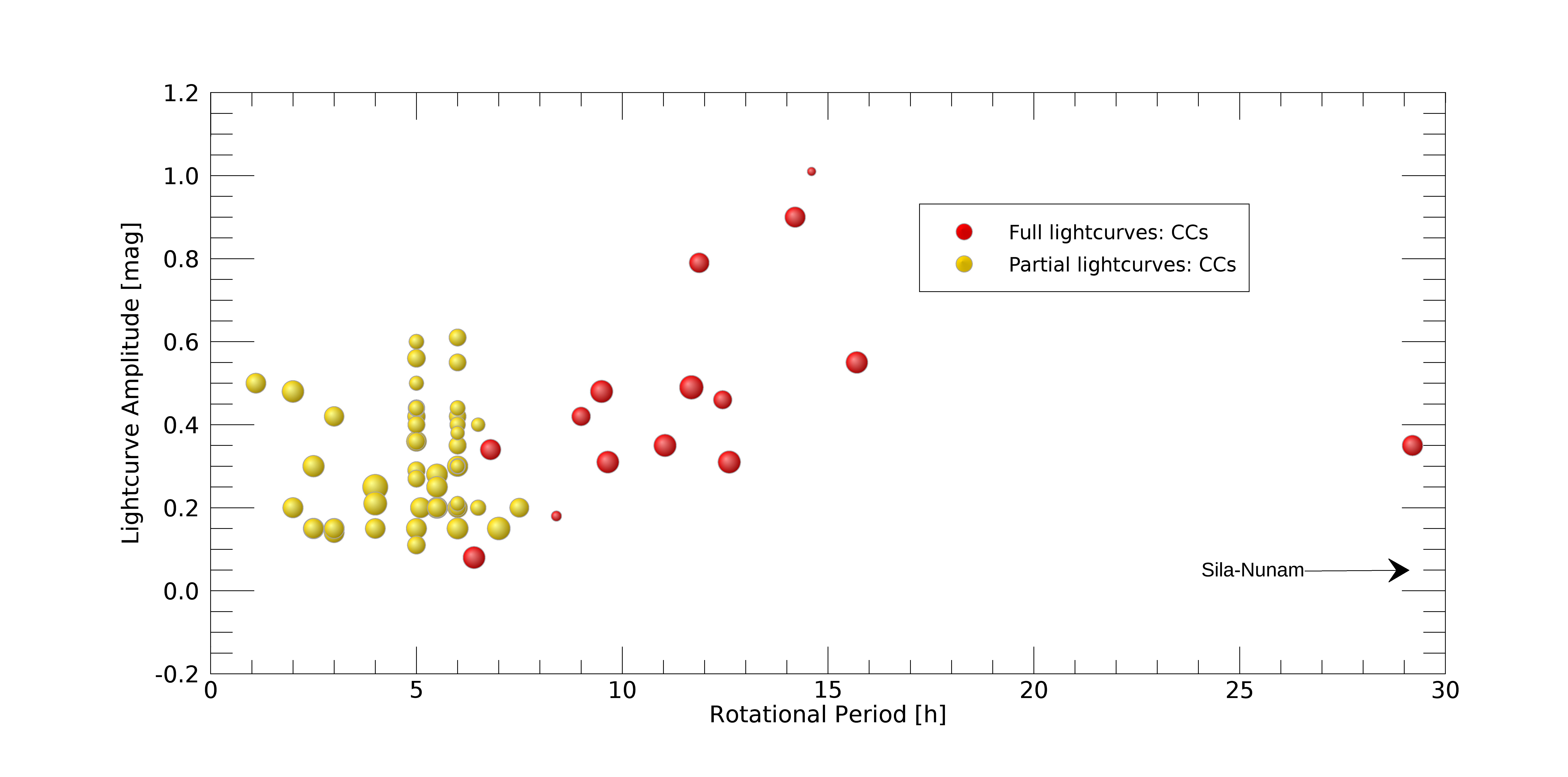}
\includegraphics[width=20cm, angle=0]{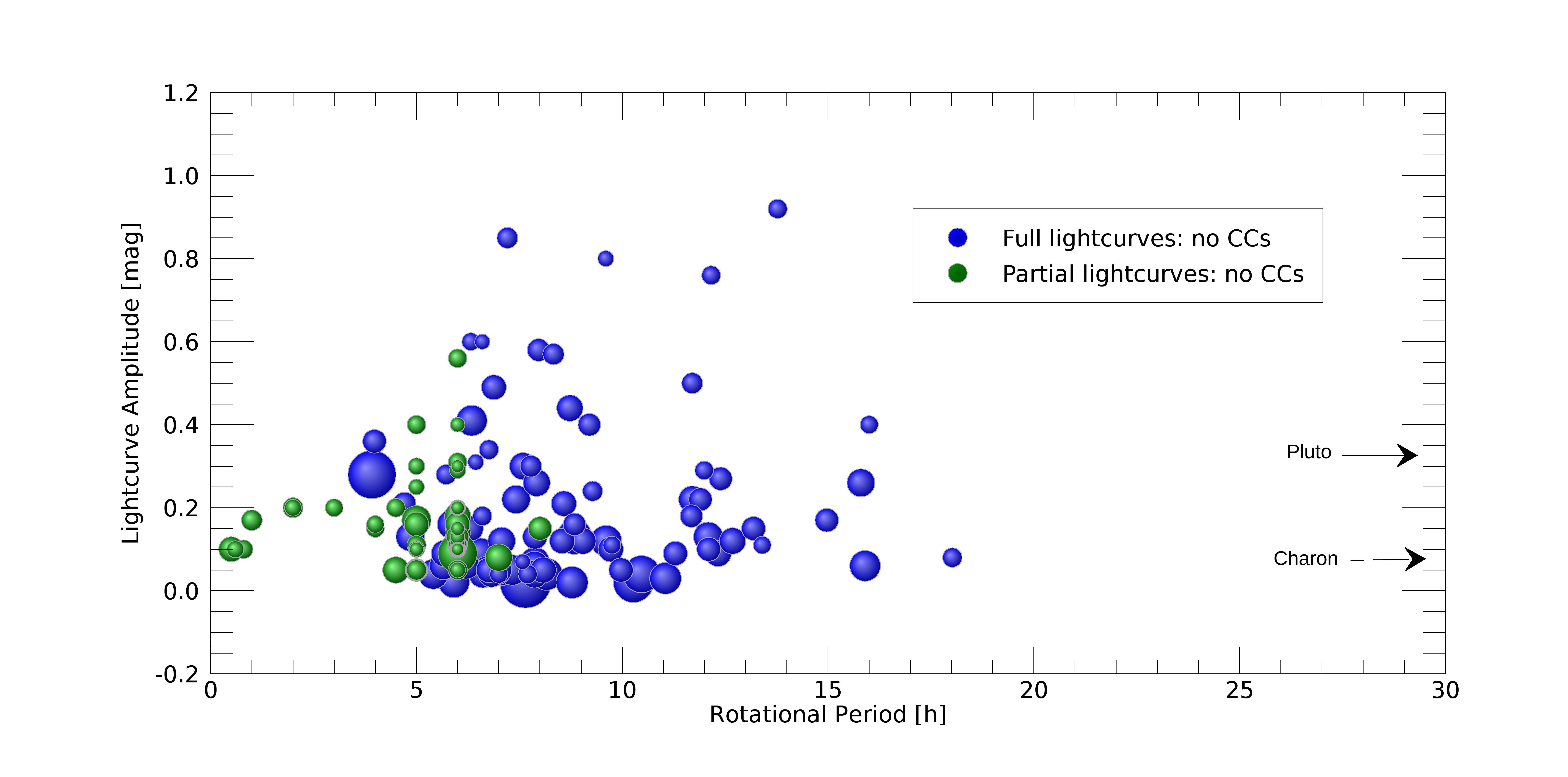}
 \caption{The upper panel summarizes the lightcurve studies for the CC population and the lower one is for the other TNOs. The bubble size indicates the size of the objects (i.e. large bubbles for large objects). The largest visible bubble is for Makemake (H=-0.3~mag, periodicity of 7.65~h). Same bubble scaling has been used for both plots allowing a direct comparison between theses two populations. Due to their very long rotational periods (out of the plot's scale), Sila-Nunam and Pluto-Charon are not plotted.}
\label{fig:AmplitudePeriod}
\end{figure*}

  \section{Contact binaries}
 
 \subsection{Definition}

The definition of contact binary systems includes objects with a peanut shape or bi-lobed shape (like comet 67P), two objects touching in one point and thus in contact and two objects with a small separation. To confirm the nature of the system/object, multiple lightcurves obtained at several epochs are required for modeling purposes. Also, multi-chord stellar occultations or even flybys can infer the shape (e.g., 2014~MU$_{69}$, \citet{Moore2018}). Therefore, we adopted the following definition: i) a lightcurve with an inverted-U shape at the maximum of brightness, a V-shape at the minimum and a peak-to-peak amplitude greater than 0.9~mag$\footnote{An object in hydrostatic equilibrium with a lightcurve amplitude greater than 0.9~mag will break and create a binary \citep{Weidenschilling1980, Leone1984}.}$ is due to a \textit{confirmed} contact binary \citep{Dunlap1969, Leone1984, Cellino1985, Sheppard2004, Lacerda2011, Lacerda2014}, ii) a lightcurve with an inverted-U shape at the maximum of brightness, a V-shape at the minimum, and a large peak-to-peak amplitude but not reaching the 0.9~mag threshold is due to a \textit{likely} contact binary \citep{Lacerda2014, Thirouin2017, ThirouinSheppard2017, Thirouin2018}. The morphology of a contact binary lightcurve can be produce by objects with other shapes as suggested by \citet{Zappala1980, Harris2018}. Thus, it is also important to take into consideration the likelihood of such options. Finally, we want to point out that the recent flyby of 2014~MU$_{69}$ clearly demonstrated the existence of contact binaries in the trans-Neptunian belt \citep{Stern2019}. 
 
\subsection{Current status in the Cold Classical population}

In about one year, the number of confirmed/likely contact binaries in the trans-Neptunian belt grew from two to nine (\citet{Sheppard2004, Lacerda2014, Thirouin2017, ThirouinSheppard2017, Thirouin2018}, and this work). \citet{Thirouin2018} showed  an abundance of Plutino contact binaries. \citet{ThirouinSheppard2017} and this work highlight the discovery of two likely contact binaries in the CC population: 2002~CC$_{249}$, and 2004~VC$_{131}$, and have hints that 2004~VU$_{75}$ and 2004~MU$_{8}$ are good candidates to this category. 

Using the formalism from \citet{Sheppard2004}, we estimate the equal-sized contact binary fraction in the CC population. In case of an object with axes as a$>$b and b=c, the lightcurve amplitude changes with the angle of the object's pole relative to the perpendicular of the line sight ($\theta$): 
\begin{equation}
\Delta_m = 2.5 \log \left(\frac{1+\tan \theta}{(b/a)+\tan \theta}\right)
\label{eq:frac1}
\end{equation}
The lightcurve amplitude of an ellipsoid (a$\geq$b=c) varies as: 
\begin{equation}
\Delta_m = 2.5 \log \left(\frac{a}{b}\right) - 1.25 \log \bigg[ \left( \left(\frac{a}{b}\right)^2 -1 \right) \sin^2 \theta +1\bigg]
\label{eq:frac2}
\end{equation}
Pole orientation aspects are important to estimate the fraction of contact binaries, and thus we will consider several cases. An object with a/b=3 will display a variability of 0.9~mag if $\theta$=10$^\circ$. The probability of observing an object from a random distribution within 10$^\circ$ of the sight line is P($\theta$$\leq$10$^\circ$)=0.17. Similarly, and as discussed in \citet{Thirouin2018}, we can estimate the probability of different $\theta$ angles using different cut-off for the amplitudes\footnote{An amplitude of 0.4/0.5/0.6/0.7~mag is for $\theta$=49/36/27/20$^\circ$}, and therefore debias the pole orientations of our objects. As mentioned, the large amplitude is only reached when the system's components are equator-on. Therefore, considering smaller amplitude for an equator-off configuration is needed.  

Using previous equations and several cut-off for the lightcurve amplitude (and so different P($\theta$)), we estimate the contact binary fraction based on our sample\footnote{Despite its large amplitude, the known resolved binary Logos-Zoe is not taken into account in our estimates as a contact binary.} and assuming equal-sized binaries. We found that f($\Delta m$$\geq$0.7~mag)$\sim$1/(42$\times$P($\theta$$\leq$20$^\circ$))$\sim$7~$\%$, and f($\Delta m$$\geq$0.5~mag)$\sim$8~$\%$ using the Equation~\ref{eq:frac1}. Based on Equation~\ref{eq:frac2}, we estimated for our sample: f($\Delta m$$\geq$0.7~mag)$\sim$6~$\%$, and f($\Delta m$$\geq$0.5~mag)$\sim$8~$\%$. Potential contact binaries reported here have an absolute magnitude ranging from 6 to 7~mag, and only taking into account objects in this size range, we found f($\Delta m$$\geq$0.5~mag; 6$\leq$H$\leq$7)$\sim$9~$\%$, and f($\Delta m$$\geq$0.5~mag; 6$\leq$H$\leq$7)$\sim$10~$\%$ with Equation~\ref{eq:frac1} and  Equation~\ref{eq:frac2}, respectively. In conclusion, the contact binary fraction in the CC population is less or about 10~$\%$ based on our entire sample and on a specific size range. Using our dataset and the literature, we calculated\footnote{We considered that only two CCs have a lightcurve amplitude $>$0.7~mag: 2002~CC$_{249}$ and 2003~BF$_{91}$. \citet{Kern2006} reported an amplitude 0.90$\pm$0.36~mag for the satellite of 2003~QY$_{90}$, but based on their very sparse lightcurve and large uncertainty, we have not taken into account this object.  } f($\Delta m$$\geq$0.7~mag)$\sim$7~$\%$ and f($\Delta m$$\geq$0.7~mag)$\sim$6~$\%$ with the Equation~\ref{eq:frac1} and  Equation~\ref{eq:frac2}, respectively. Assuming a cut-off\footnote{Seven CCs have a lightcurve amplitude $>$0.5~mag, 2002~CC$_{249}$, 2003~BF$_{91}$, 2004~VC$_{131}$, 2013~SM$_{100}$, 2013~UN$_{13}$, 2015~RA$_{280}$, and 2015~RB$_{280}$. We did not take into account uo3l88 because the lightcurve presents a large dispersion. } at 0.5~mag, we obtained with both equations the same result f($\Delta m$$\geq$0.5~mag)$\sim$15~$\%$. 

Despite the hints that 2004~VU$_{75}$ and 2004~MU$_{8}$ are maybe contact binaries, we did not include them in our previous estimates as we do not have their full lightcurves (and thus a secure lightcurve amplitude estimate). However, as both objects have an amplitude larger than 0.4~mag in a few hours, we can assume that their full lightcurves will likely be larger than 0.5~mag, and we can include them in our f($\Delta m$$\geq$0.5~mag) estimate. For our sample only, we obtained: f($\Delta m$$\geq$0.5~mag)=16$\%$, and 17\% with Equation~\ref{eq:frac1}, and Equation~\ref{eq:frac2} (respectively). With our sample and the literature, the fraction is the same for both equations, f($\Delta m$$\geq$0.5~mag)=19$\%$. 
Two objects from \citet{Alexandersen2018} are good candidates for follow-up observations based on their potential large amplitude, 2015~RO$_{281}$ and 2013~UL$_{15}$. Assuming that these objects have a full amplitude larger than 0.4~mag, we estimated that f($\Delta m$$\geq$0.4~mag)=21$\%$, and 25\% with Equation~\ref{eq:frac1}, and Equation~\ref{eq:frac2}. Others objects have $\Delta m$$>$0.4~mag, but as their lightcurves are noisy, they are not considered in our previous estimates. Finally, we emphasize that these fractions are lower limits and more full lightcurves are required for several objects in order to infer their shape as well as continue to build a representative sample of the CC population. Also, we are only considering equal-sized binaries for our estimate, and thus the fraction will increase by adding the non equal-sized binaries. For the purpose of this section, we assumed that none of the flat lightcurve are due to contact binaries with a pole-on orientation. Contact binaries with very long rotational periods undetectable over our observing blocks are not considered. Therefore, as already said the previous percentages are lower limits. 

\begin{figure*}
\includegraphics[width=20cm, angle=0]{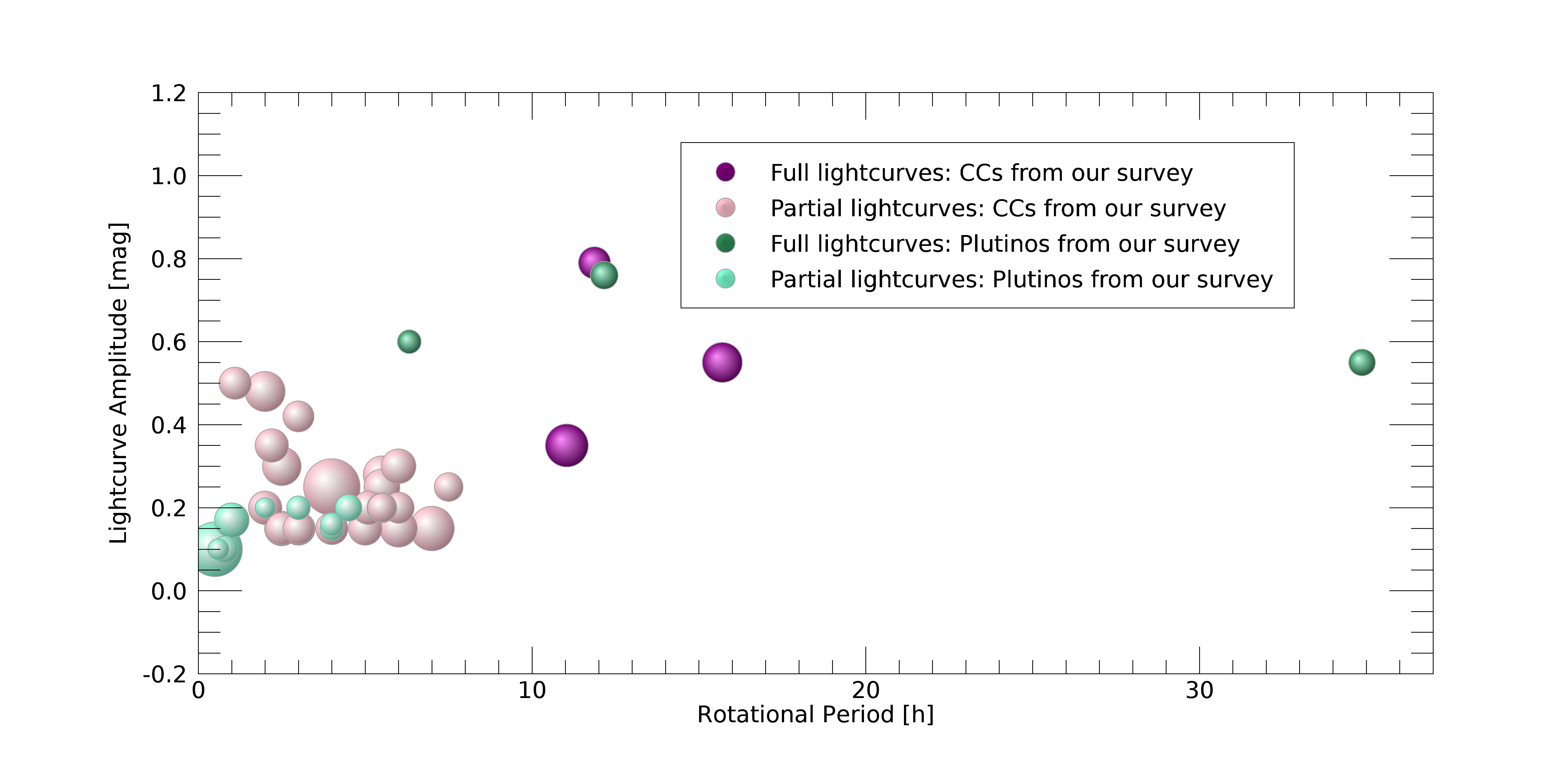}
 \caption{Partial/full lightcurves obtained with our surveys of the CC and Plutino populations. The potential Plutino contact binaries are smaller than the ones found in the CC population. We observed an handful of Plutinos compared to the CCs, and found an abundance of contact binaries.   }
\label{fig:AmplitudePeriod2}
\end{figure*}

\subsection{Cold Classicals versus Plutinos}

The contact binary population in the Kuiper belt has been estimated to up to 30$\%$ \citep{Sheppard2004, Lacerda2014}. However, we find that only 10-25$\%$ of the CCs could be contact binaries. Therefore, it seems that there is a deficit of contact binaries in the CC population. On the other hand, we found an excess of them with an estimate up to 40-50$\%$ in the Plutino population \citep{Thirouin2018}. Such a find (despite the still low statistical number) is interesting, especially because the opposite tendency is noticed with the resolved wide binaries: a deficit of resolved wide binaries in the Plutinos and an excess in the CC population \citep{Noll2008, Thirouin2018}. Also, it is interesting to mention that the size of the likely contact binaries found in these two sub-populations is different. In the Plutino population, the likely contact binaries have an absolute magnitude around 7~mag, whereas they are larger with an absolute magnitude around 6~mag in the CC group (Figure~\ref{fig:AmplitudePeriod2}, except for the potential contact binary 2004~VU$_{75}$ with H=6.7~mag). For both studies, we used the same observing strategy for partial/full lightcurve, and we also focused on a large range of object's sizes allowing a comparison of these two sub-populations (Figure~\ref{fig:AmplitudePeriod2}). However, as the CCs are further away compared to the Plutinos, we have the tendency to observe larger CCs on average. So, observing smaller fainter CCs may find more contact binaries like in the Plutino population. One possible explanation for the contact binary fractions is linked to the formation/evolution of these two populations. As said, the CC population was likely formed in-situ and never suffered any strong dynamical evolution whereas the resonance populations have not been formed where they are today and have been pushed outwards during the migration of Neptune. Therefore the formation/evolution of theses two sub-populations is different. Assuming that all planetesimals formed as binary systems, the different fraction of contact binaries are likely an outcome of the higher velocity dispersion and more intense and longer dynamical interactions the resonance populations likely encountered after formation. To confirm such a find, more contact binaries have to be found and we also need to test other resonances to infer if the high contact binary fraction is present in all resonances or only the 3:2. 
 
\section{Context for New Horizons: (486958) 2014~MU$_{69}$}

The second target of the \textit{NASA New Horizons} spacecraft is a small Cold Classical TNO with an absolute magnitude H=11.1~mag. Only two CCs in this size range have been observed for lightcurve variability: 2003~BF$_{91}$ and 2003~BG$_{91}$ and thus can be used as comparison for 2014~MU$_{69}$. Both have slow rotations with periods of 8.4 and 14.6~h and 2003~BF$_{91}$ displays a large amplitude of about 1~mag whereas 2003~BG$_{91}$ has a moderate amplitude of 0.18~mag. Both lightcurves have been obtained with the \textit{HST}, present a large dispersion and binning was needed to produce the lightcurves. Unfortunately both objects have not been observed since 2003 (orbital arcs of 13 and 92~days) and thus are likely lost. Despite the very limited sample of very small CCs to compare 2014~MU$_{69}$ to, we can use the rest of the CC population for extrapolation. In fact, by observing a large number of CCs over diverse size range, we can infer the rotational and physical properties of this population and extrapolate to smaller sizes. So, far we have shown that the CCs tend to rotate slowly and are more deformed than the other TNOs. Also, there is an increase of lightcurve amplitude with decreasing size, suggesting that the small CCs are more deformed that the large ones. If 2014~MU$_{69}$ follows similar trends, we have to expect a slow rotator with a deformed shape. Results from a stellar occultation by 2014~MU$_{69}$ seems to indicate that the shape is complex and thus confirm our trend. However, based on \textit{HST} data it seems that the lightcurve of 2014~MU$_{69}$ is flat \citep{Benecchi2017, Benecchi2018}. A reasonable explanation to reconcile the occultation and lightcurve data is to consider that 2014~MU$_{69}$ has a (nearly) pole-on orientation which was confirmed by the flyby \citep{Zangari2019, Showalter2019}. Based on the preliminary results from the flyby, 2014~MU$_{69}$ has a potential rotational period of 15$\pm$1~h \citep{Stern2019}. Therefore, 2014~MU$_{69}$ seems to follow all the trends reported in this work.

Based on a stellar occultation and based on the flyby results, 2014~MU$_{69}$ is a contact binary. In this work and \citet{ThirouinSheppard2017}, we report the discovery of two likely contact binaries and some hints for two more in the CC population. By taking into account our sample and the literature, we estimate that 10-25$\%$ of the CC population could be contact binaries, suggesting that 2014~MU$_{69}$ is one of the few CC contact binaries. So far, the likely CC contact binaries are ``large'' (H$\sim$6~mag, except for 2004~VU$_{75}$), and thus it is interesting that we are not finding smaller CCs to be contact binaries as in the 3:2 population. However, we should also consider that the shape of these systems can be different with size: contact binaries with 2 separated or in contact objects at large sizes and a peanut-shape at smaller sizes.   

%________________________________________________________________

\section{Summary and Conclusions}  

Over the past three years, we used the Discovery Channel and the Magellan telescopes to study the rotational properties of the dynamically Cold Classical trans-Neptunian objects. Based on our 42 complete/partial lightcurves and the literature, we derived information about the shape and the rotational period distributions of the Cold Classical
 population. Our results are:

\begin{itemize}
\item Our first results from our survey dedicated to the rotational and physical properties of the Cold Classical population are presented. This survey is the first one entirely dedicated to this sub-population of the trans-Neptunian belt, and provides context for the second flyby of the \textit{NASA New Horizons} mission. 
\item We report the discovery of one new likely contact binary in the Cold Classical population: 2004~VC$_{131}$ and we have evidence that 2004~MU$_{8}$ and 2004~VU$_{75}$ are maybe also contact binaries. We estimate that the Cold Classical population has only 10-25$\%$ of contact binaries, compare to the 40-50$\%$ found in the 3:2 resonance \citep{Thirouin2018}. This estimate is a lower limit and assumes equal-sized binaries \citep{Sheppard2004}. The likely fraction of contact binaries will increase if also considering non equal-sized contact binary systems.
\item Objects in the Cold Classical population display a larger variability than the other TNOs, suggesting that they are more elongated or deformed than the rest of the trans-Neptunian population. About 65$\%$ of the other TNOs have an amplitude below 0.2~mag but only 36$\%$ of the Cold Classicals have a low variability. We also noticed a higher amplitude at smaller sizes which is not noticed in the other TNO sample. Because these tendencies are not present in the rest of the trans-Neptunian population, they are probably primordial characteristics of the Cold Classical population. 
\item Similarly, the Cold Classicals seem to rotate slower than the other TNOs with a mean rotational period of 9.48$\pm$1.53~h compared to the 8.45$\pm$0.58~h for the rest of the TNOs. Once again, this slow rotation can be a primordial characteristic of the Cold Classical population.
\item We perform a search for correlation/anti-correlation between rotational and physical parameters using the sparse and full lightcurves, and by using several size range. We report a strong correlation between rotational period and lightcurve amplitude in the Cold Classical group (not noticed in the rest of the TNOs). There is no clear explanation yet for this trend. 
\item We also report the discovery of a new nearly equal size wide binary, 2014~LQ$_{28}$ with a  magnitude difference of about 0.4~mag between the two components. With H=5.7~mag, 2014~LQ$_{28}$ follow the trend that all large Cold Classicals are resolved binaries \citep{Noll2014}. 
\item Our survey also provides context for the second flyby of the New Horizons mission. Based on early results presented during press conferences by the New Horizons team, 2014~MU$_{69}$ is a contact binary with a potential rotational period of about 15-16~h \citep{Stern2019}. Therefore, 2014~MU$_{69}$ is a slow rotator as the rest of the population. The shape of 2014~MU$_{69}$ is not unusual in the trans-Neptunian belt as we already found several confirmed/likely contact binaries through their lightcurves. However, we do not find a lot of contact binaries in the Cold Classical group.  
\end{itemize}
%
%________________________________________________________________

 \acknowledgments

We thank the referee for her/his careful reading of this paper and useful comments. 
Thanks to Larry Wasserman and Nuno Peixinho for their help with ASURV. We also acknowledge Chad Trujillo for some images obtained with the Lowell's Discovery Channel Telescope. Thanks to M. Bannister and J.-M. Petit for the official designation of the OSSOS objects. This paper includes data gathered with the 6.5~m Magellan-Baade Telescope located at las Campanas Observatory, Chile. This research is based on data obtained at the Lowell Observatory's Discovery Channel Telescope (DCT). Lowell is a private, non-profit institution dedicated to astrophysical research and public appreciation of astronomy and operates the DCT in partnership with Boston University, the University of Maryland, the University of Toledo, Northern Arizona University and Yale University. Partial support of the DCT was provided by Discovery Communications. LMI was built by Lowell Observatory using funds from the National Science Foundation (AST-1005313). Authors acknowledge the DCT and Magellan staffs. Authors also acknowledge support from the National Science Foundation (NSF), grant number AST-1734484 awarded to the ``Comprehensive Study of the Most Pristine Objects Known in the Outer Solar System".   \software{ASURV \citep{Isobe1986, Isobe1990, LaValley1990, LaValley1992}}

\clearpage

 \startlongtable

\begin{deluxetable*}{lccc|ccccccc}
\tablecaption{\label{ObsLog} Orbital parameters of CCs observed for this work: semi-major axis (a), inclination (i), and eccentricity (e) from the MPC. Our observing circumstances are also reported.   }
\tablewidth{0pt}
\tablehead{
TNO  & a  & e  & i & Date                     & $\#$  &  $\Delta$ &   r$_h$  & $\alpha$ &  Filter & Telescope\\
        & [AU]  &   &  [$^{\circ}$] &  UT  &   &    [AU]  &  [AU] & [$^{\circ}$] &   & 
}
   \startdata
 (58534) 1997~CQ$_{29}$ &  	45.407	&0.118& 	2.9  & 03/18/2017 &  3   & 42.113 &  43.103  & 0.1 &   VR     & DCT     \\ 
 Logos-Zoe &    &     &  &    & &          &     & &  &     \\\hline
  2000~CL$_{104}$   & 44.447	& 0.074&1.2  & 03/08/2016   &   6  & 42.709 & 43.700 & 0.1 &  VR &  Magellan   \\ 
     &&&    &  03/09/2016  &   7  & 42.708 & 43.700 & 0.1 &  VR &  Magellan \\ \hline
      (138537) 2000~OK$_{67}$   &46.581&	0.140&	4.9 & 07/27/2015  &   15  & 39.458  & 40.121   & 1.1 & VR &  DCT     \\
     &&&            & 08/20/2015  &  4   & 39.211 &  40.120 & 0.6 & VR &  DCT     \\
      &&&          & 08/21/2015   &   9  & 39.204 & 40.120  & 0.6 &    VR &  DCT  \\
	&&&	& 09/03/2015  & 15 & 39.134 & 40.119 & 0.3&   VR &  DCT  \\\hline
	 2000~OU$_{69}$ & 43.401 &0.054  & 4.4& 08/19/2015  & 8 & 40.115 &  41.120  &  0.2 &  VR &  DCT\\ \hline
	 2001~QS$_{322}$ &43.995 & 0.039 & 0.2 & 09/06/2016  &  19 & 41.502 & 42.417 & 0.6 & VR &  DCT  \\ \hline
 (363330) 2002~PQ$_{145}$ & 43.982 & 0.043 & 3.1 & 08/21/2015  &  20   & 44.665 &  45.674 & 0.1 & VR &  DCT  \\  
		  &&&   & 09/03/2015  & 9 & 44.707 & 45.674 & 0.4 & VR &  DCT  \\  
           &&&       & 09/04/2015  &  6 &  44.711 & 45.674 &0.4& VR &  DCT   \\  \hline
 (149348) 2002~VS$_{130}$ & 44.812 & 0.120 & 3.0 & 12/01/2015  &  8 & 41.581 & 42.563 & 0.1& VR &  DCT  \\  
           &&&      & 12/03/2015  & 4 & 41.579 & 42.563 & 0.1& VR &  DCT   \\ 
 		&&& & 12/04/2015  &  12 & 41.576 &42.564 & 0.1 & VR &  DCT     \\  \hline
 2003~QE$_{112}$ &43.118 &0.040  &4.2 & 10/01/2016  &  7  & 43.734 & 44.639  &0.6 &  VR &  Magellan   \\ \hline
 2003~QJ$_{91}$  &44.407 & 0.038 & 2.5 & 10/01/2016  &  7  &43.586 & 44.393 & 0.8 &  VR &  Magellan   \\ \hline
 2003~QY$_{111}$  & 43.269 & 0.039 &2.9 & 10/28/2017&  14& 41.533 & 42.463 & 0.5 & VR & DCT   \\\hline
 2003~SN$_{317}$ &42.430&0.045 &1.5& 08/22/2015  & 4 & 41.222 &41.900 &1.0&  VR &  DCT  \\ \hline
 2003~YU$_{179}$ &46.546 &0.156 &4.9& 02/14/2016  &19 & 39.470 & 40.424  &0.4&  VR &  DCT   \\\hline
  (444018) 2004~EU$_{95}$ &44.443& 0.048 &2.8& 05/17/2018 & 9 &  41.546   & 42.526         &0.3 &VR &  Magellan   \\
								&&&	& 05/18/2018  &  1   & 41.550 & 42.526      &0.4 & VR&  Magellan  \\  \hline
 2004~HD$_{79}$  &46.295&0.027&1.3&  04/23/2017 &  6  & 46.309 & 47.175 & 0.6 &    r' & Magellan  \\
                &&& & 04/24/2017 &12 &46.300 & 47.175  &0.6& r' &  Magellan  \\  \hline
 (469610) 2004~HF$_{79}$&43.623&0.035 & 1.5 & 04/22/2017 & 3  & 41.373 &42.221 & 0.7 &  r' & Magellan   \\ 
               &&&           &04/24/2017 & 11 & 41.364 & 42.220& 0.7 &  r' & Magellan  \\  \hline
(444025) 2004~HJ$_{79}$ &44.253&0.047&3.3& 05/17/2018  &  4   & 43.428 & 44.412  & 0.3 &VR &  Magellan  \\
								&&& & 05/18/2018  &  9   & 43.431  & 44.412  & 0.3 & VR &  Magellan  \\
								&&&	 & 05/19/2018  & 3    &43.436&   44.412   &0.3 &VR & Magellan  \\ 
      &&&	 & 02/02/2019  & 3    & 44.405  &  44.381    &1.3 &VR & Magellan  \\  
       &&&	 & 02/03/2019  & 2    &  44.386  &   44.380    &1.3 &VR & Magellan  \\ 
         &&&	 & 02/28/2019  &  5    &  43.959  &   44.377   & 1.2 &VR & Magellan  \\ 
           &&&	 & 03/01/2019  &  6    & 43.944   &    44.377  &1.2  &VR & Magellan  \\ 
            &&&	 & 03/02/2019  &  5    &  43.928  &  44.377    & 1.2 &VR & Magellan  \\ 
          \hline
 2004~HP$_{79}$ &48.029 & 0.191 & 2.2&05/22/2018           &5 & 37.883 &38.877 &0.3 & VR & DCT \\ \hline
  2004~MT$_{8}$ &43.380 &0.036 &2.2& 05/16/2018  &  6   & 44.539 &   44.926    & 1.2 &VR &  Magellan   \\  
					&&& & 05/17/2018  &  2   & 44.522 &  44.926     &1.2 & VR &  Magellan \\
					&&& & 05/19/2018  &  4   & 44.492  &  44.926     & 1.2 & VR &  Magellan  \\  \hline
 2004~MU$_{8}$&45.131 &0.075 &3.6 & 05/18/2018  &  11   & 47.259 &  47.638     &1.1 & VR &  Magellan \\ 
 					 &&&& 06/13/2018  &   10  & 46.891  &   47.635  & 0.8 & VR  &  DCT    \\  
 					 \hline
 2004~OQ$_{15}$&43.945&0.129 &9.7 & 07/02/2017 & 13   &38.382& 39.280 & 0.7&   VR &DCT   \\ \hline
 2004~PV$_{117}$ &46.348 &0.159 &4.3& 09/06/2016  &16& 39.493 &  40.455& 0.4 & VR &  DCT \\ \hline
 2004~PX$_{107}$ &43.854 &0.060 &3.0 & 07/03/2017 & 5 & 40.991 & 41.838 &0.8 &   VR & DCT   \\ \hline
 	 2004~PY$_{107}$ & 44.447 & 0.101 & 1.6 & 08/12/2018  & 4  & 41.093&  42.105 &  0.1&  VR &  DCT   \\  
\hline
 2004~VC$_{131}$&43.728 &0.070&0.5 & 10/28/2017  &27 &39.850  &  40.759 & 0.6 &  VR &  DCT   \\
						 &&& & 11/21/2017  & 4  &  39.772 & 40.760  & 0.0  & VR   & Magellan    \\
	      &&& & 11/22/2017  & 3  &39.773& 40.760& 0.0  & VR  & Magellan    \\
	  &&& & 11/23/2017  & 5 & 39.773 &  40.760 & 0.1  & VR & Magellan    \\			
	  \hline
	 2004~VU$_{75}$ & 43.543 & 0.136 & 3.3 & 08/12/2018  &  9 &43.343&  43.624 &  1.3&  VR &  DCT   \\  
	  &&& & 10/06/2018 & 4 & 42.741  & 43.643 & 0.6  & VR   &   DCT   \\	
	  &&& & 11/09/2018 & 8 & 42.675  & 43.655 & 0.2  & VR   &   DCT   \\	 
	  	  &&& & 11/12/2018 & 5 & 42.686   & 43.656 & 0.3  & VR   &   DCT   \\	 
	  	  	  &&& & 11/13/2018 & 14 &  42.690   & 43.656 &  0.3 & VR   &   DCT   \\	
        &&& & 12/08/2018 &  4 &  42.886 & 43.665  & 0.8 & VR  &   Magellan   \\	 
        &&& & 12/09/2018 &  4 & 42.897  &  43.665 & 0.8 & VR   &   Magellan   \\	 
        &&& & 12/10/2018 &  4 & 42.908  & 43.666  & 0.8 & VR  &   Magellan   \\	 
        &&& & 12/11/2018 &  4 & 42.921  & 43.666  &  0.9& VR   &   Magellan   \\	 
        &&& & 12/12/2018 &  4 &  42.933 & 43.667  & 0.9 & VR  &   Magellan   \\	 
	  	  	  \hline
 2005~EX$_{297}$ &44.053  &0.115 & 4.8& 03/14/2016  &   12  &43.237& 44.198 &0.3 & VR &  DCT  \\ \hline
  2005~JP$_{179}$ &43.219 &0.029 &2.1 & 05/22/2018  & 2  & 41.486  & 42.469  &0.3  &   VR &  DCT    \\
        &&&	 & 02/02/2019  & 3    & 42.589 &  42.485   &1.3 &VR & Magellan  \\    \hline
 2005~PL$_{21}$  &46.750 &0.153 &4.7& 10/01/2016  &  7  & 43.618   &  44.415  & 0.8&    VR &  Magellan   \\ \hline
 2010~TF$_{192}$ &43.144 &0.022 &2.3& 10/28/2017 & 6 & 42.480  &  43.390& 0.5&  VR &  DCT  \\ \hline
 2010~TL$_{182}$ &43.695 &0.056 &1.6& 09/06/2016  & 17  & 40.600   & 41.339   & 1.0 &   VR &  DCT  \\ \hline
 2011~BV$_{163}$&44.013 &0.100 &4.5 & 02/02/2017  & 10 &  38.786 & 39.771   & 0.1 & VR &  DCT  \\\hline
  2012~DA$_{99}$ &43.025 &0.039 &3.2& 05/16/2018  &   5  & 40.580 &  41.377      & 0.9& VR&  Magellan   \\ 
                            &&&& 05/17/2018  &  1   &    40.590&  41.377     & 0.9& VR & Magellan   \\        &&&	 & 02/02/2019  & 3    & 40.956  &  41.369    &1.2 &VR & Magellan  \\   
  &&&	 & 02/28/2019  &  8    & 40.597  &  41.369    & 0.9 &VR & Magellan  \\  
    &&&	 & 03/01/2019  &   8   & 40.586  &  41.369    & 0.9 &VR & Magellan  \\ 
      &&&	 & 03/02/2019  &   2   & 40.576  &  41.369    & 0.8 &VR & Magellan  \\  \hline
2012~DZ$_{98}$ &42.098&0.027 &2.8& 05/18/2018  &   6  & 42.003 &    42.079& 1.0& VR &  Magellan \\ 
					&&&& 05/19/2018  &  3   & 42.014 &    42.709   & 1.0& VR  & Magellan   \\  \hline 
 2013~AQ$_{183}$ &46.330 &0.159 &2.6&   02/02/2017 &  8   & 37.965 &  38.944  & 0.2 &    VR &  DCT  \\
		&&& & 03/18/2017  &  16   & 38.165&  38.944  &  0.9&    VR &  DCT  \\\hline
 2013~EM$_{149}$ &45.564 &0.061  &2.6& 05/17/2018 & 9 & 42.185    & 43.168 &   0.3    & VR &  Magellan    \\  
 					&&&& 05/18/2018  &   1  &42.193 &  42.169     &0.3 & VR &Magellan   \\ \hline
 2013~FA$_{28}$ &44.407 &0.043 &1.5& 02/02/2017  & 11 & 44.669 &  44.964  & 1.2 &   VR &  DCT  \\
       &&&	 & 02/02/2019  & 4    & 44.644  &  44.890    &1.2 &VR & Magellan  \\  
       &&&	 & 02/03/2019  &3     &  44.629  &   44.890    &1.2 &VR & Magellan  \\ 
   &&&	 & 03/01/2019   &4     &  44.226  &  44.888    & 0.9 &VR & Magellan  \\
   &&&	 &  03/02/2019  &5     &  44.213  & 44.888    & 0.9 &VR & Magellan  \\ \hline
 2014~GZ$_{53}$   &44.177&0.042 &5.9& 05/18/2018   &   8   & 41.356 &  42.321 & 0.4 & VR & Magellan   \\ 
 					  &&&& 05/19/2018  &  4   & 41.360 &  42.321     & 0.4& VR &  Magellan  \\  \hline 
 2014~LQ$_{28}$   &43.662 &0.096 &1.3& 09/28/2016  &  8  &38.803 &  39.790  &0.2  &    VR&  Magellan   \\ \hline
 2014~LR$_{28}$   &44.161&0.052 &1.5& 10/01/2016  &  7  & 45.445& 46.116  &0.9 &  VR &  Magellan  \\ \hline
 2014~LS$_{28}$  &43.614 &0.068 &3.8&  04/22/2017  &  4 & 41.230  & 42.159  &0.5 &   r' &  Magellan \\ 
		 &&&& 04/23/2017   &  6 & 41.227 &  42.159 & 0.5 & r' &  Magellan    \\ 
		&&&& 04/24/2017   & 8 & 41.220   &  42.159&  0.5 & r'  &  Magellan   \\ 
\hline
 2014~OA$_{394}$ &46.816 &0.187 & 4.4& 09/28/2016  & 8  & 37.438 &38.275  & 0.8 & VR&  Magellan  \\\hline
 2014~OM$_{394}$  &44.001 &0.078 &2.4& 09/28/2016  &  8  & 45.743 & 46.741   & 0.1  &VR &  Magellan  \\\hline
\enddata
\end{deluxetable*}

\begin{deluxetable*}{lccc|cc|c}
\tablecaption{\label{Summary_photo} We report our findings with the object's periodicity (P), the full lightcurve amplitude ($\Delta m$). The zero phases without light-time correction ($\phi_0$). Absolute magnitudes (H from the MPC) used to estimate the diameters (D). We also indicate if the object is a known or not resolved binary \citep{Noll2008b, Stephens2006}. Some objects have not been observed for companion search (to our knowledge), thus it is unknown if they are binary or not and we used a question mark to identify them.    }
\tablewidth{0pt}
\tablehead{
TNO &  P. & $\Delta m$ &   $\phi_{0}$ & H & D &  Resolved Binary?\\
        & [h]  & [mag]  &  [2450000+ JD] &[mag] & [km]   &no/yes/? \\
        &    &    &   &  & 0.04/0.20  &                 }
 \startdata
 (58534) 1997~CQ$_{29}$   &$>$1.1 & $>$0.5  & 7830.68212 & 6.6 & 318/142 &Yes  \\ 
2000~CL$_{104}$   & $>$5.5 & $>$0.2 &7455.58917 & 6.2 &382/171& No\\ 
  (138537) 2000~OK$_{67}$   &$>$6 & $>$0.15 & 7230.85274 &  6.2 &382/171& No \\  
 2000~OU$_{69}$   & $>$2.5 & $>$0.15 &  7253.68460 & 6.6 &318/142& No\\ 
  2001~QS$_{322}$ &  $>$6 & $>$0.3 & 7637.73440& 6.4 &349/156 & ?\\  
 (363330) 2002~PQ$_{145}$   &  ... & $\sim$0.1&7255.70248 & 5.5&528/236&No \\ 
  (149348) 2002~VS$_{130}$ & ... & $\sim$0.1& 7357.93353 &6.3 &365/163&?\\ 
  2003~QE$_{112}$ & ... &$\sim$0.1 & 7662.51963 &6.6&318/142 &? \\ 
 2003~QJ$_{91}$   & $>$6 &$>$0.2&   7662.50988  &6.7&304/136& ? \\ 
 2003~QY$_{111}$  &  $>$5.5 & $>$0.2&  8054.64931  & 6.9&277/124 &  ?  \\ 
 2003~SN$_{317}$ &...& $\sim$0.1  &  7256.94671 &6.5&333/149& ?\\ 
 2003~YU$_{179}$  & $>$5.5&$>$0.2 & 7432.60937  &6.8&290/130&?\\ 
(444018) 2004~EU$_{95}$    & ... & $\sim$0.1 &8255.49186 &7.0&265/118&?  \\
 2004~HD$_{79}$  &$>$7&$>$0.15&  7866.84916& 5.7 &481/215&  ? \\
  (469610) 2004~HF$_{79}$  &... & $\sim$0.15& 7865.89661 &6.3& 365/163&   ?\\ 
   (444025) 2004~HJ$_{79}$    & $>$7.5 & $>$0.20 & 8255.50327 &6.9&277/124& ?\\
    2004~HP$_{79}$  &$>$3 &$>$0.15 & 8260.72835  & 6.7 & 304/136 & ?   \\
      2004~MT$_{8}$  & $>$2&  $>$0.2 & 8254.83578&6.5&333/149&   ?\\
   2004~MU$_{8}$    &$>$2& $>$0.48& 8256.84049 &6.0&419/188 &Likely contact binary? \\ 
 2004~OQ$_{15}$ & ...&$\sim$0.1& 7936.84200& 6.8&290/130& ?  \\
  2004~PV$_{117}$  &...&$\sim$0.1&  7637.63756&6.5&333/149&?\\ 
  2004~PX$_{107}$  &...& $\sim$0.1  & 7937.87280 &7.2 &241/108&  No \\  
   2004~PY$_{107}$ &...  & $\sim$0.1 & 8342.76970    &6.4 &349/156 & ?  \\ 
 2004~VC$_{131}$ & 15.7 & 0.55$\pm$0.04 &  8054.68366  &6.0&419/188& Likely contact binary\\
   2004~VU$_{75}$$^{a}$ &  $>$3 & $>$0.42 & 8342.83471  & 6.7 &  304/136& Likely contact binary?\\ 
  2005~EX$_{297}$  &... &$\sim$0.1&  7461.63724 &6.1& 400/179&?\\  
  2005~JP$_{179}$    &... & $\sim$0.08 & 8260.78454 &6.4&349/156&? \\ 
 2005~PL$_{21}$  & $>$4 &$>$0.15 & 7662.51333 &6.6&318/142&?  \\  
  2010~TF$_{192}$    & $>$2.5& $>$0.3 &8054.90005&  6.1&400/179&?\\  
 2010~TL$_{182}$  &$>$5.5 & $>$0.25 & 7637.76776 & 6.3&365/163&?\\  
 2011~BV$_{163}$  &$>$2.5 & $>$0.15 &  7786.85122& 6.4 &349/156&?\\ 
 2012~DA$_{99}$    &...&$\sim$0.1 & 8254.49418 &6.5&333/149&? \\ 
2012~DZ$_{98}$    &$>$5.1&$>$0.2& 8256.48601 &6.5&333/149&? \\
2013~AQ$_{183}$  &$>$5 &$>$0.15&  7786.83242&6.5 &333/149&?\\
2013~EM$_{149}$    & ...& $\sim$0.1& 8255.49759 &6.8&290/130& ?\\
 2013~FA$_{28}$   & ... & $\sim$0.1 &   7786.96917&6.0&419/188&?\\ 
  2014~GZ$_{53}$    & ...& $\sim$0.1 &8256.53896 &6.0&419/188&? \\
  2014~LQ$_{28}$ A  &...  & $\sim$0.08 &   7659.59329 & 5.7 &481/215& Yes \\ 
    2014~LQ$_{28}$ B  & ...  & $\sim$0.11 &  ...  & ...&...&...\\ 
    2014~LR$_{28}$  & $>$4 & $>$0.25&   7662.50470& 5.0& 665/297 &?\\ 
 2014~LS$_{28}$    & 11.04 & 0.35 & 7865.83999& 5.8&460/206&?\\ 
 2014~OA$_{394}$   &$>$3 & $>$0.15 & 7659.58800&6.5&333/149&?\\
 2014~OM$_{394}$    & ... &$\sim$0.1 & 7659.60258&5.9&439/196& ?  \\
 \hline 
 \enddata
 \tablenotetext{a}{Potential rotational periods of 8.46~h, 10.2~h, or 12.9~h.}
\end{deluxetable*}
%\end{longrotatetable}

\clearpage
\startlongtable
%\begin{longrotatetable}
\begin{deluxetable*}{lccccc}
%\tabletypesize{\tiny}
%\rotate
\tablecaption{\label{Summary_CCs} Summary of the published lightcurve studies of the dynamically Cold Classical TNOs.    }
\tablewidth{0pt}
\tablehead{
TNO & P. (single)   & P. (double) & $\Delta m$  & H & Reference \\ 
        & [h] &  [h]& [mag] & [mag] &   \\ }
\startdata
(19255) 1994~VK$_{8}$$^{d}$ & 3.9/4.3/4.7/5.2 & 7.8/8.6/9.4/10.4 & 0.42 & 7.0 & RT99\\
 & 4.75 & - & - & ... & CB99\\
 (58534) 1997~CQ$_{29}$ & - & - & $\sim$0.8 & 6.6 & N08\\
Logos-Zoe$^{a}$ & &  &   &  &  \\
(79360) 1997~CS$_{29}$   & - & - & $<$0.08 & 5.3 & SJ02\\
Sila-Nunam$^{a}$ &  & & &  &  \\
  & - & - & $<$0.22 & ... & RT99\\
  & 150.1488 & 300.2388 & 0.120$\pm$0.012/0.044$\pm$0.010 & ... & R14\\
  & - & - & 0.14$\pm$0.07 & ... & G12, BS13\\
  (66652) 1999~RZ$_{253}$ &- & - & $<$0.05 & 5.9 & LL06\\
  Borasisi-Pabu$^{a}$ &  &  &  &  & \\
    & 6.4$\pm$1.0 & - & 0.08$\pm$0.02 & ... & K06\\
    (80806) 2000~CM$_{105}$$^{a}$& - & $>$3 & $<$0.14 & 6.6 & LL06\\
    (88611) 2001~QT$_{297}$  & - & - & $<$0.15 & 5.8 & O03\\
     Teharonhiawako$^{a}$ &  &   &   &   &  \\
      & 5.50$\pm$0.01 or 7.10$\pm$0.02 & 11.0$\pm$0.02 or 14.20$\pm$0.04 & (0.32 or 0.30)$\pm$0.04 & ... & K06\\
(88611B) 2001~QT$_{297}$~B  & 4.75 & - & 0.6 & ... & O03\\ 
 Sawiskera$^{a}$ &  & &   & &  \\
 & 4.749$\pm$0.001 & 9.498$\pm$0.02 & 0.48$\pm$0.05 & ... & K06\\
(275809) 2001~QY$_{297}$$^{a}$ & 5.84 & 11.68 & 0.49$\pm$0.03 & 5.4 & T12\\
  & 12.2$\pm$4.3 & - & 0.66$\pm$0.38 & ... & K06\\
(126719) 2002~CC$_{249}$$^{b}$ & -& 11.87$\pm$0.01 & 0.79$\pm$0.04 & 6.6 & TS17\\
 2002~GV$_{31}$& - & 29.2 & 0.35$\pm$0.06 & 6.4 & P15 \\
2002~VT$_{130}$$^{a}$ & - & $>$4 & $>$0.21 & 5.6 & T14 \\
2003~BF$_{91}$$^{c}$ & \textbf{9.1}/7.3 & - & 1.09$\pm$0.25 & 11.7 & TB06\\
2003~BG$_{91}$$^{c}$ & \textbf{4.2}/4.5/4.6/4.9 & - & 0.18$\pm$0.075 & 10.7 & TB06\\
2003~BH$_{91}$$^{c}$ &  2.8 & - & $<$0.15 & 11.9 & TB06\\
2003~FM$_{127}$  &  6.22$\pm$0.02 & - & 0.46$\pm$0.04 & 7.1  & K06 \\
2003~QY$_{90}$ A$^{a}$ &  3.4$\pm$1.1 & - & 0.34$\pm$0.12 & 6.4 & KE06 \\
2003~QY$_{90}$ B$^{a}$& 7.1$\pm$2.9 & - & 0.90$\pm$0.36 & ... & KE06 \\
2003~QY$_{111}$& - & $>$2.5 & 0.72 & 6.9 & SS09\\
2005~EF$_{298}$$^{a}$ & \textbf{4.82}/6.06 & \textbf{9.65}/12.13 & 0.31$\pm$0.04 & 5.9 & BS13\\
(303712) 2005~PR$_{21}$$^{a}$ & - & $>$5.5 & $<$0.28 & 6.2 & BS13 \\ 
2013~SM$_{100}$  & - & $>$5 & $>$0.60 & 8.5 & A18 \\
2013~UC$_{18}$  & - & $>$5 & $>$0.44 & 8.3 & A18 \\
(505476) 2013~UL$_{15}$  & - & $>$5 & $>$0.36  & 6.6 & A18 \\  
2013~UP$_{15}$  & - & $>$5 & $>$0.29 & 7.5 & A18 \\
2013~UR$_{22}$  & - & $>$5 & $>$0.44  & 7.8 & A18 \\ 
2013~UY$_{16}$  & - & $>$5 & $>$0.36 & 7.6 & A18 \\
2013~UN$_{15}$  & - & $>$5& $>$0.56 & 7.3 & A18 \\
2013~UW$_{16}$  & - &  $>$5& $>$0.11& 7.3 & A18 \\ 
2013~UW$_{17}$  & - & $>$5  & $>$0.40 &7.6 & A18 \\
2015~RA$_{280}$ & - & $>$6 & $>$0.61  & 7.6 & A18 \\ 
2015~RB$_{280}$  & - &$>$6 & $>$0.55  & 7.6 & A18 \\ 
2015~RB$_{281}$  & - & $>$5 & $>$0.42  & 7.4 & A18 \\ 
2015~RC$_{280}$  & - & $>$6.5& $>$0.40  & 9.0 & A18 \\ 
2015~RE$_{280}$ & - & $>$6 & $>$0.20  & 7.9 & A18 \\ 
2015~RH$_{280}$ & - &$>$6 & $>$0.38  & 9.0 & A18 \\ 
2015~RH$_{281}$  & - & $>$6 & $>$0.44  & 8.4 & A18 \\ 
2015~RK$_{281}$  & - &$>$6 & $>$0.21  & 8.6 & A18 \\ 
2015~RO$_{281}$ & - & $>$6 & $>$0.35  & 7.5 & A18 \\
2015~RP$_{281}$ & - & $>$6 & $>$0.42  & 7.7 & A18 \\ 
2015~RQ$_{280}$ & - &$>$6& $>$0.30  & 8.8 & A18 \\ 
2015~RT$_{279}$  & - &$>$6 & $>$0.40  & 8.2 & A18 \\ 
2015~RW$_{279}$  & - & $>$6.5 & $>$0.20  & 8.2 & A18 \\ 
2015~RZ$_{279}$ & - & $>$5 & $>$0.27  & 7.6 & A18 \\ 
uo3l88$^{f}$  & - & $>$5 & $>$0.50  & 8.3 & A18 \\ 
uo5t55$^{f}$  & - & $>$6 & $>$0.30 & 8.7 & A18 \\
\hline
\enddata
\tablenotetext{a}{Known resolved binary systems. In some cases, the primary and the satellite have been observed separately and a lightcurve for each is available, and thus we indicate both values individually.}
\tablenotetext{b}{Likely a contact binary \citep{ThirouinSheppard2017}.}
\tablenotetext{c}{The lightcurves of 2003~BG$_{91}$, 2003~BF$_{91}$, and 2003~BH$_{91}$ were obtained with \textit{HST} \citep{Trilling2006}. The lightcurve of 2003~BH$_{91}$ presents a very high dispersion and a rotational period of 2.8~h seems unlikely \citep{Sheppard2008, Thirouin2013}, and \citet{Trilling2006} were not confident about this result. Therefore, 2003~BH$_{91}$ is not considered for the purpose of this work.}
\tablenotetext{d}{Thanks to HST observations, no companions have been detected for 1994~VK$_{8}$ \citep{Noll2008b}.}
 \tablenotetext{e}{References list:  CB99: \citet{Collander1999}; RT99: \citet{Romanishin1999}; SJ02: \citet{Sheppard2002}; O03: \citet{Osip2003}; KE06: \citet{Kern2006}; K06: \citet{Kern2006_phd}; LL06: \citet{Lacerda2006}; TB06: \citet{Trilling2006}; N08: \citet{Noll2008b}; SS09: \citet{SantosSanz2009}; G12: \citet{Grundy2012}; T12: \citet{Thirouin2012}; BS13: \citet{Benecchi2013}; R14: \citet{Rabinowitz2014}; T14: \citet{Thirouin2014}; P15: \citet{Pal2015}; TS17: \citet{ThirouinSheppard2017}; A18: \citet{Alexandersen2018}.} 
 \tablenotetext{f}{uo3l88 and uo5t55 are not fully characterized by the OSSOS survey yet and thus have not be submitted to the MPC and have no official designation.}
 \end{deluxetable*}
%\end{longrotatetable}

\clearpage

 \startlongtable
\begin{deluxetable}{lcccc}
\tablecaption{\label{Corr} Orbital parameters of CCs observed for this work: semi-major axis (a), eccentricity (e), inclination (i), perihelion distance (q), and aphelion distance (Q) from the Minor Planet Center (November 2018). Our observing circumstances are also reported. Sila-Nunam is excluded/included in our samples as it is a tidally locked system. Two OSSOS objects are not fully characterized yet, and so they are not included in our search with orbital elements, but they are included in the samples for the $\Delta m$ vs. P.  \\
}
\tablewidth{0pt}
\tablehead{
Correlated values & Sample &  $\rho$ & SL & Nb \\
& & & [\%] & }
   \startdata
\textit{Full} & &   &  &  \\
\textit{lightcurves}    &&   &  &  \\
   $\Delta m$ vs. P & with Sila & 0.358 & 84 & 16 \\
   \hline
      $\Delta m$ vs. P$^{a}$ & no Sila & 0.596 & 98 & 15 \\
   \hline
      $\Delta m$ vs. P & Binaries$^{b}$ & 0.651 & 95 &10 \\
 \hline
         $\Delta m$ vs. i &with Sila & -0.126 & 38   & 16 \\
           \hline
 $\Delta m$ vs. i &no Sila & -0.151 & 43   & 15 \\
           \hline
 $\Delta m$ vs. i &Binaries& 0.061 & 16   & 10 \\
           \hline
      $\Delta m$ vs. e &with Sila  & -0.090 & 29   & 16 \\
           \hline
      $\Delta m$ vs. e &no Sila  & -0.285 & 72   & 15 \\
           \hline
            $\Delta m$ vs. e &Binaries& 0.224 & 55   & 10 \\
           \hline
         $\Delta m$ vs. a &with Sila &-0.099 & 31   & 16 \\
           \hline 
 $\Delta m$ vs. a &no Sila &-0.108 & 32   & 15 \\
           \hline         
 $\Delta m$ vs. a &Binaries& -0.073 & 19  & 10 \\
           \hline        
           $\Delta m$ vs. H &with Sila & 0.335  & 81   & 16\\
               \hline
             $\Delta m$ vs. H &no Sila & 0.212  & 58   & 15\\
               \hline             
      $\Delta m$ vs. H &Binaries& 0.556 & 91   & 10 \\
           \hline
         $\Delta m$ vs. q  & with Sila &0.088 & 28 & 16 \\
      \hline
          $\Delta m$ vs. q  & no Sila &0.275 & 71 & 15 \\
      \hline     
       $\Delta m$ vs. q &Binaries& -0.191 & 48   & 10 \\
           \hline
        $\Delta m$ vs. Q  & with Sila &-0.115 & 35  & 16 \\
         \hline 
        $\Delta m$ vs. Q  & no Sila &-0.240 & 64 & 15 \\
         \hline 
          $\Delta m$ vs. Q &Binaries& 0.159 & 40   & 10 \\
           \hline
           P vs. e &with Sila& -0.256  & 68  & 16 \\
           \hline
     P vs. e & no Sila& -0.097 &  28  & 15 \\
\hline
   P vs. e &Binaries& -0.360  & 72  & 10 \\
           \hline
         P vs. i & with Sila & -0.101 &  30  & 16 \\
		\hline
		 P vs. i & no Sila& -0.079 &  23  & 15 \\
\hline
   P vs. i &Binaries& -0.140  & 33  & 10 \\
           \hline
         P vs. a & with Sila& 0.135 &  40  & 16 \\
\hline
         P vs. a & no Sila& 0.175 &  49  & 15 \\
\hline
   P vs. a &Binaries& -0.152  & 35  & 10 \\
           \hline
         P vs. H & with Sila & -0.093 &  28  & 16 \\
                  \hline
         P vs. H & no Sila& 0.103 &  30  & 15 \\
\hline
                     P vs. H &Binaries& -0.147  & 34  & 10 \\
           \hline
         P vs. q  & with Sila &0.339 & 81 & 16 \\
      \hline
              P vs. q & no Sila& 0.197 &  54  & 15 \\
\hline 
         P vs. q &Binaries& 0.396  & 77  & 10 \\
           \hline
         P vs. Q  & with Sila &-0.094 & 28  & 16 \\
         \hline
          P vs. Q & no Sila& 0.047 &  14  & 15 \\
\hline
            P vs. Q &Binaries& -0.311  & 65  & 10 \\
\hline
\hline
    \textit{Full+partial} &  &   &  &  \\
    \textit{lightcurves} &  &   &  &  \\
          $\Delta m$ vs. P & All, with Sila  & -0.076 & 45 & 64\\   %
            \hline
               $\Delta m$ vs. P & All, no Sila  & -0.047 & 29 & 63\\   %
            \hline        
   $\Delta m$ vs. P &  H$\leq$6, no Sila & 0.142 & 35 & 11 \\  %
            \hline
 $\Delta m$ vs. P &  H$\leq$6, with Sila & -0.089 & 23& 12 \\  %
            \hline
             $\Delta m$ vs. P &  6$<$H$\leq$8 & 0.051  & 25  & 41   \\  %
            \hline
               $\Delta m$ vs. P & 8$<$H$\leq$12  & 0.056 & 15 & 13  \\  %
             \hline
     $\Delta m$ vs. i & All & -0.194 & 88 & 64 \\  %
             \hline
               $\Delta m$ vs. i & H$\leq$6     &  -0.087 & 23 &12 \\  %
            \hline
   $\Delta m$ vs. i & 6$<$H$\leq$8   & -0.270 & 91 & 41\\   %
            \hline
    $\Delta m$ vs. i & 8$<$H$\leq$12   & -0.408 &80 & 11  \\  %
           \hline
         $\Delta m$ vs. e &All& 0.062 &38 & 64 \\  %
             \hline    
     $\Delta m$ vs. e & H$\leq$6   & 0.129 &  33  & 12 \\  %
            \hline
   $\Delta m$ vs. e & 6$<$H$\leq$8   & 0.022 & 11  & 41 \\   %
            \hline
            $\Delta m$ vs. e & 8$<$H$\leq$12   & -0.472 & 86 & 11 \\  %
            \hline 
      $\Delta m$ vs. a &All  & 0.005 & 3 & 64 \\  %
             \hline 
       $\Delta m$ vs. a &  H$\leq$6 &0.100 & 26  & 12 \\  %
             \hline
    $\Delta m$ vs. a &  6$<$H$\leq$8 & -0.040  & 20 & 41  \\  %
            \hline     
        $\Delta m$ vs. a & 8$<$H$\leq$12  & -0.135  &  33  & 11 \\  %
           \hline  
    $\Delta m$ vs. H  &All& 0.273 & 97 & 64 \\  %
             \hline  
       $\Delta m$ vs. H & H$\leq$6 & 0.186 & 46  & 12 \\  %
             \hline       
        $\Delta m$ vs. H & 6$<$H$\leq$8 & 0.186 &  76 & 41 \\  %
             \hline   
       $\Delta m$ vs. H & 8$<$H$\leq$12 & 0.068 &  17 & 11  \\  %
             \hline         
        $\Delta m$ vs. q &All& -0.042 & 26 & 64 \\  %
             \hline  
       $\Delta m$ vs. q & H$\leq$6 &0.020 & 5  & 12 \\  %
             \hline       
        $\Delta m$ vs. q & 6$<$H$\leq$8 & -0.026 &  13 & 41 \\  %
             \hline   
       $\Delta m$ vs. q & 8$<$H$\leq$12 & 0.472 &  86 & 11  \\  %
             \hline                      
       $\Delta m$ vs. Q &All& 0.040 &25 &64 \\  %
             \hline  
       $\Delta m$ vs. Q & H$\leq$6 &0.230 & 55  & 12 \\  %
             \hline       
        $\Delta m$ vs. Q & 6$<$H$\leq$8 & 0.001 &  1 & 41 \\  %
             \hline   
       $\Delta m$ vs. Q & 8$<$H$\leq$12 & -0.472 &  86 & 11  \\  %
             \hline            
        P vs. e &All, with Sila & -0.270 & 97  & 64 \\  %
            \hline 
       P vs. e &All, no Sila & -0.207 & 90  &63 \\  %
            \hline 
             P vs. e &   H$\leq$6, with Sila &  -0.449  & 86  & 12\\  %
    \hline 
                 P vs. e &   H$\leq$6, no Sila &  -0.191  & 45  & 11 \\  %
    \hline 
            P vs. e &  6$<$H$\leq$8  & -0.075 &  37   & 41\\  %
    \hline 
            P vs. e &  8$<$H$\leq$12 & -0.342  &  72 & 11 \\  %
            \hline
         P vs. i &All, with Sila & -0.045 &  28  & 64 \\  %
    \hline 
             P vs. i &All, no Sila &  -0.047 &  29  & 63 \\  %
    \hline 
             P vs. i &   H$\leq$6, with Sila & -0.087    &23 & 12 \\  %
    \hline 
                 P vs. i &   H$\leq$6, no Sila & -0.121    & 30 & 11 \\  %
    \hline 
            P vs. i &  6$<$H$\leq$8  & 0.203 & 80 & 41 \\ %
    \hline 
             P vs. i &  8$<$H$\leq$12 &  -0.444  & 84 & 11 \\  %
    \hline 
         P vs. a & All, with Sila & -0.200 & 89  & 64 \\  %
    \hline 
             P vs. a & All, no Sila & -0.183 &  85  & 63 \\  %
    \hline 
            P vs. a &   H$\leq$6, with Sila& -0.181 &  45  & 12 \\  %
     \hline 
                 P vs. a &   H$\leq$6, no Sila &-0.087 & 22  & 11 \\  %
     \hline 
             P vs. a &  6$<$H$\leq$8  & -0.147 &  65 & 41 \\  %
    \hline 
            P vs. a &  8$<$H$\leq$12 &  -0.611  & 95 & 11 \\  %
    \hline
             P vs. H &All, with Sila & -0.410 &  99  & 64 \\  %
   \hline 
             P vs. H &All, no Sila & -0.364 &  99  & 63 \\  %
   \hline 
                P vs. H &   H$\leq$6, with Sila & -0.252 & 60    & 12 \\  %
    \hline 
                    P vs. H &   H$\leq$6, no Sila & 0.018 & 5    & 11 \\  %
    \hline 
             P vs. H &  6$<$H$\leq$8  & -0.375  &  98   & 41\\  %
    \hline 
           P vs. H &  8$<$H$\leq$12 &  0.417  & 81 & 11 \\%
   \hline
             P vs. q &All, with Sila & 0.138 &  73 & 64 \\  %
   \hline 
                P vs. q &All, no Sila & 0.071 &  42  & 63 \\  %
   \hline 
     P vs. q &   H$\leq$6, with Sila & 0.435 & 85   & 12 \\  %
    \hline 
         P vs. q &   H$\leq$6, no Sila & 0.210 & 49    & 11 \\  %
    \hline
       P vs. q &  6$<$H$\leq$8  & -0.076  & 37  & 41 \\  %
    \hline 
       P vs. q &  8$<$H$\leq$12 &  0.073  & 18 & 11 \\%
   \hline
            P vs. Q &All, with Sila &-0.255 &  97  & 64 \\  %
   \hline 
                P vs. Q &All, no Sila & -0.204 & 89  &63 \\  %
   \hline 
     P vs. Q &   H$\leq$6, with Sila & -0.313 & 70    & 12 \\  %
    \hline 
         P vs. Q &   H$\leq$6, no Sila & -0.110 & 27   & 11 \\  %
    \hline
       P vs. Q &  6$<$H$\leq$8  & -0.115  &  53   & 41 \\  %
    \hline 
       P vs. Q &  8$<$H$\leq$12 &  -0.489  & 88 & 11 \\%
   \hline
\hline
\enddata
\tablenotetext{a}{Without Sila-Numan and 2002~GV$_{31}$: $\rho$=0.718, SL=99$\%$. }
\tablenotetext{b}{Contact and resolved binaries.}
\end{deluxetable}
%\end{longrotatetable}
 
 \clearpage
 
\section*{Appendix A}

\begin{figure*}
\includegraphics[width=9cm, angle=0]{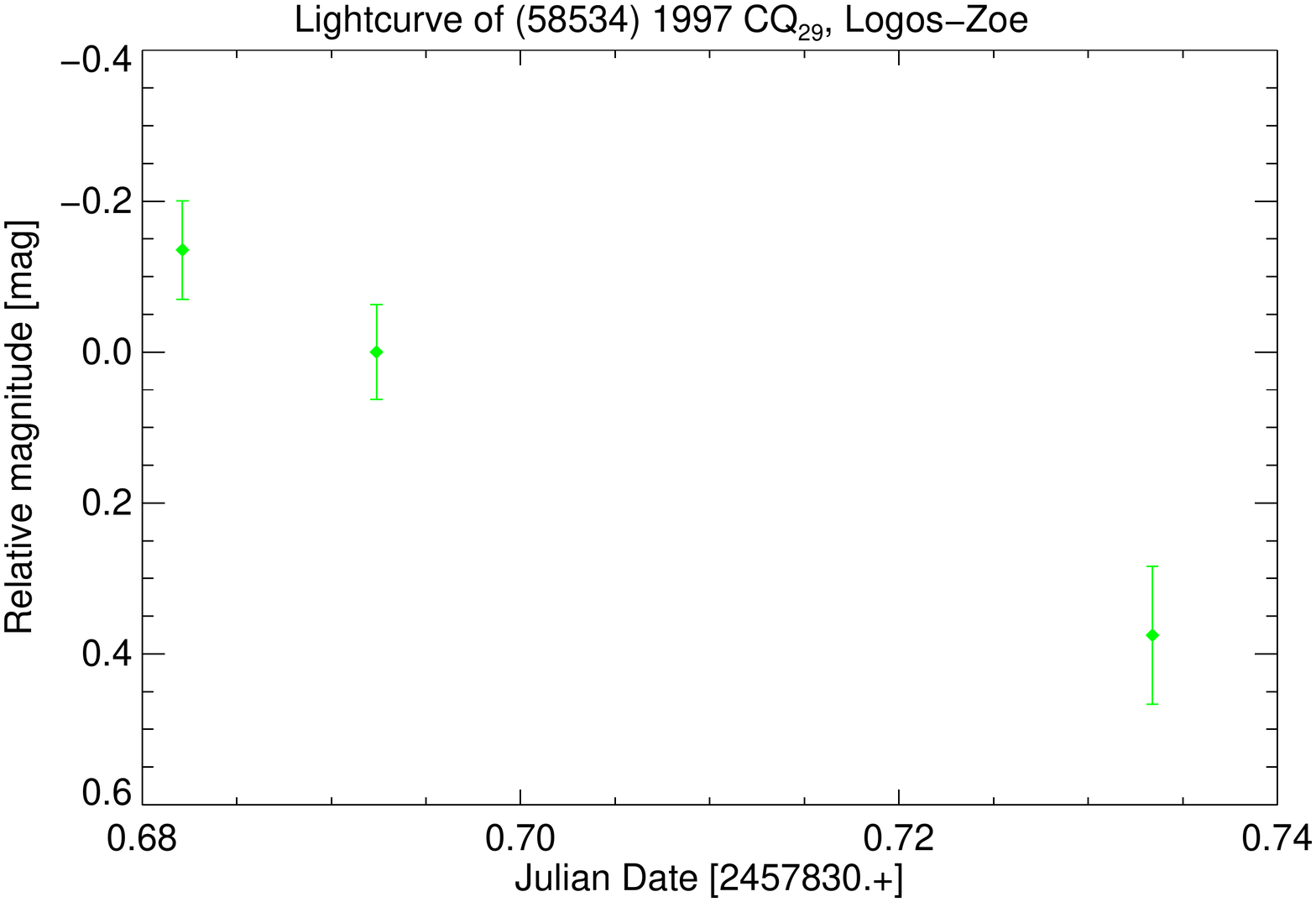}
 \includegraphics[width=9cm, angle=0]{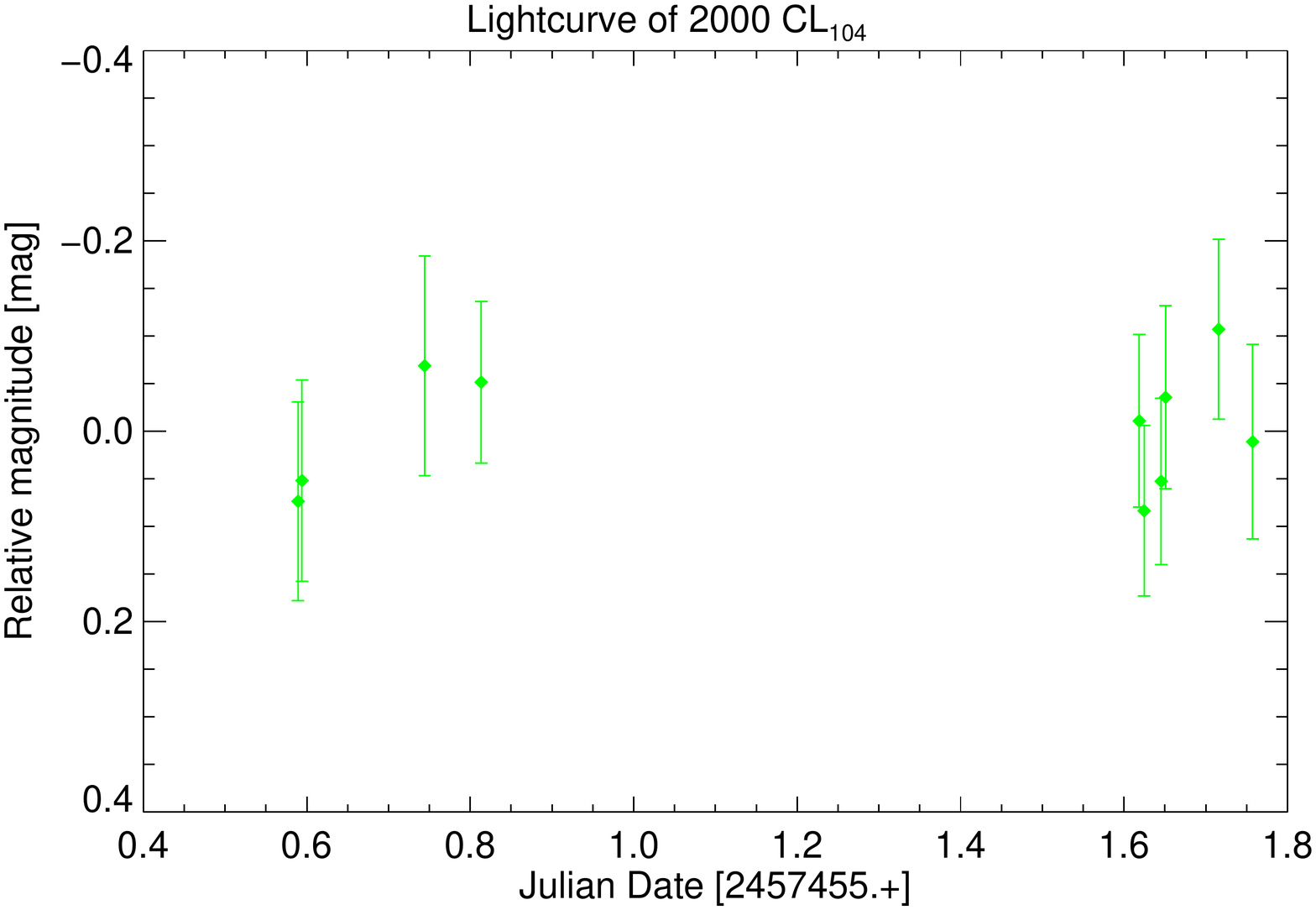}
 \includegraphics[width=9cm, angle=0]{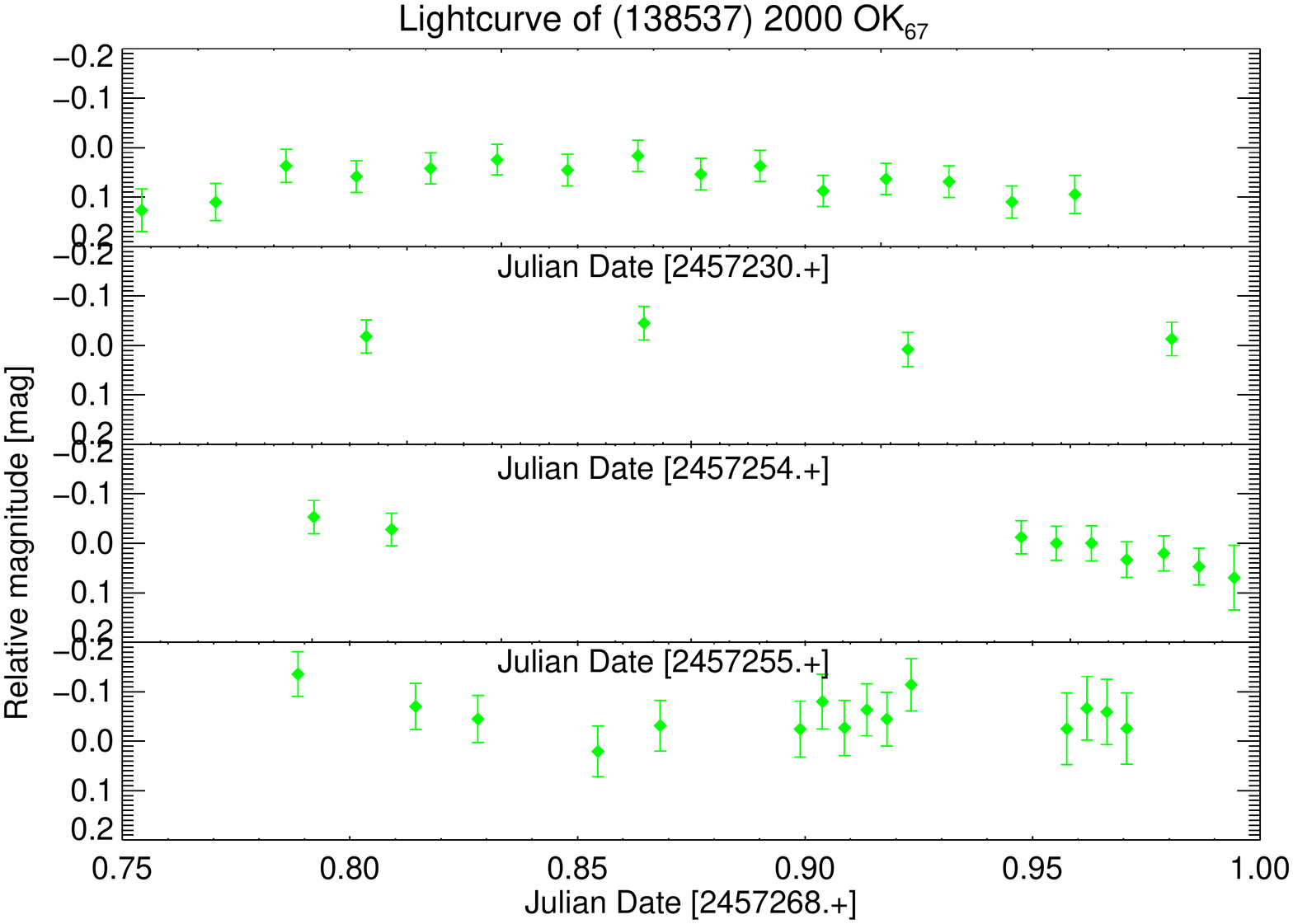}
 \includegraphics[width=9cm, angle=0]{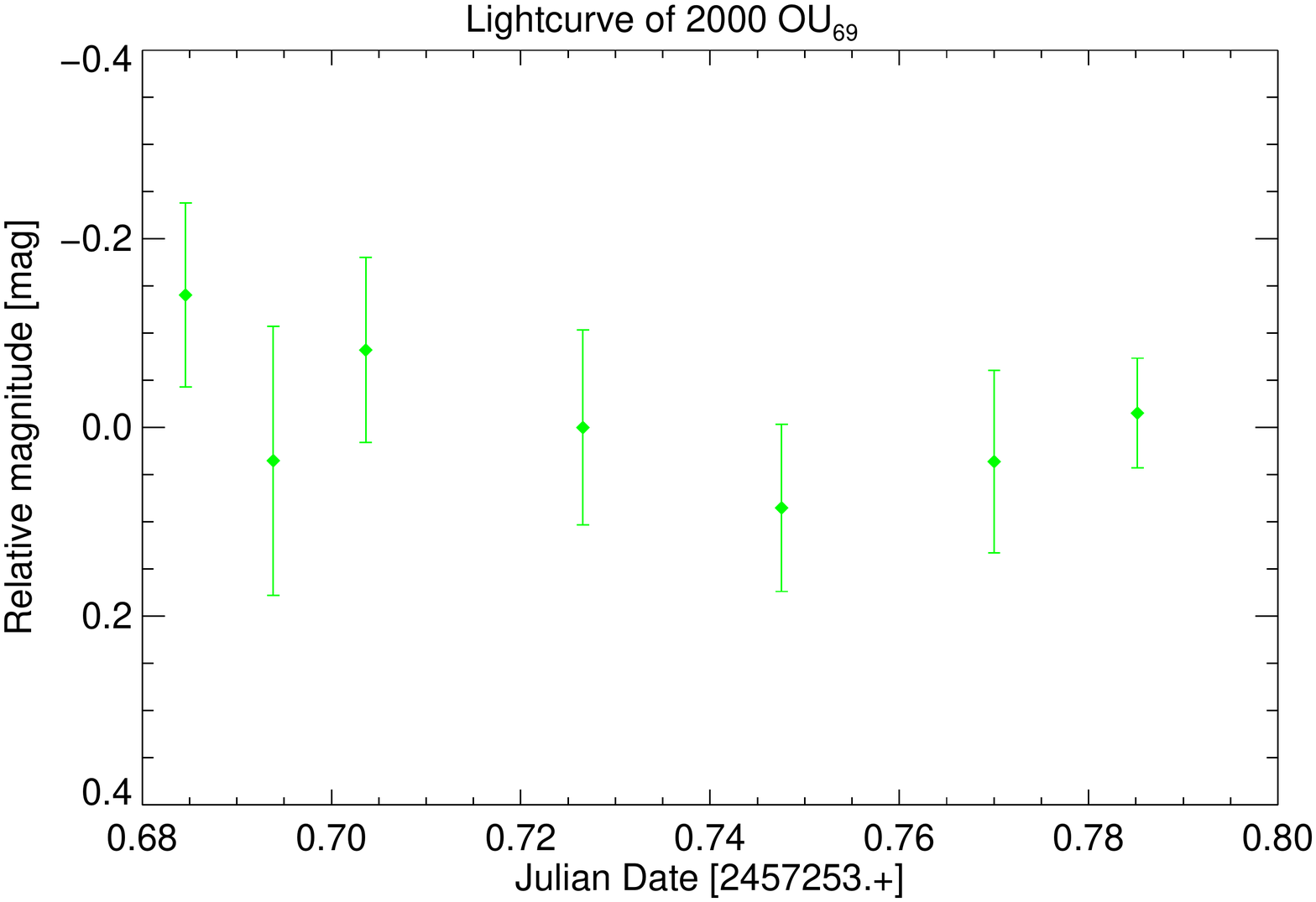}
 \includegraphics[width=9cm, angle=0]{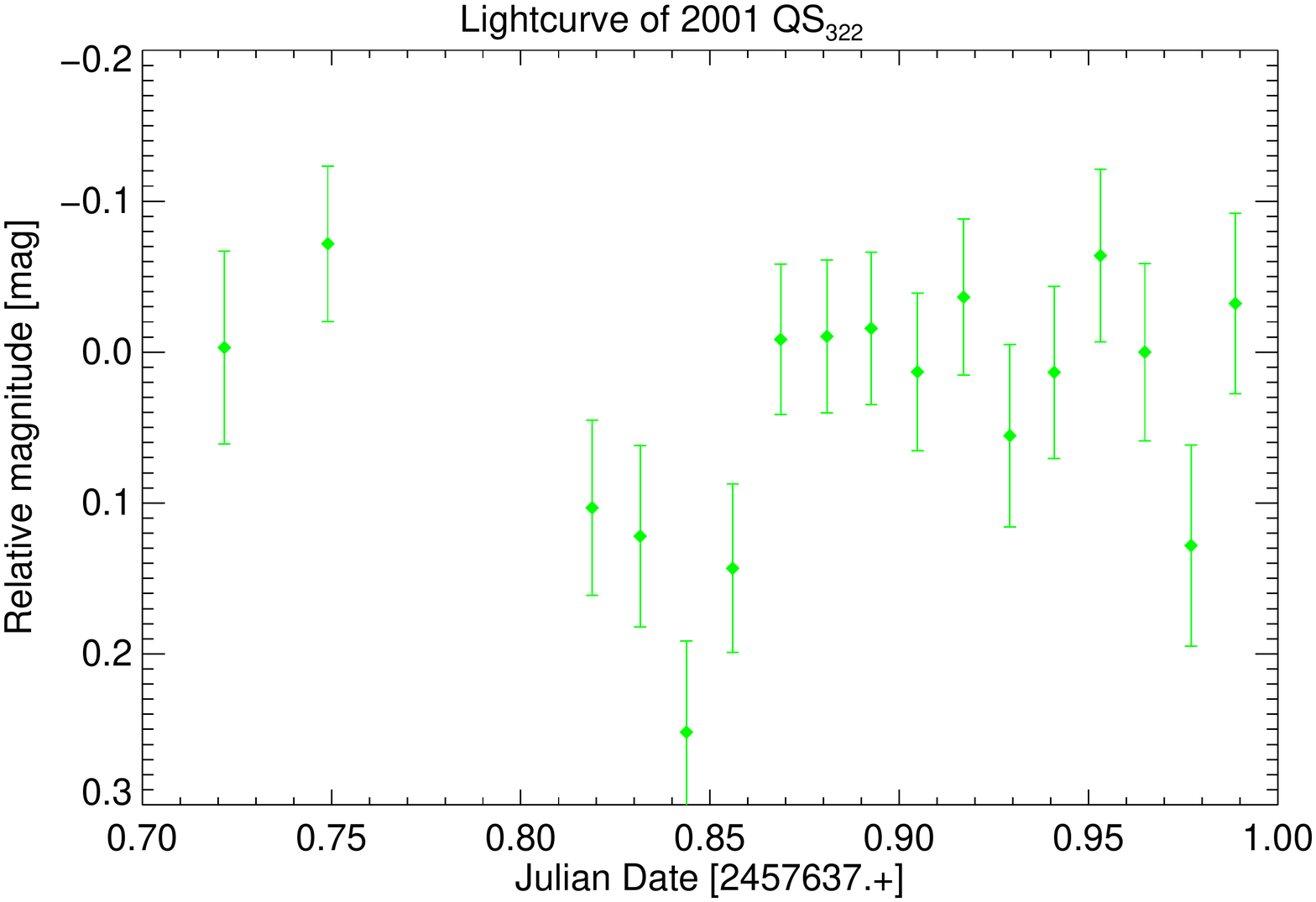}
 \includegraphics[width=9cm, angle=0]{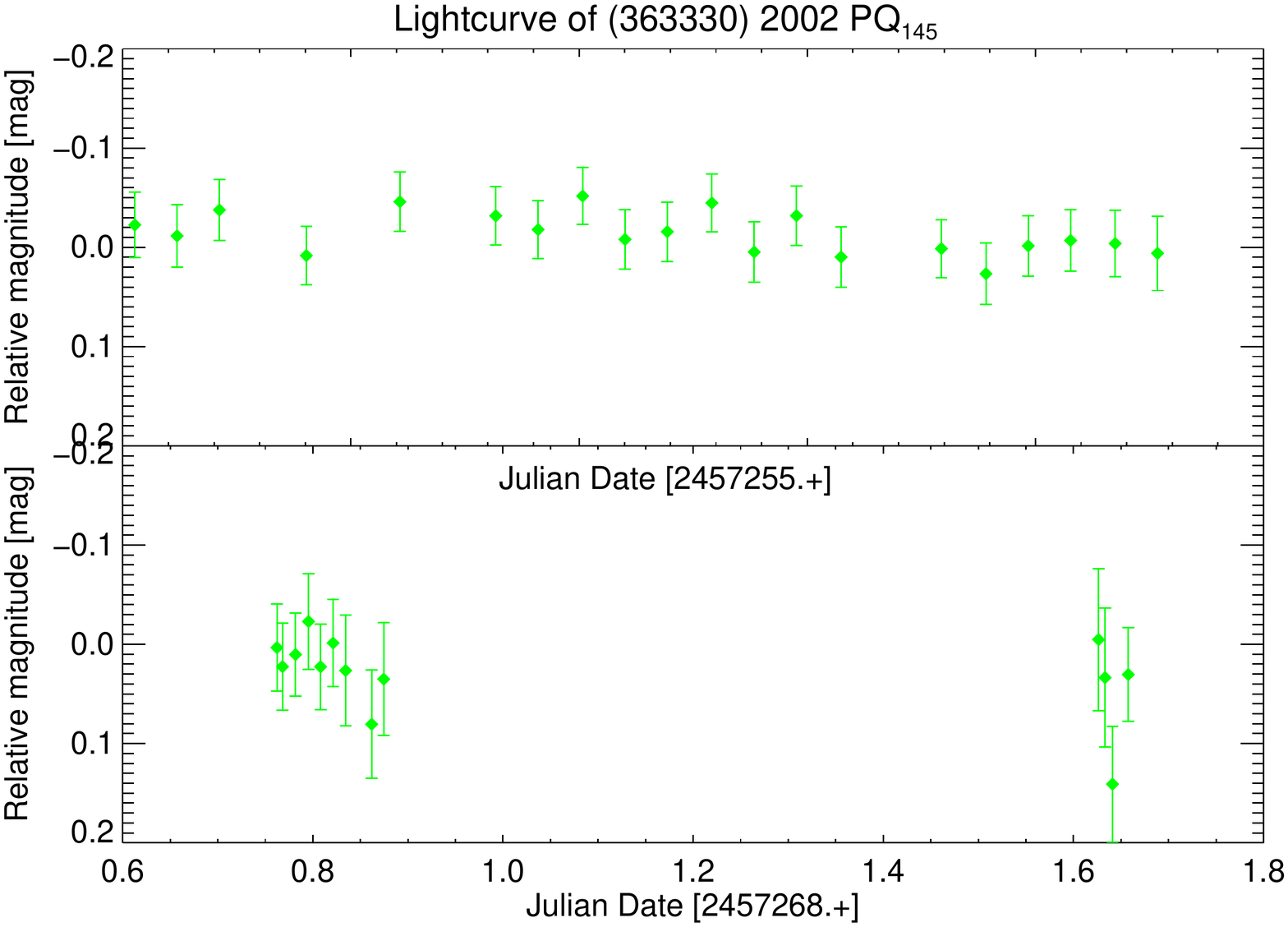}
\caption{ Partial lightcurves of several dynamically Cold Classical TNOs.  }
\label{fig:LC}
\end{figure*}

\begin{figure*}
  \includegraphics[width=9cm, angle=0]{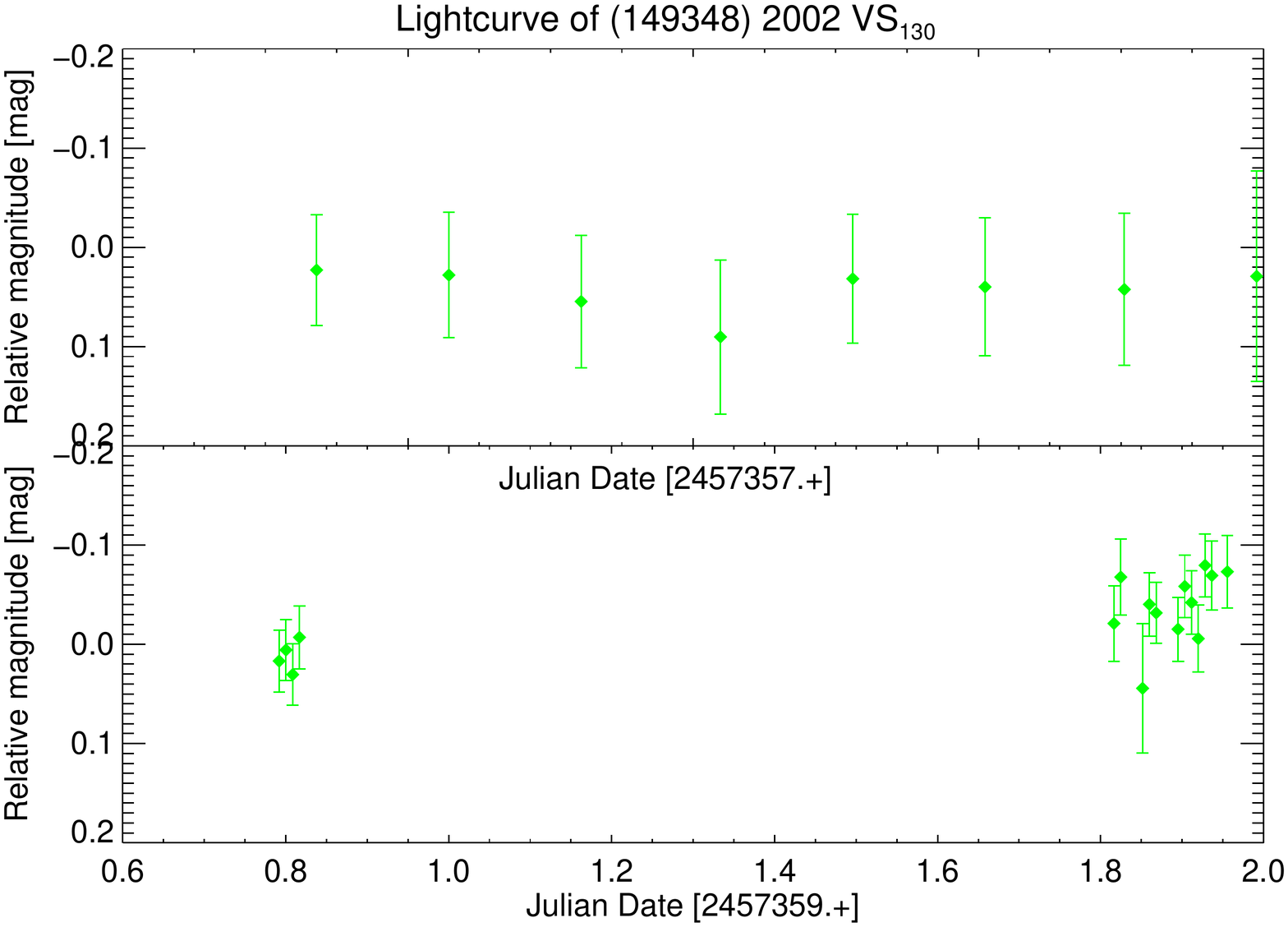}
 \includegraphics[width=9cm, angle=0]{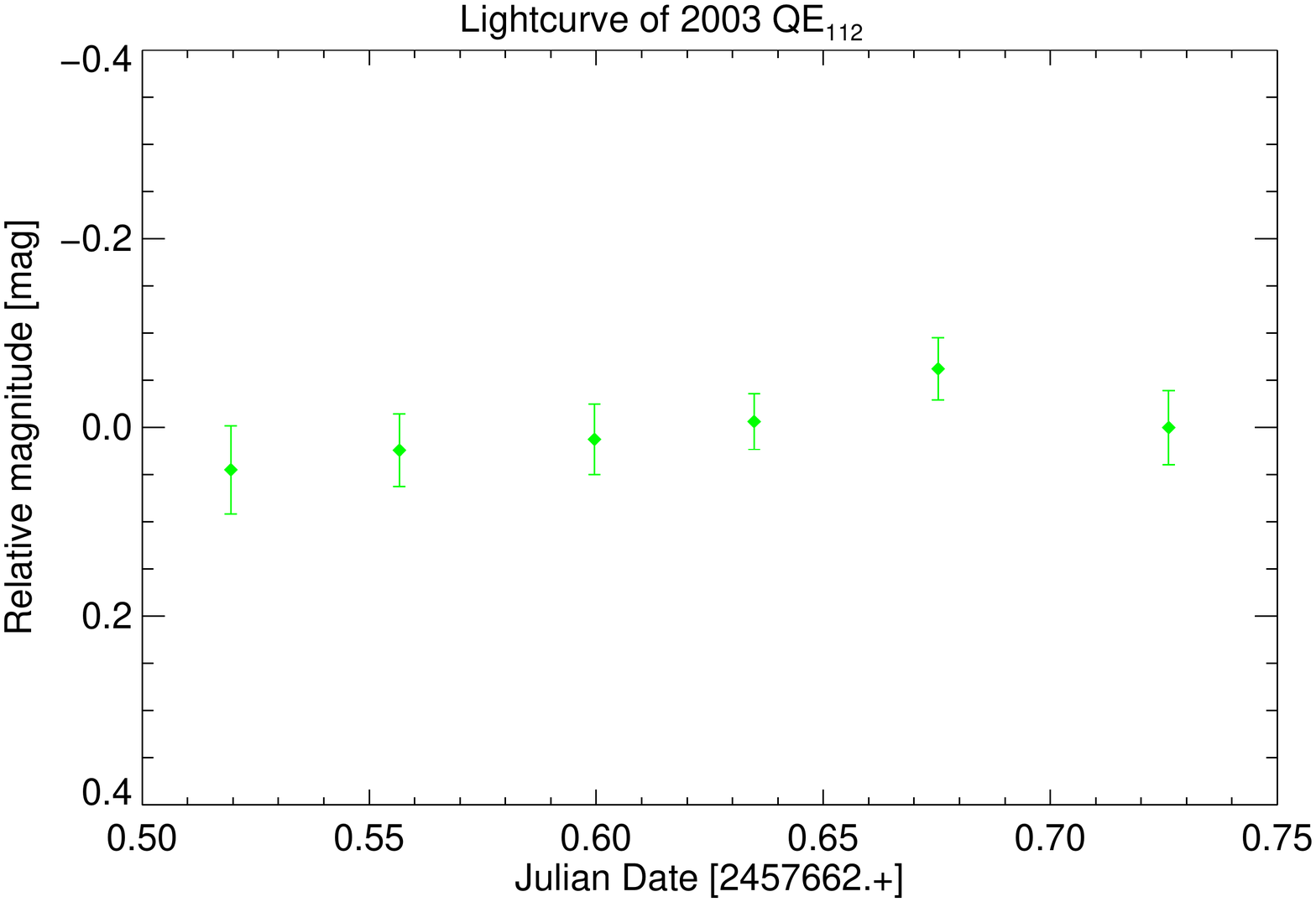}
  \includegraphics[width=9cm, angle=0]{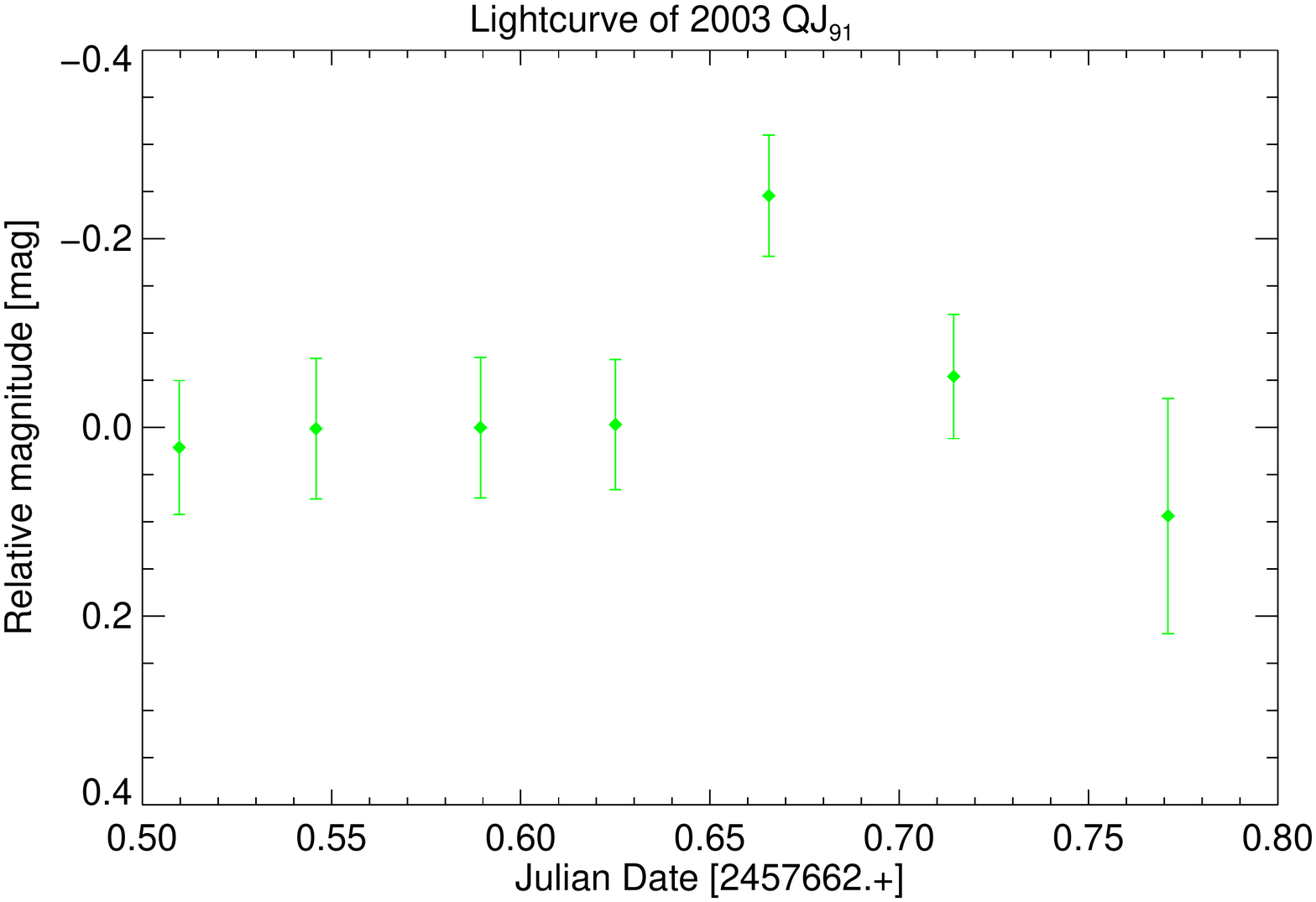}
 \includegraphics[width=9cm, angle=0]{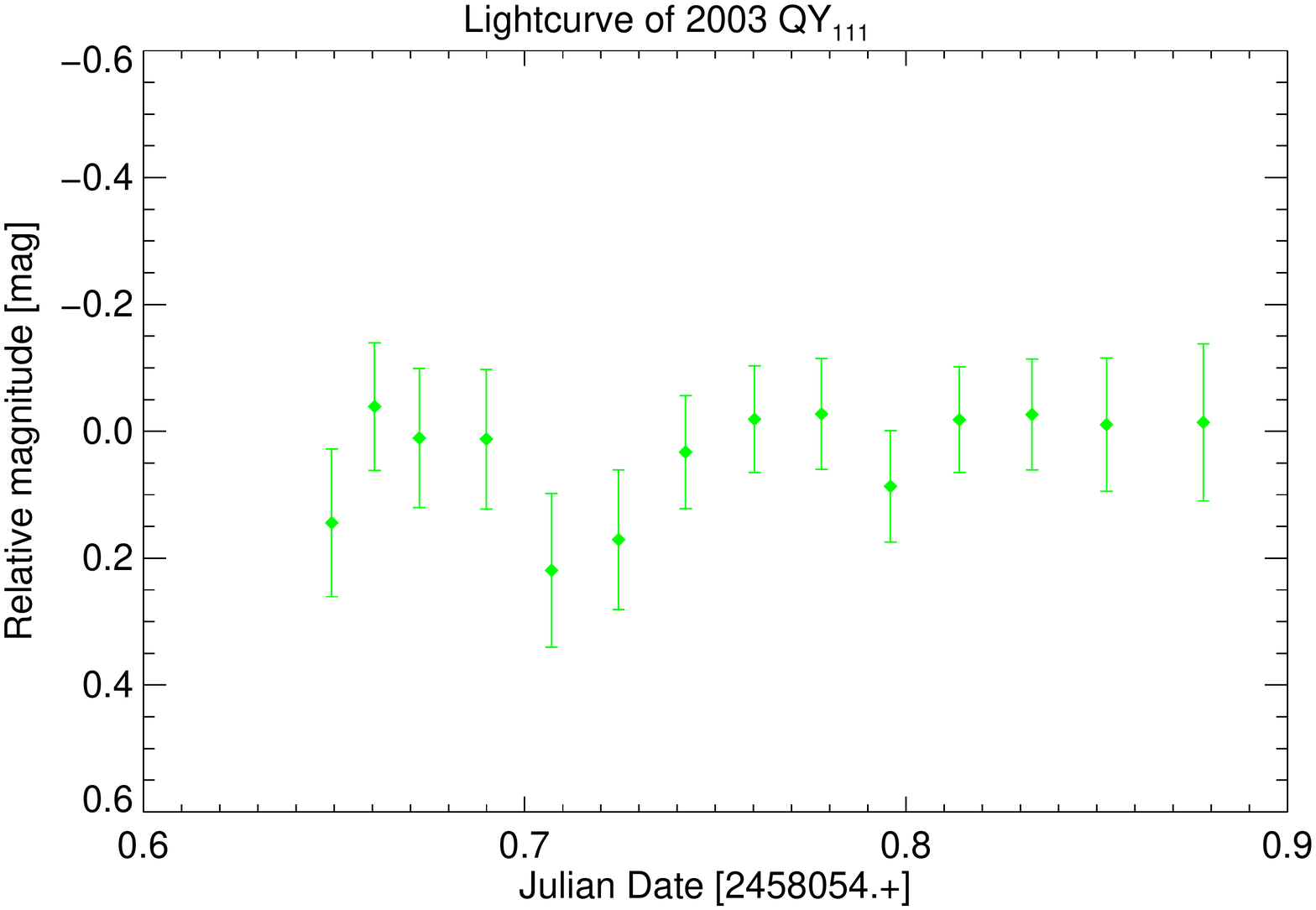}
  \includegraphics[width=9cm, angle=0]{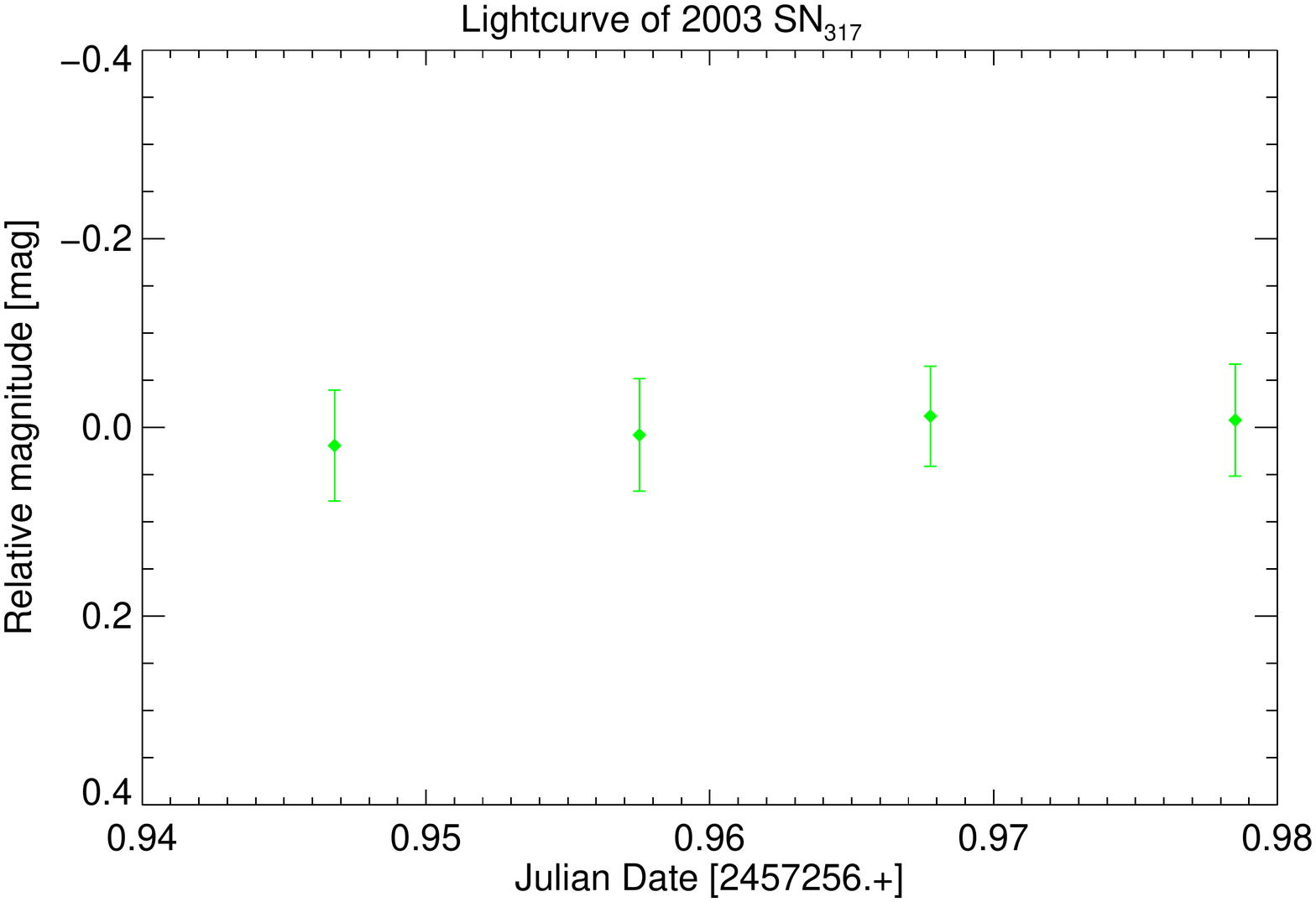}
   \includegraphics[width=9cm, angle=0]{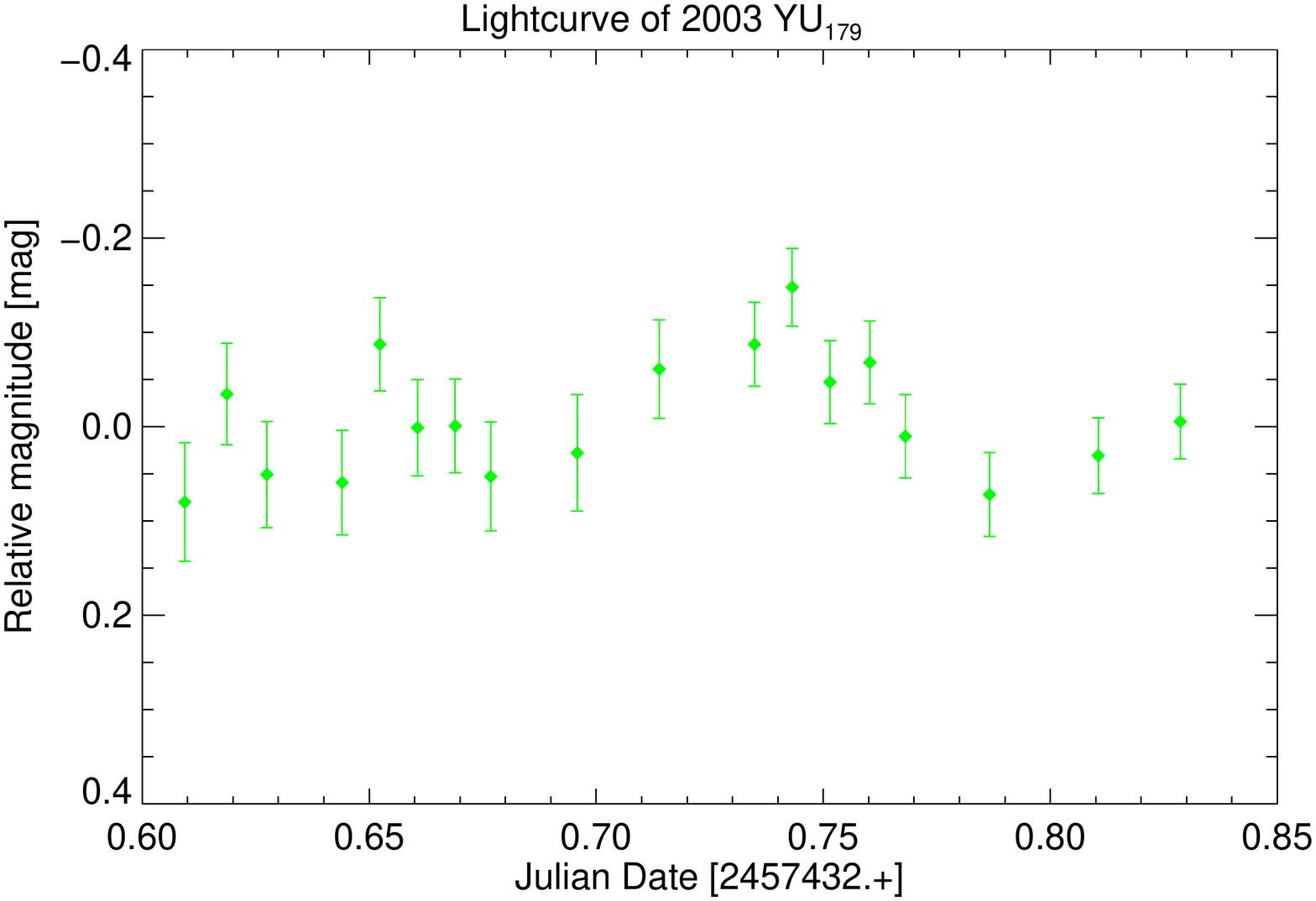}
\caption{Continued. }
\label{fig:LC}
\end{figure*}

\begin{figure*}
 \includegraphics[width=9cm, angle=0]{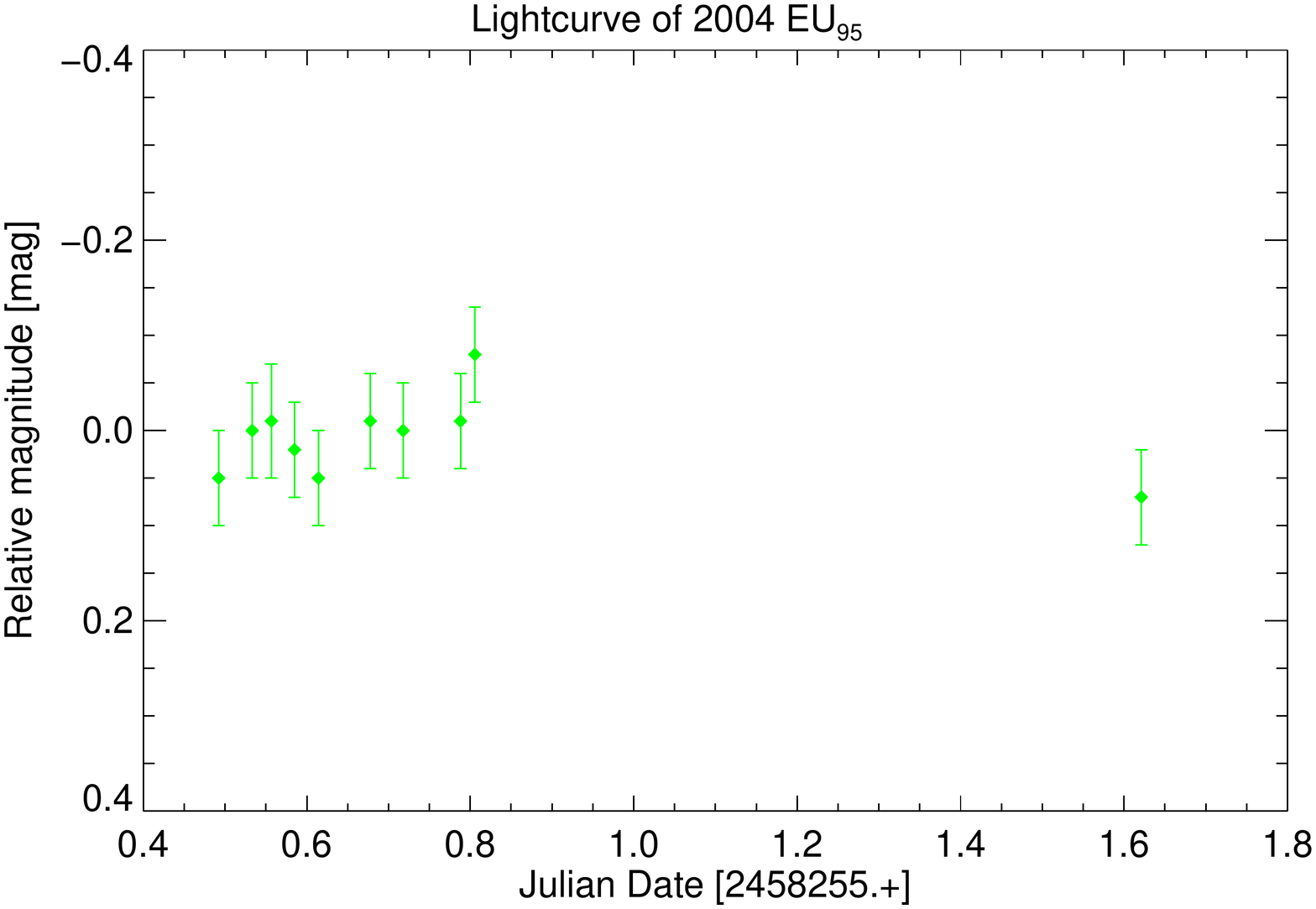}
  \includegraphics[width=9cm, angle=0]{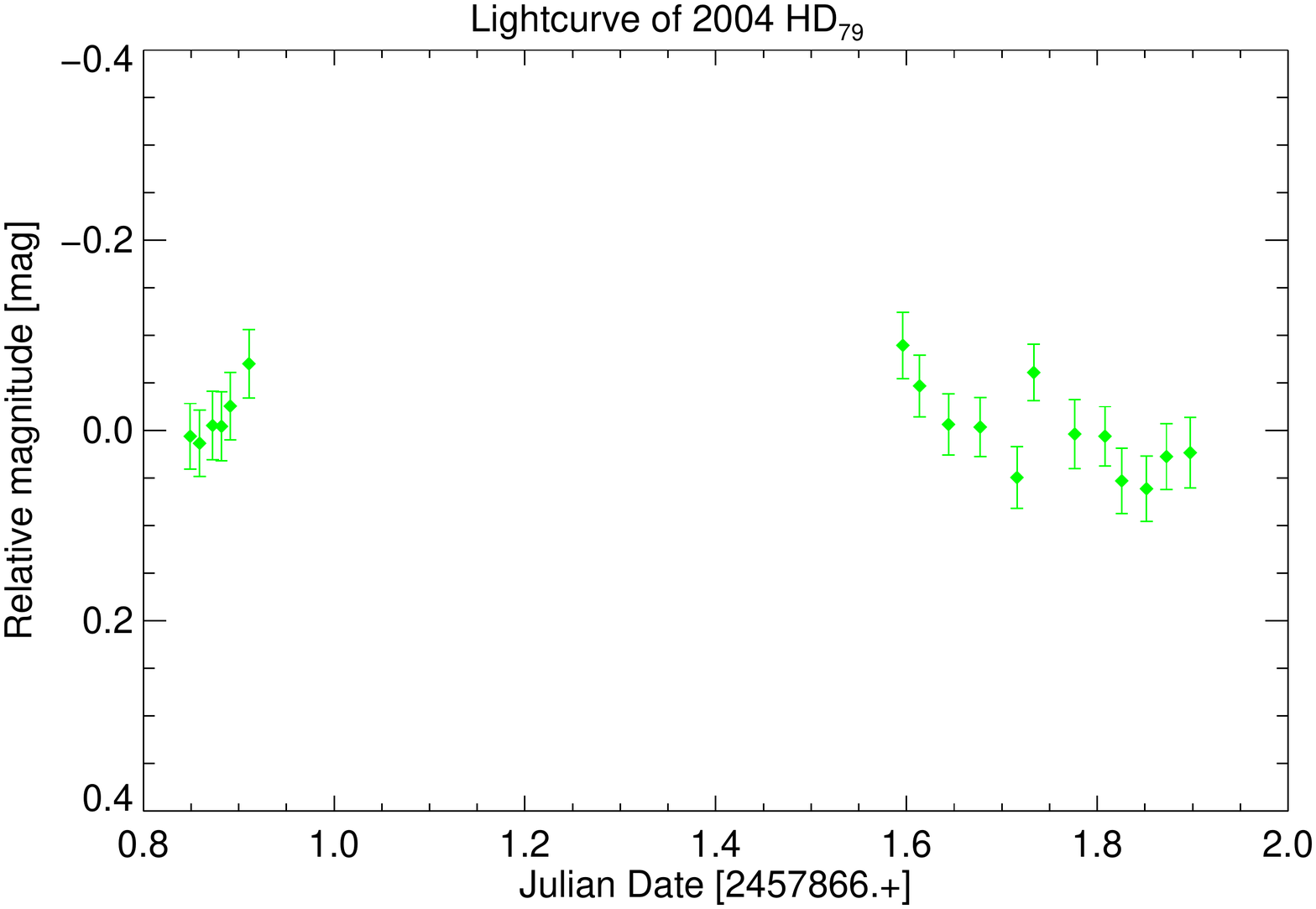}
  \includegraphics[width=9cm, angle=0]{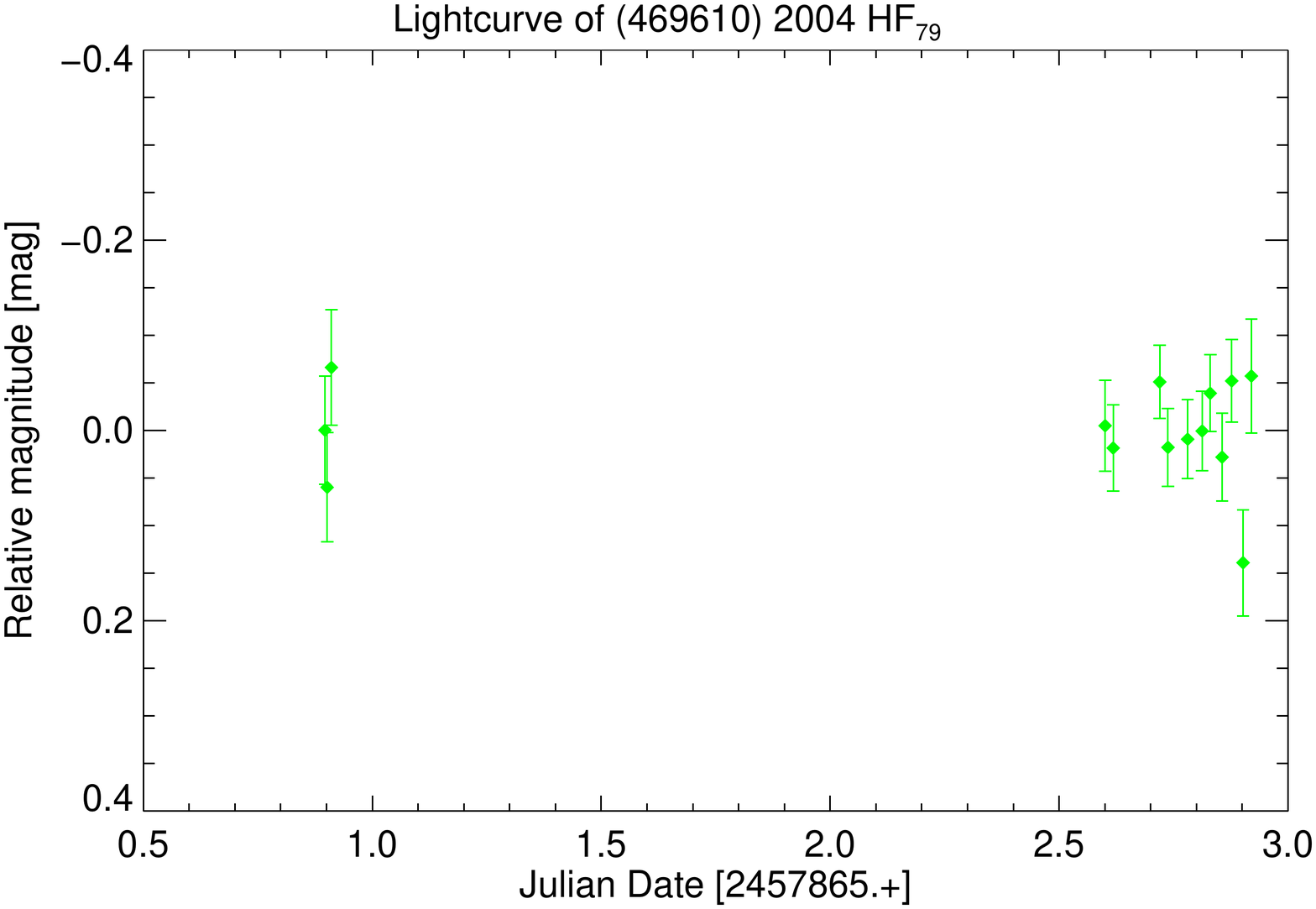}
    \includegraphics[width=9cm, angle=0]{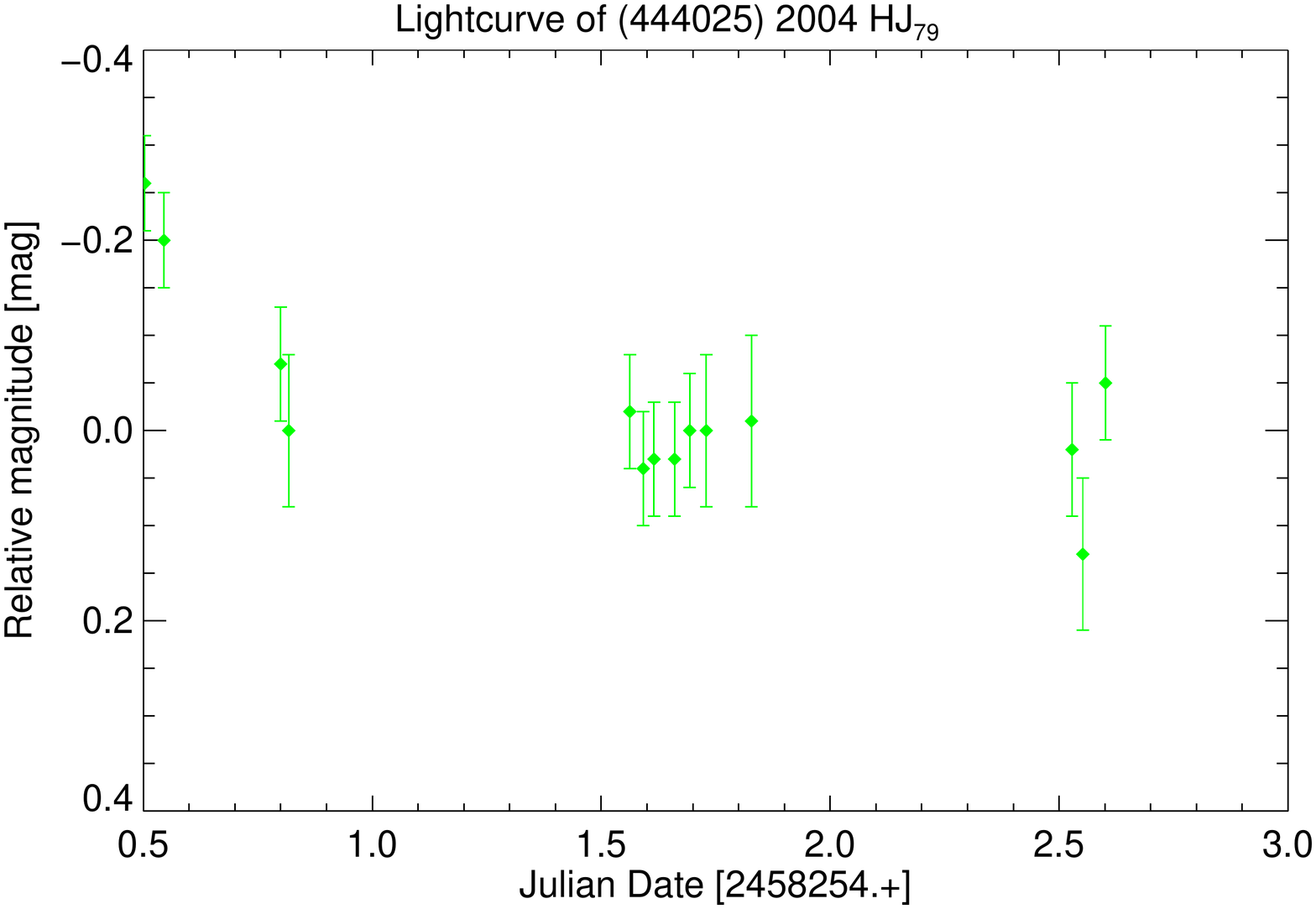}
        \includegraphics[width=9cm, angle=0]{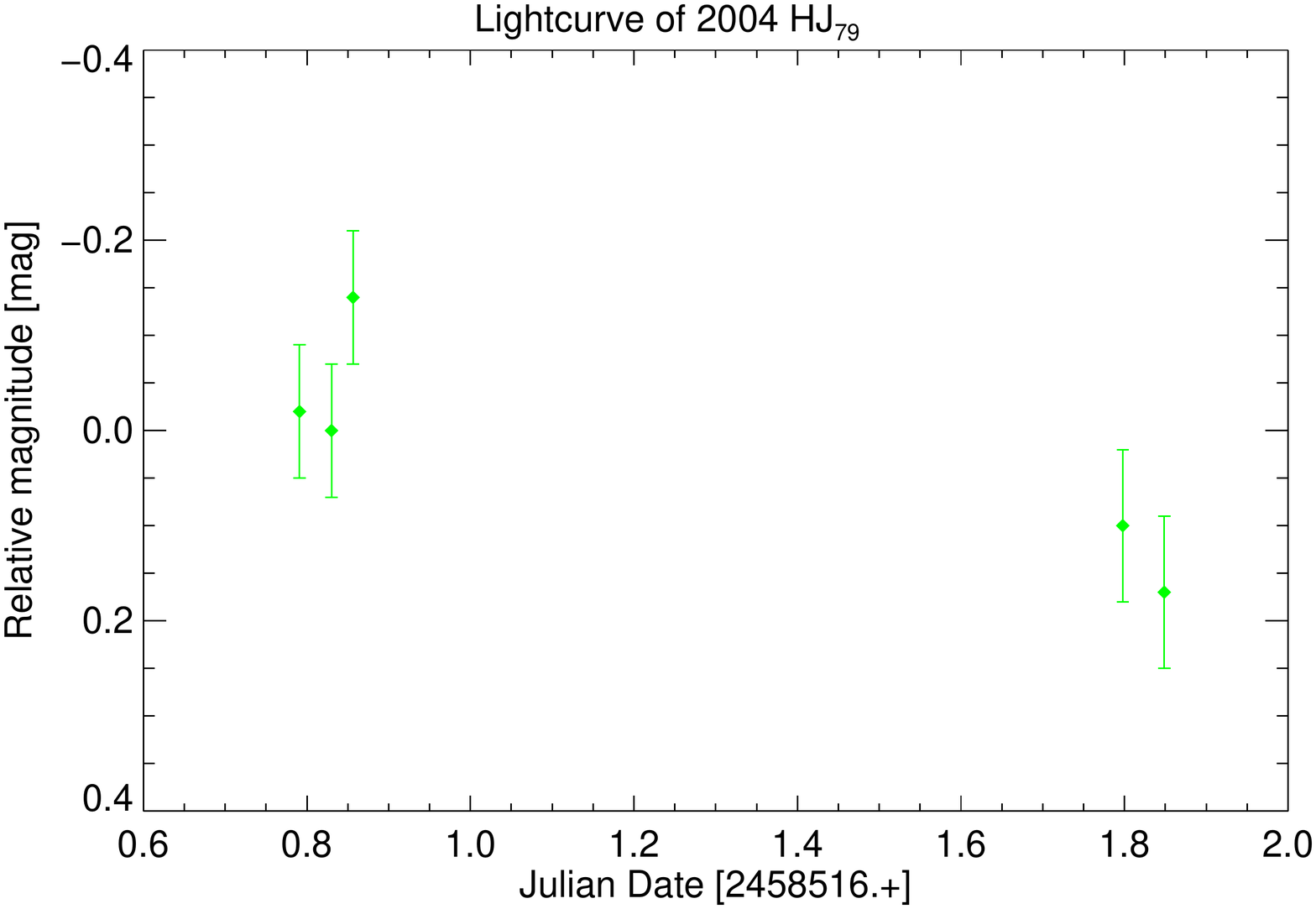}
     \includegraphics[width=9cm, angle=0]{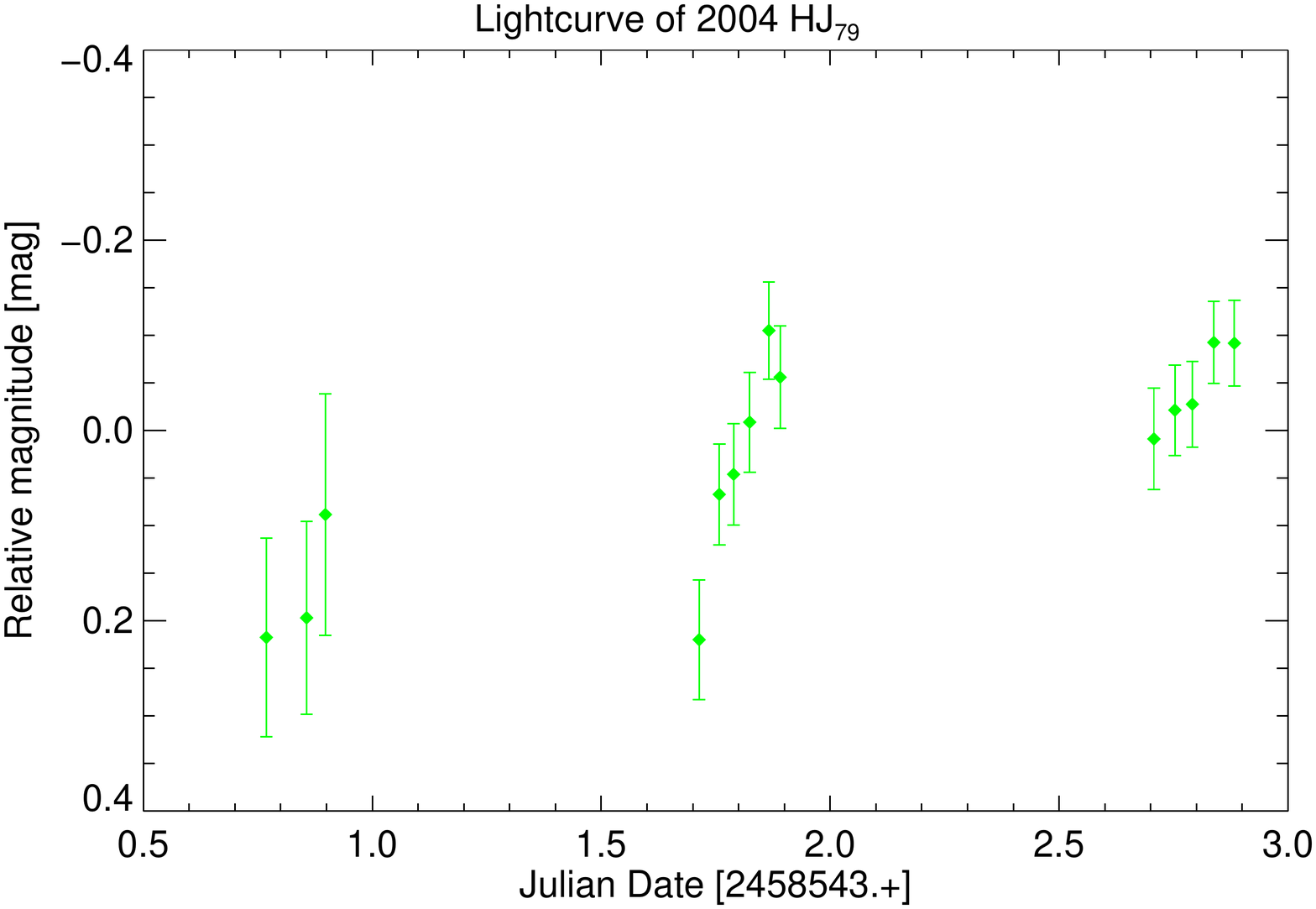}
\caption{Continued. }
\label{fig:LC}
\end{figure*}

\begin{figure*}
        \includegraphics[width=9cm, angle=0]{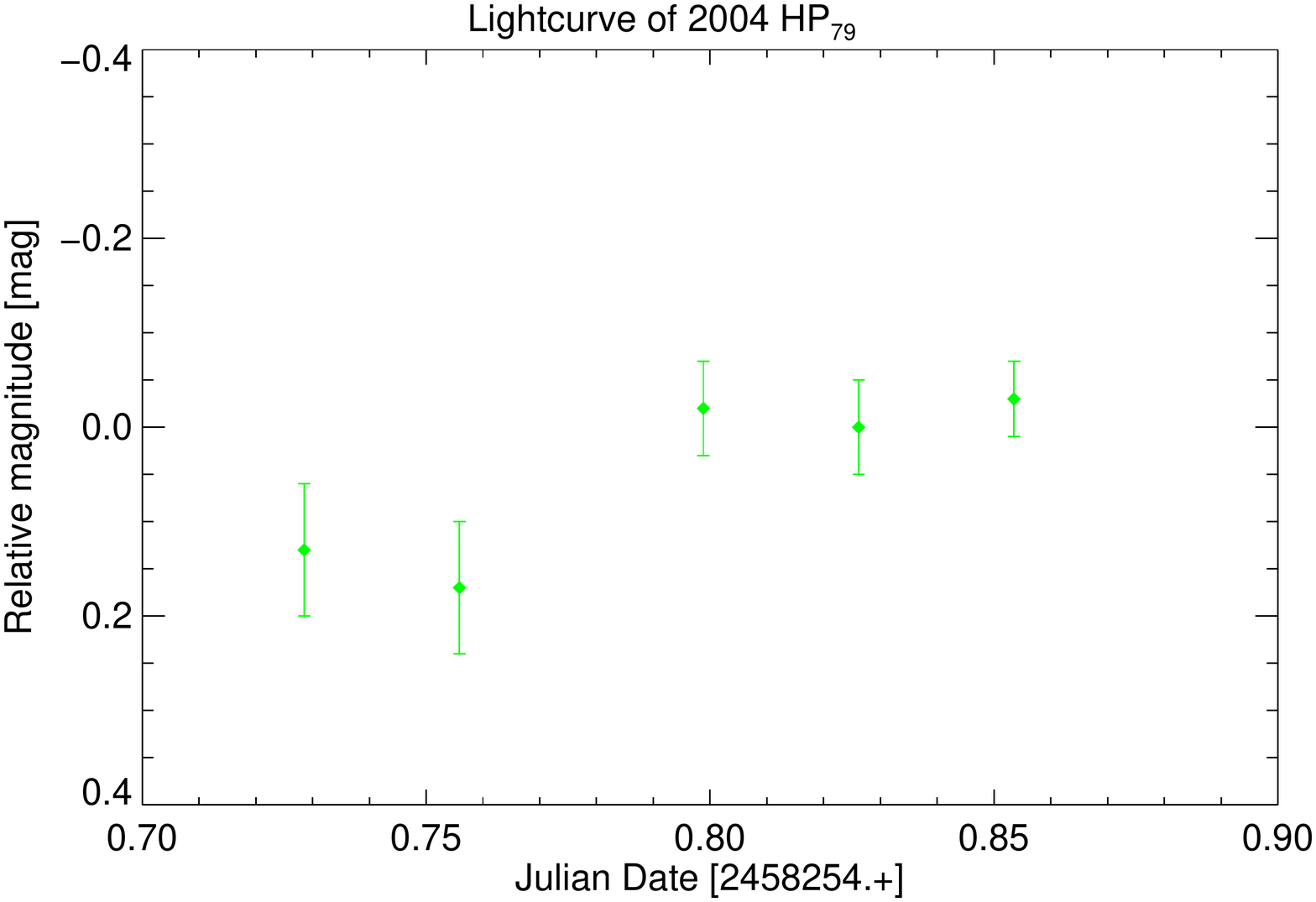}
 \includegraphics[width=9cm, angle=0]{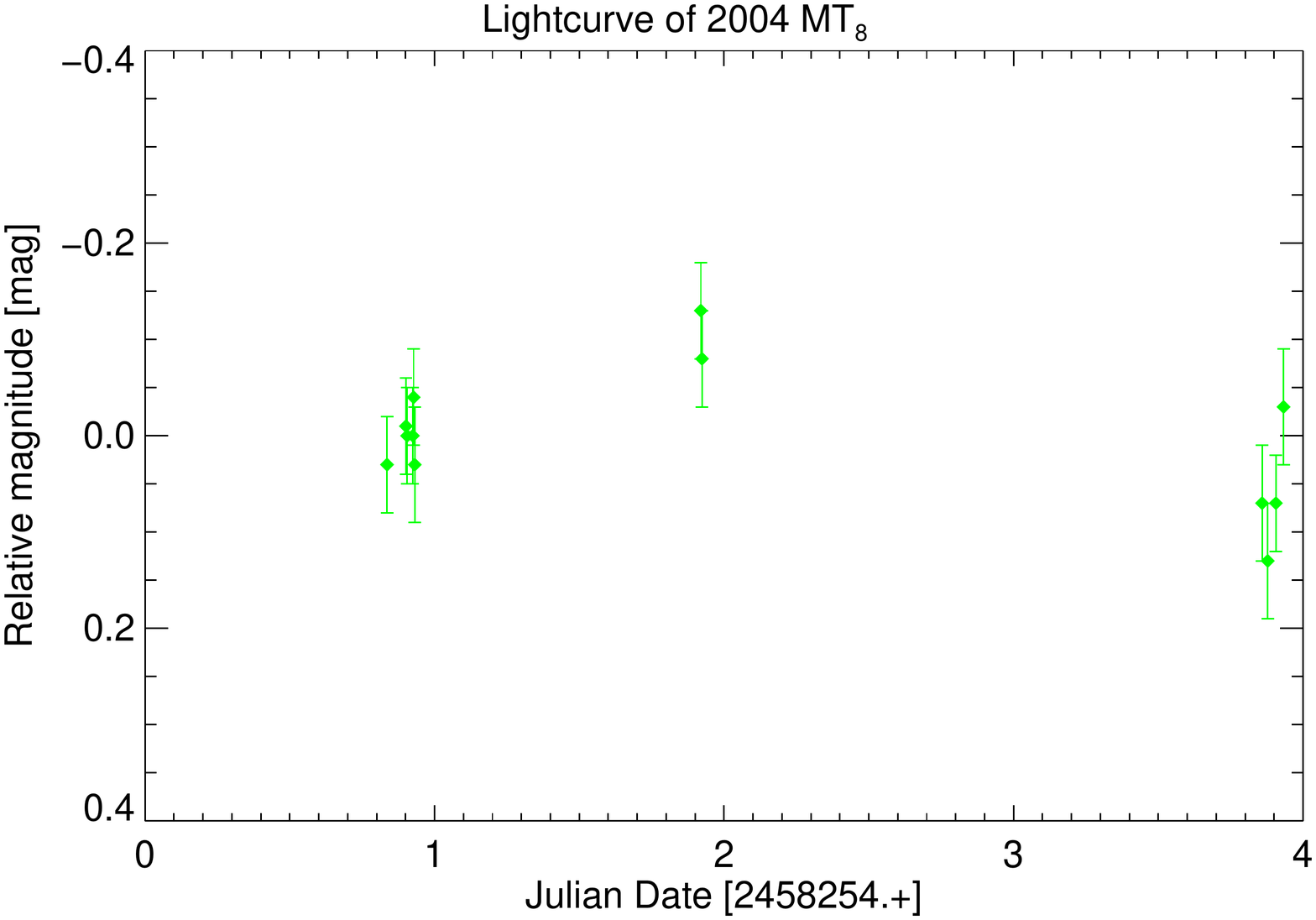}
 \includegraphics[width=9cm, angle=0]{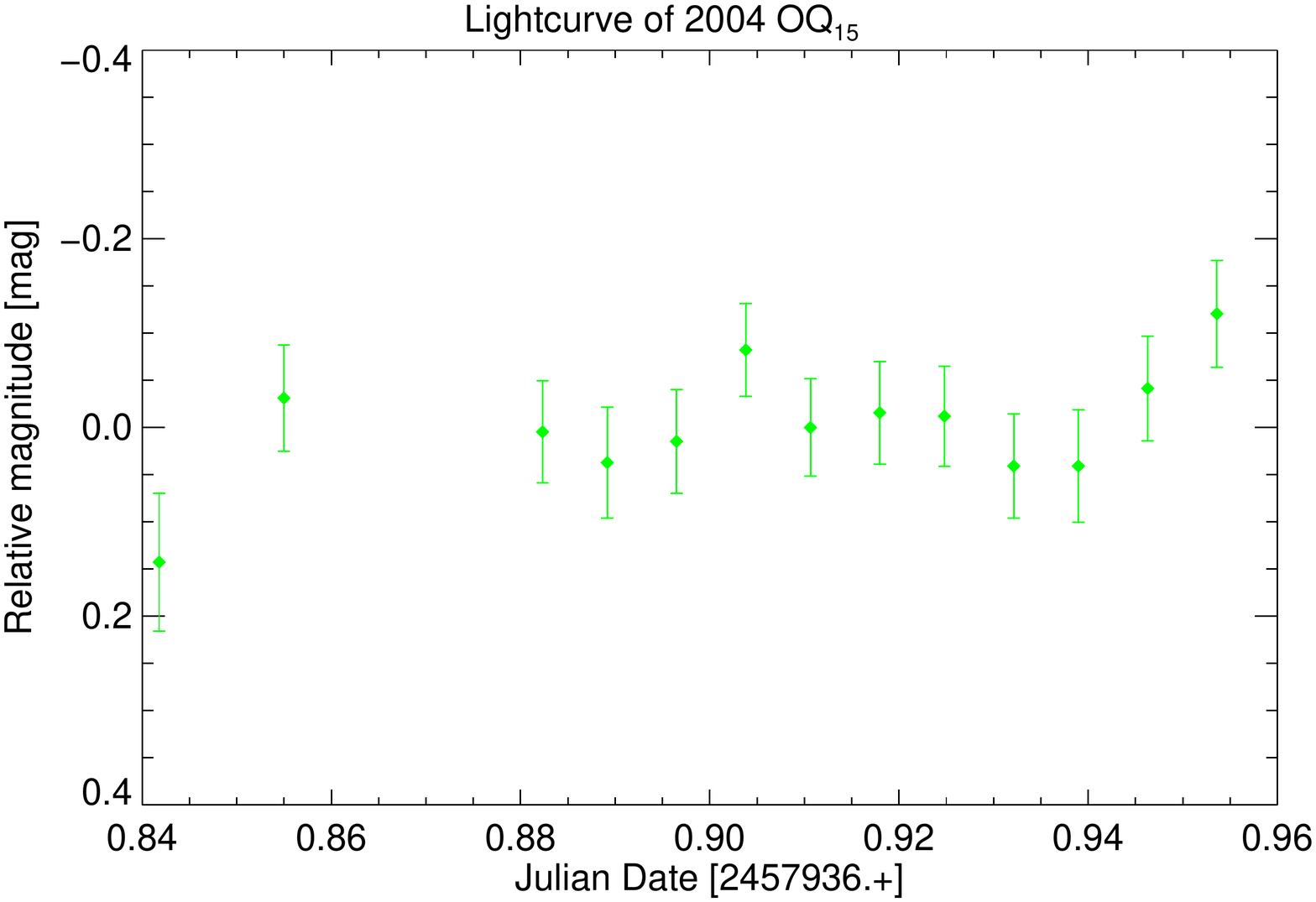}
 \includegraphics[width=9cm, angle=0]{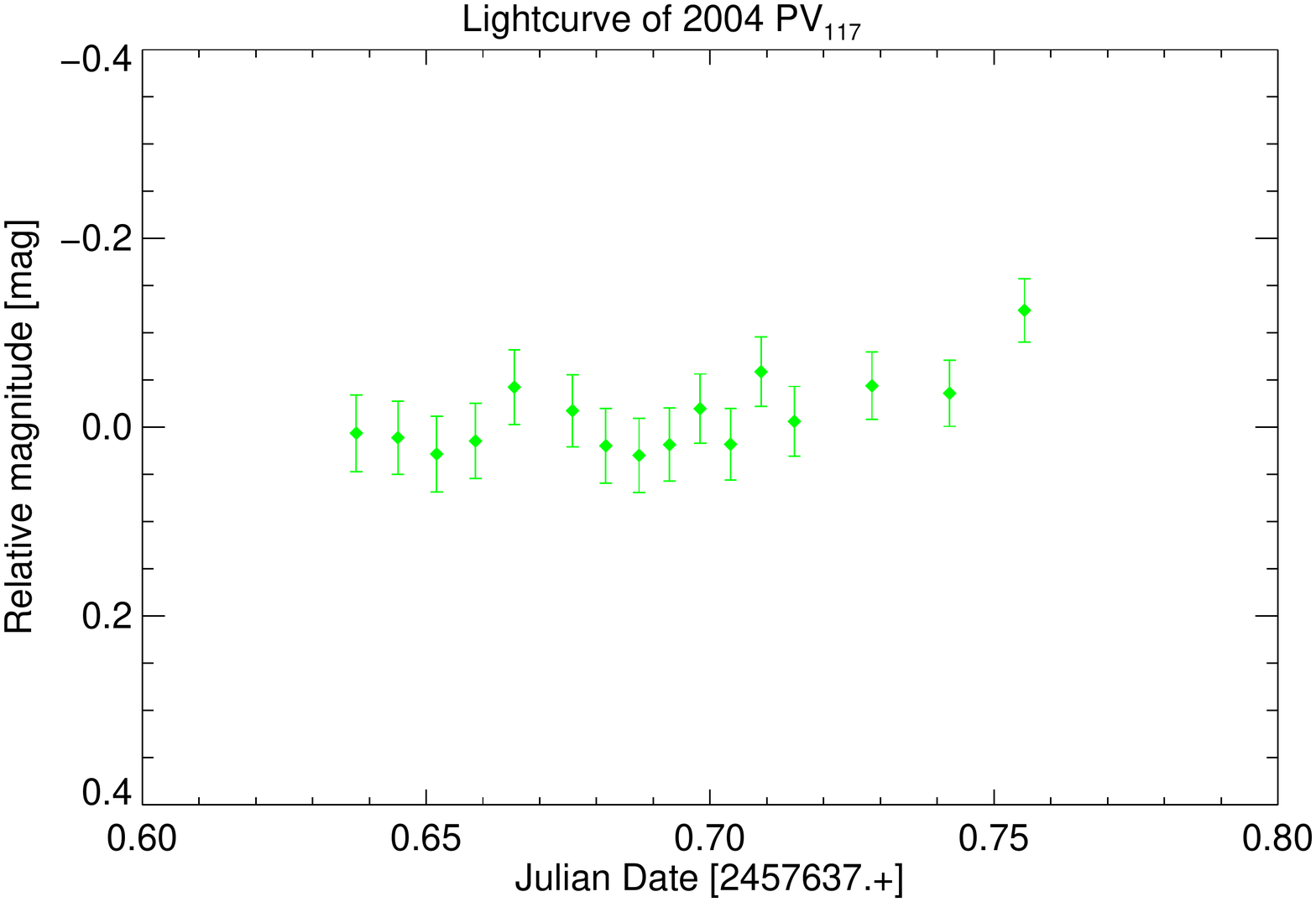}
  \includegraphics[width=9cm, angle=0]{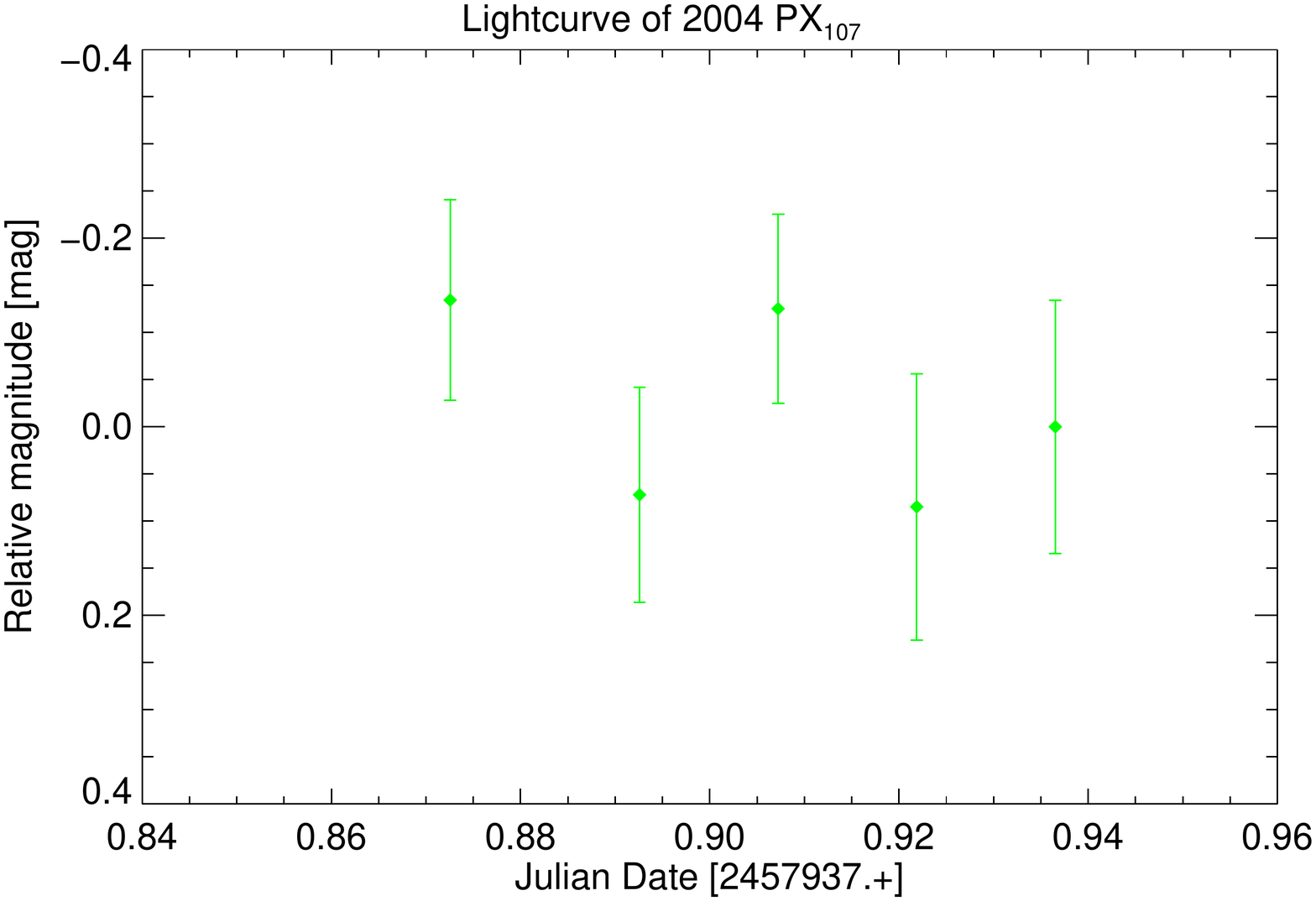}
    \includegraphics[width=9cm, angle=0]{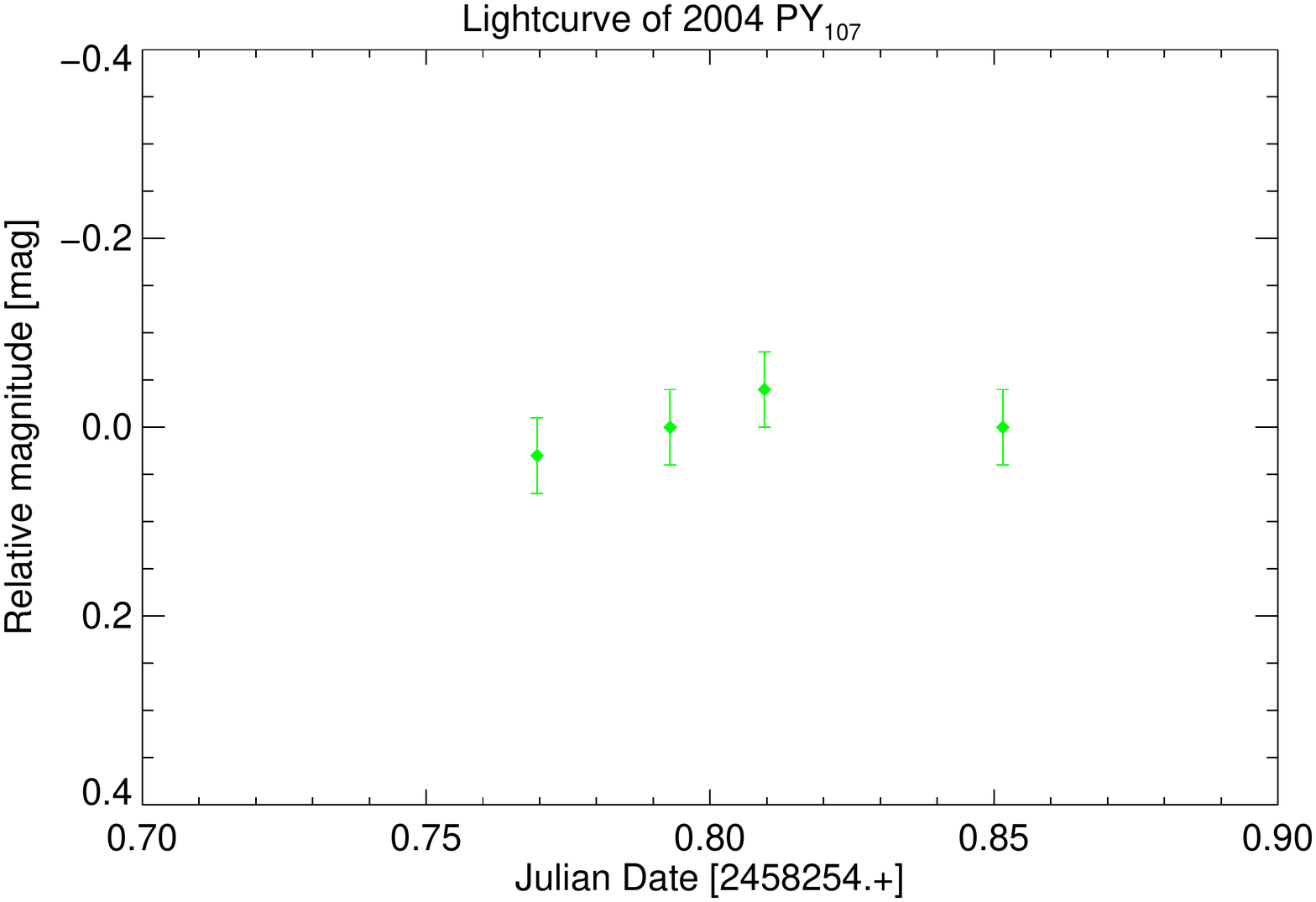}
\caption{Continued. }
\label{fig:LC}
\end{figure*}
 
\begin{figure*}
  \includegraphics[width=9cm, angle=0]{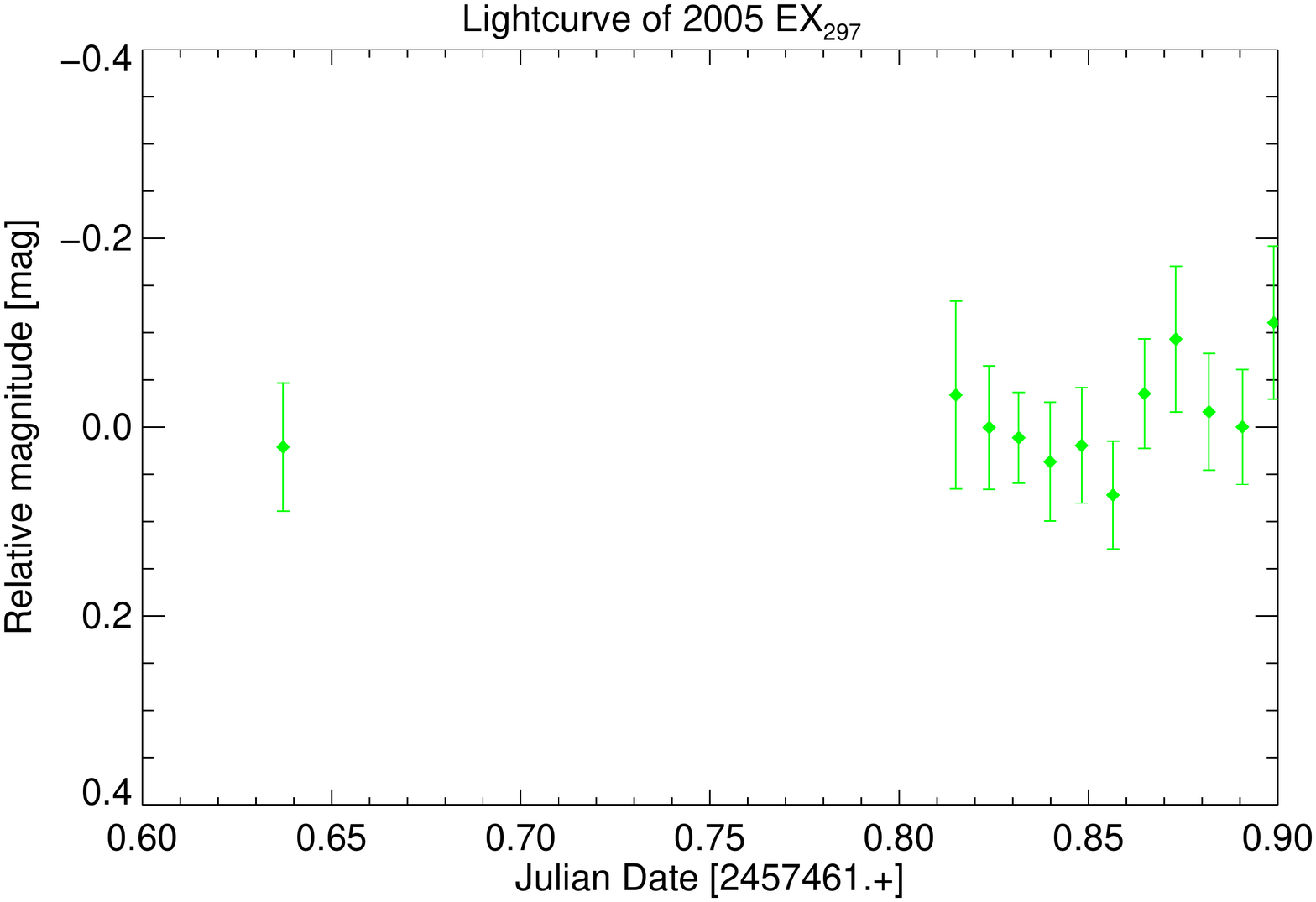}
    \includegraphics[width=9cm, angle=0]{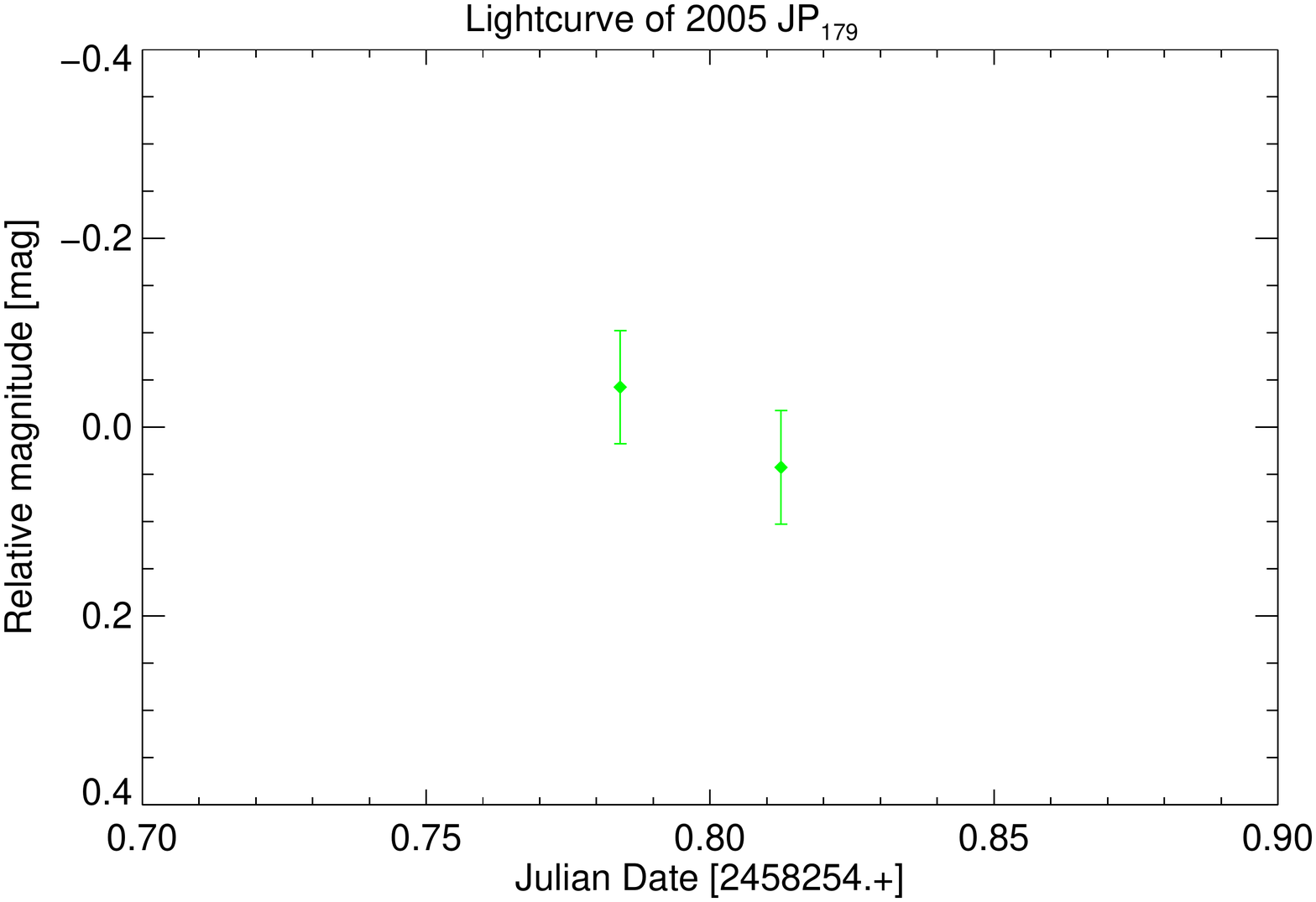}
        \includegraphics[width=9cm, angle=0]{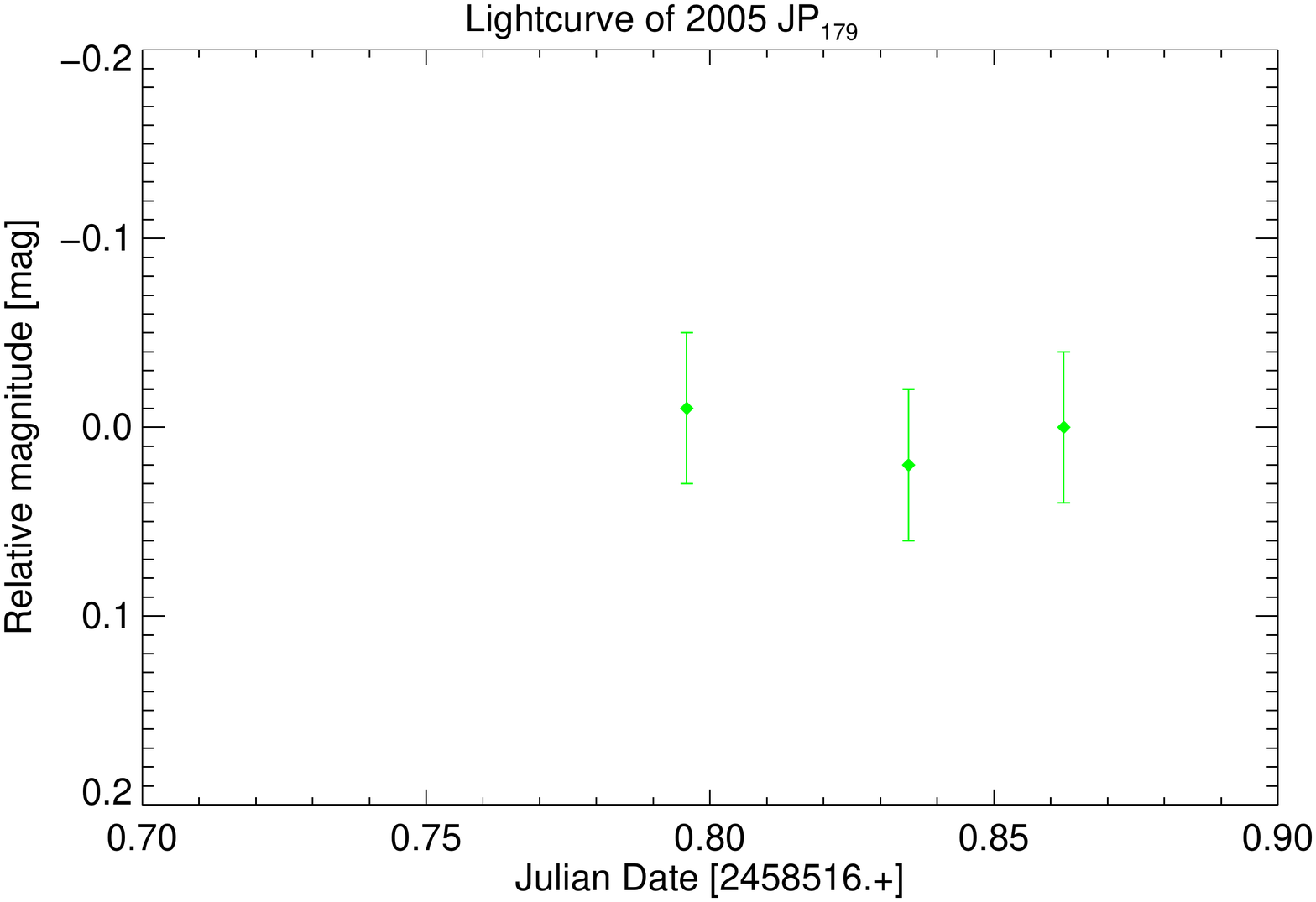}
 \includegraphics[width=9cm, angle=0]{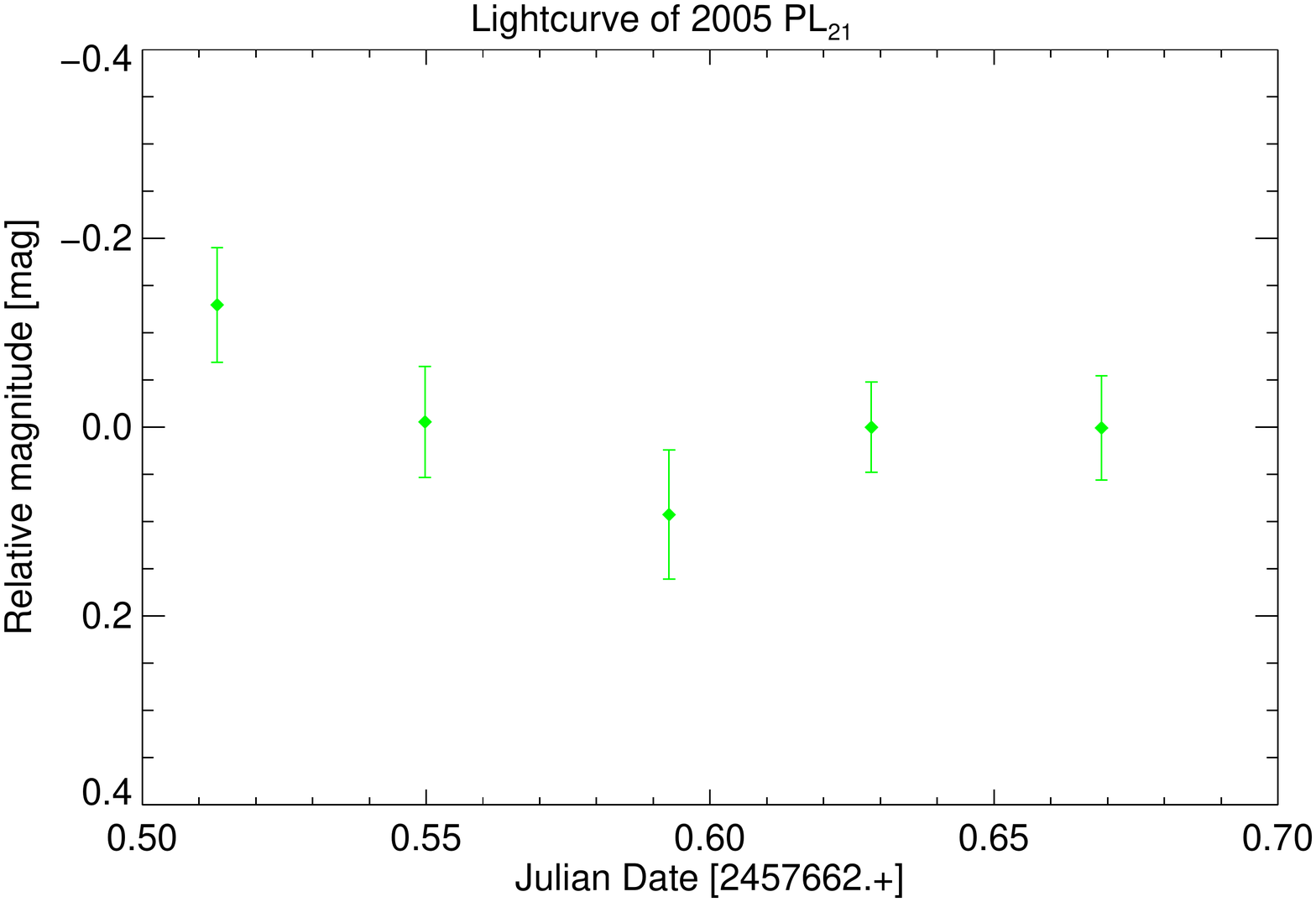}
 \includegraphics[width=9cm, angle=0]{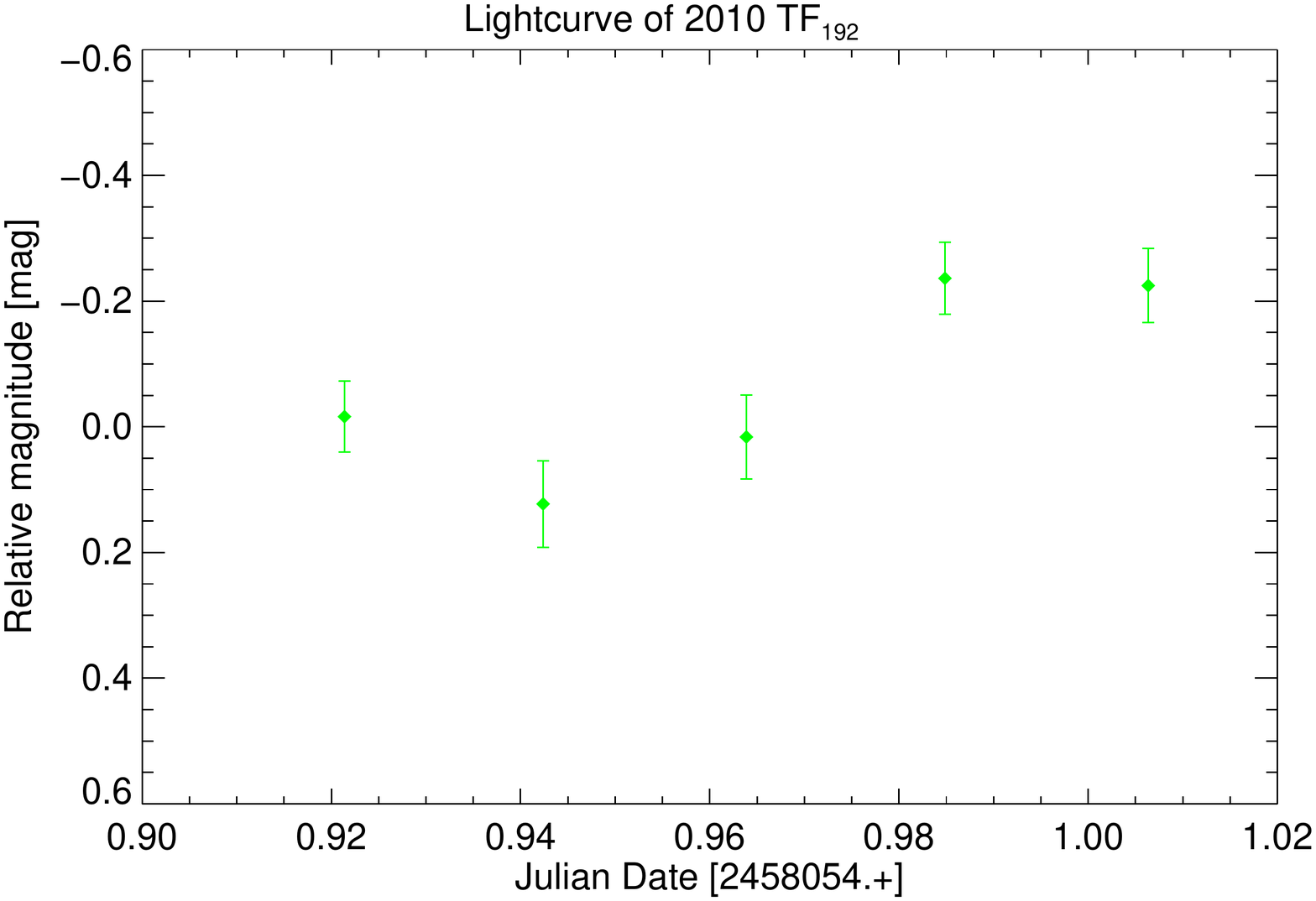}
  \includegraphics[width=9cm, angle=0]{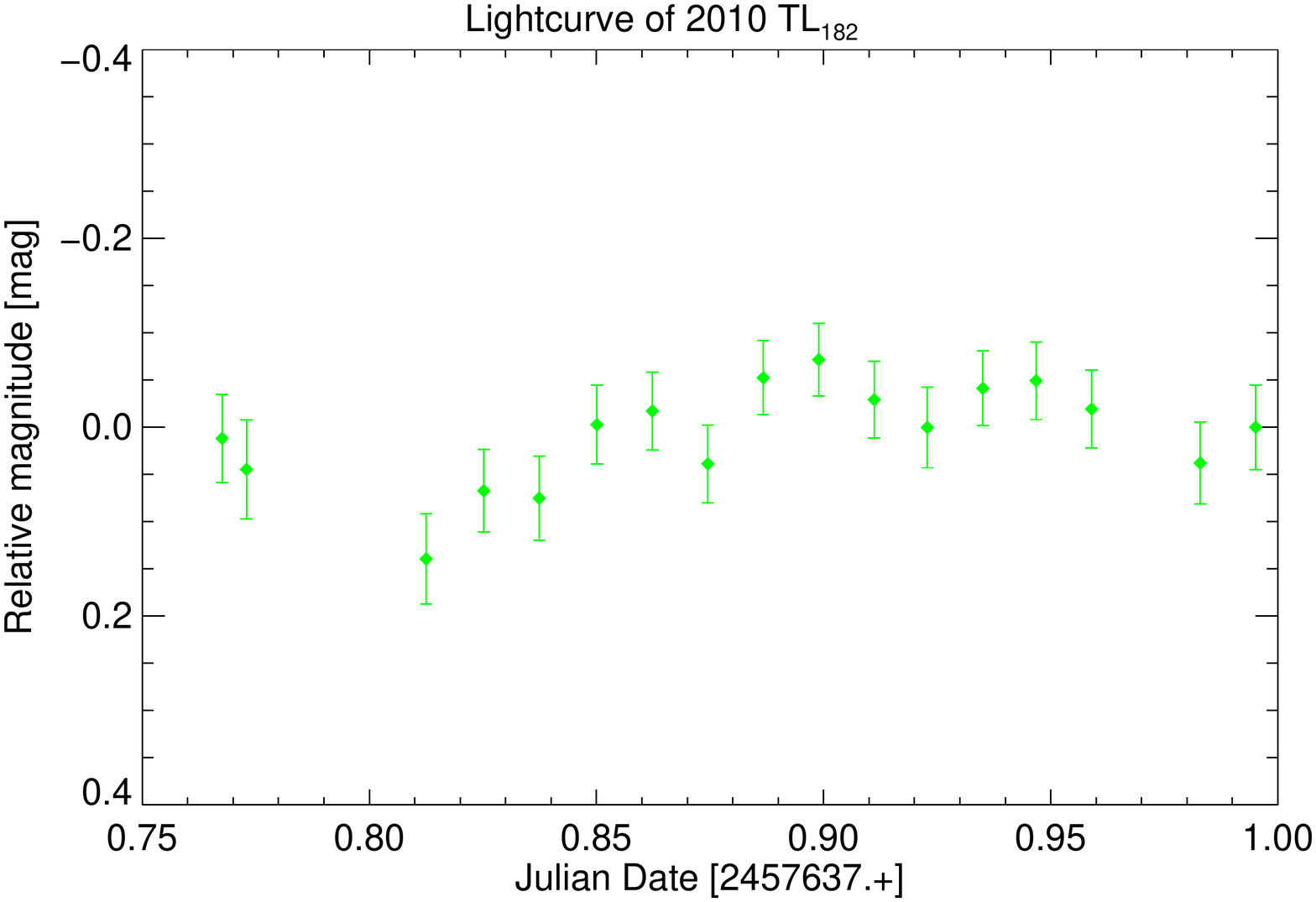}
\caption{Continued. }
\label{fig:LC}
\end{figure*}
 
 \begin{figure*}
   \includegraphics[width=9cm, angle=0]{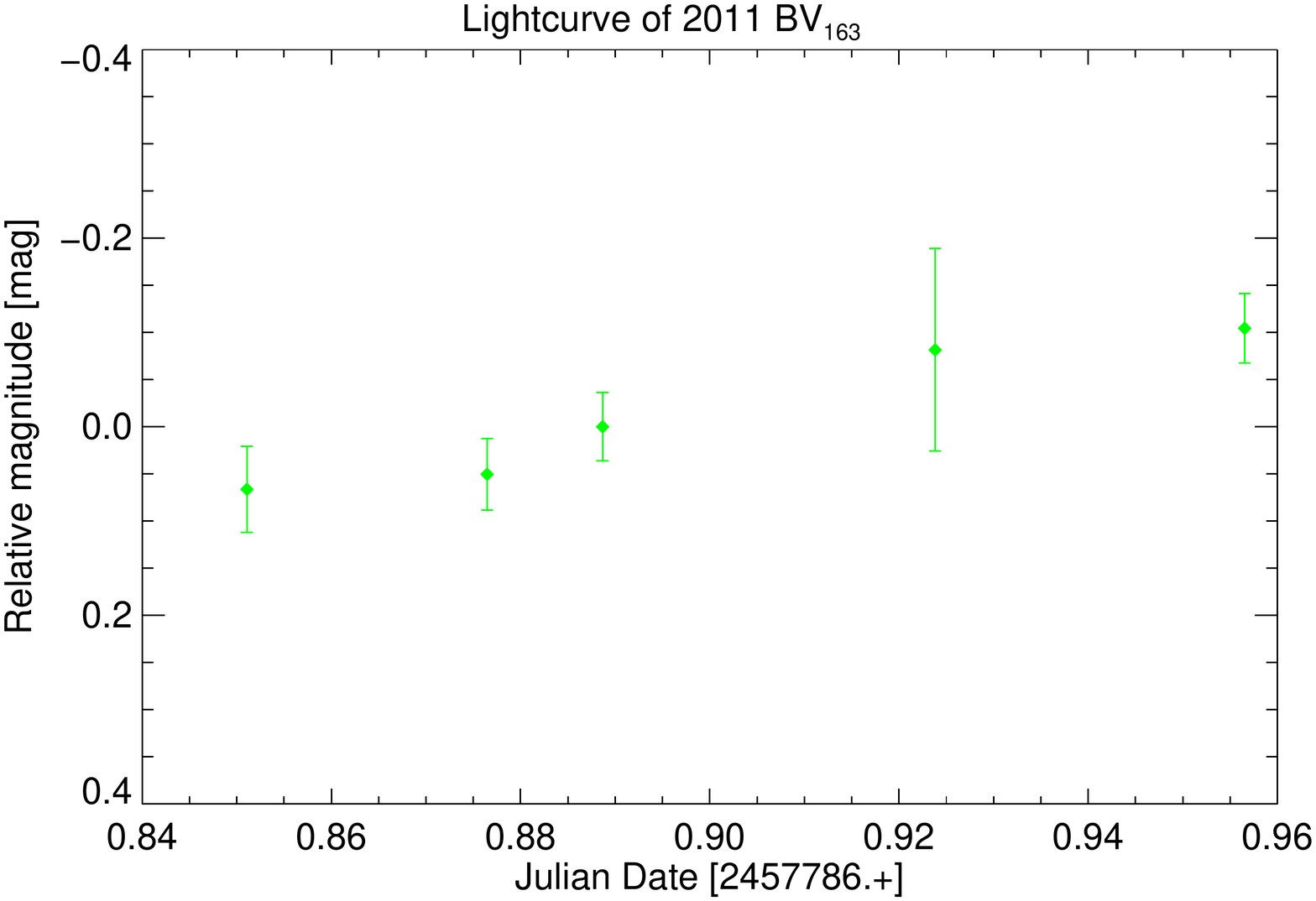}
       \includegraphics[width=9cm, angle=0]{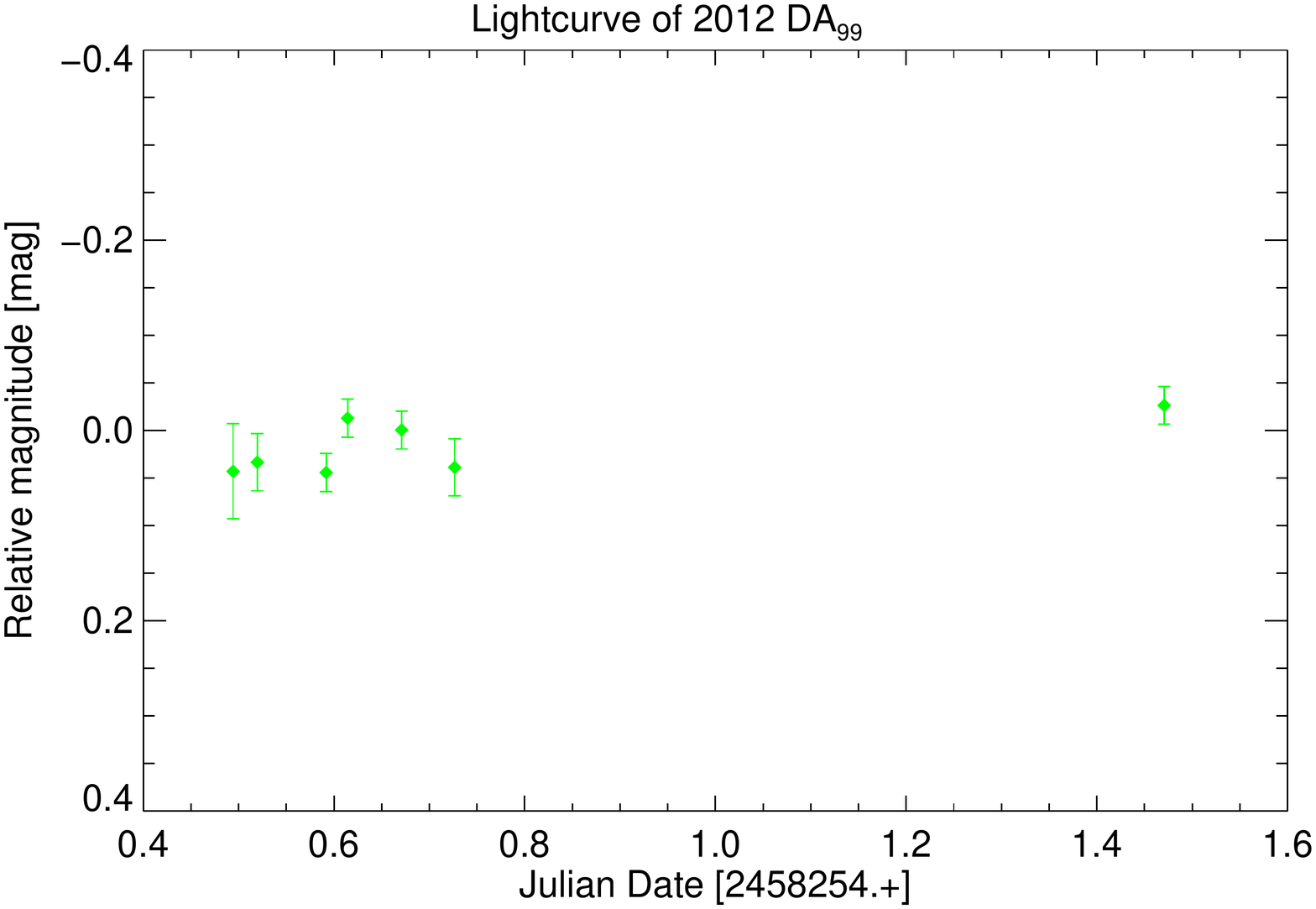} 
              \includegraphics[width=9cm, angle=0]{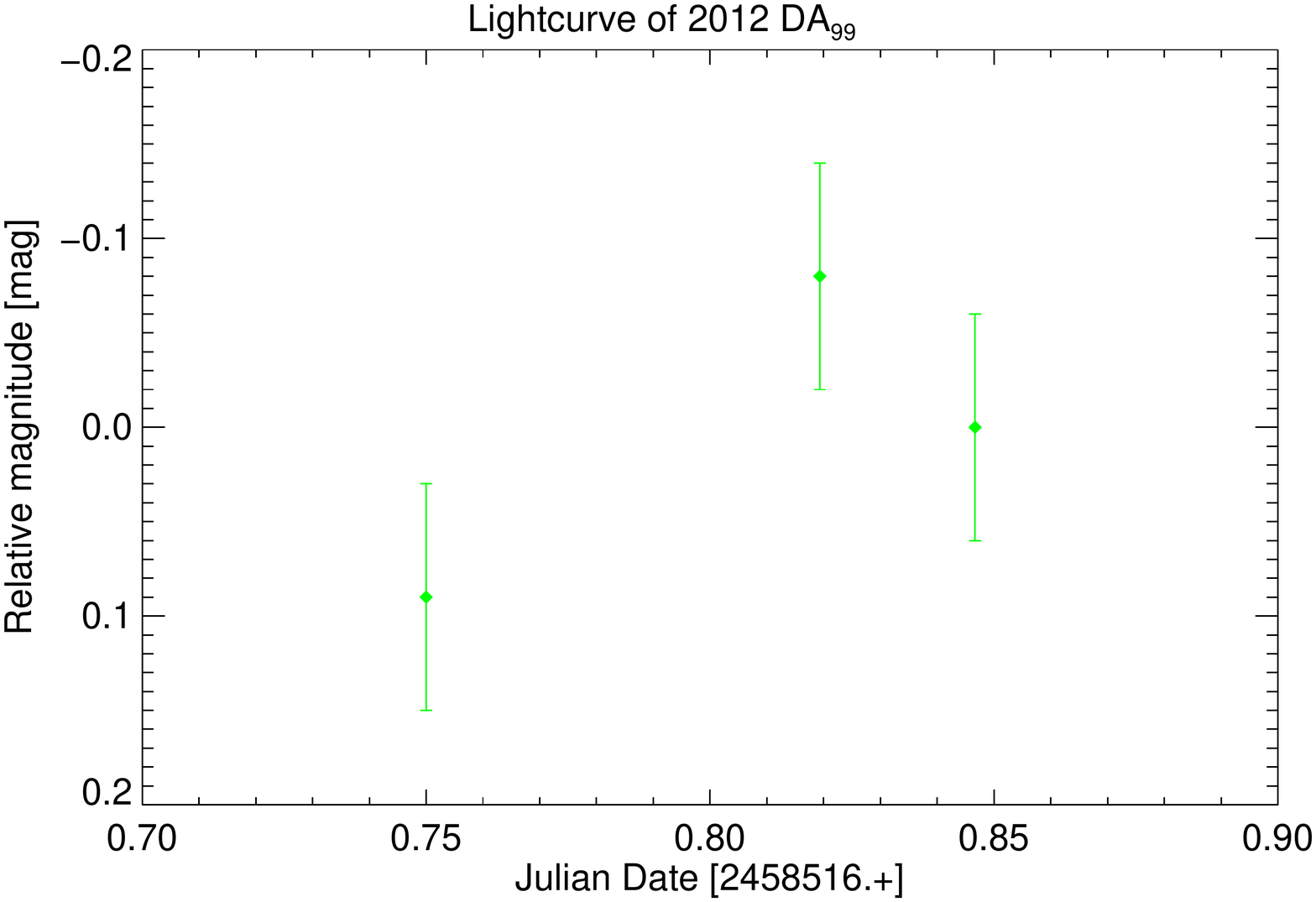} 
  \includegraphics[width=9cm, angle=0]{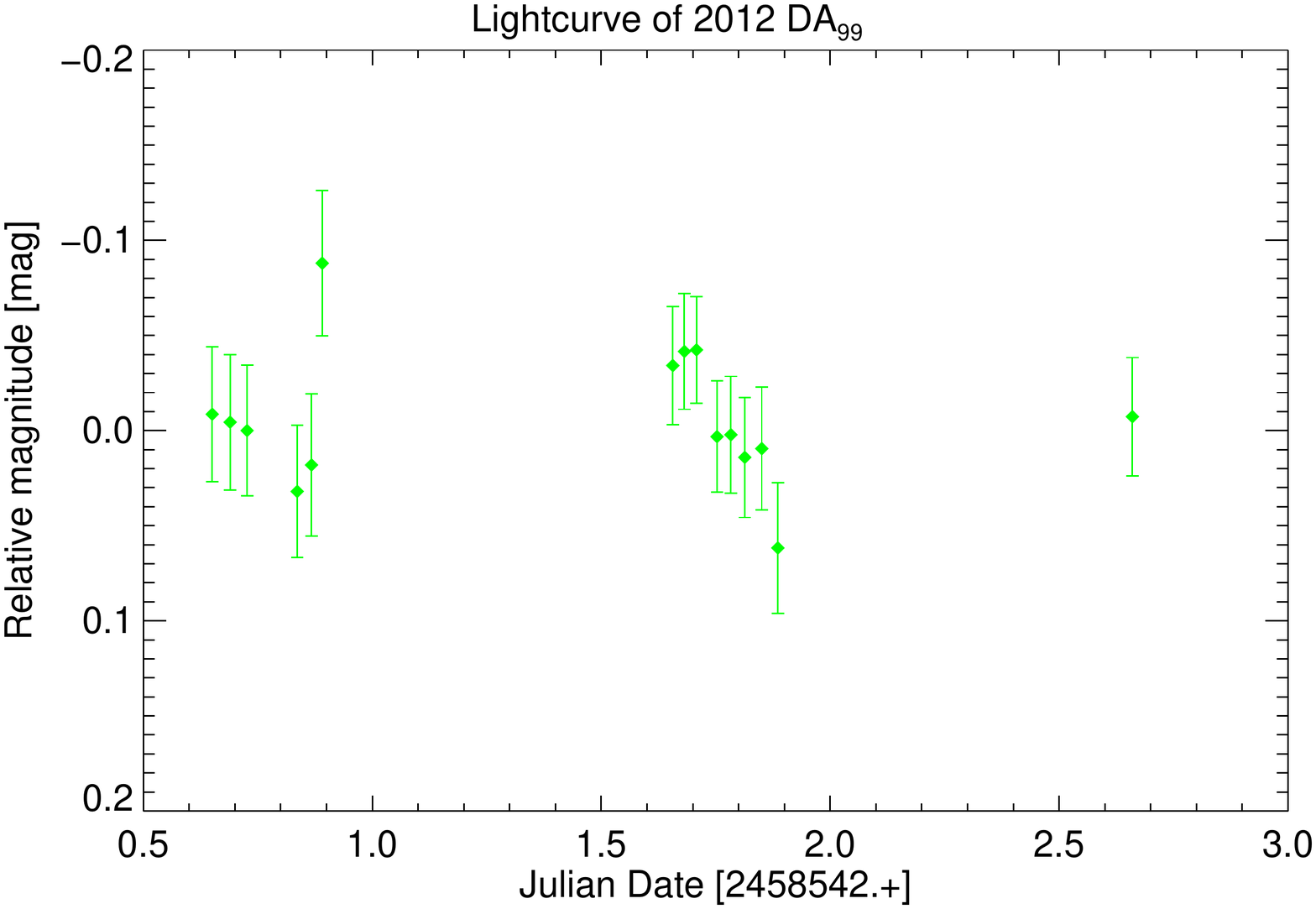} 
       \includegraphics[width=9cm, angle=0]{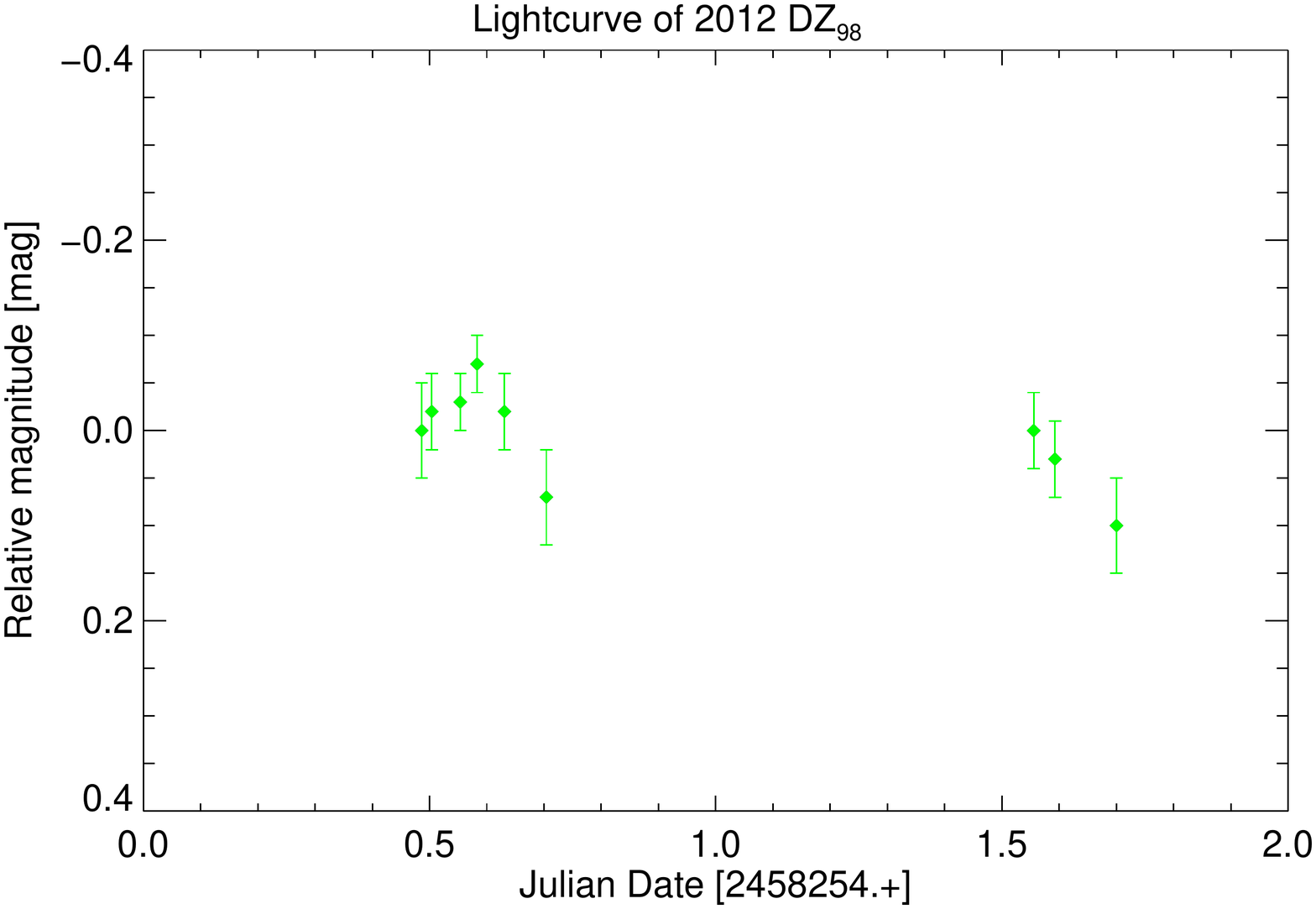}
     \includegraphics[width=9cm, angle=0]{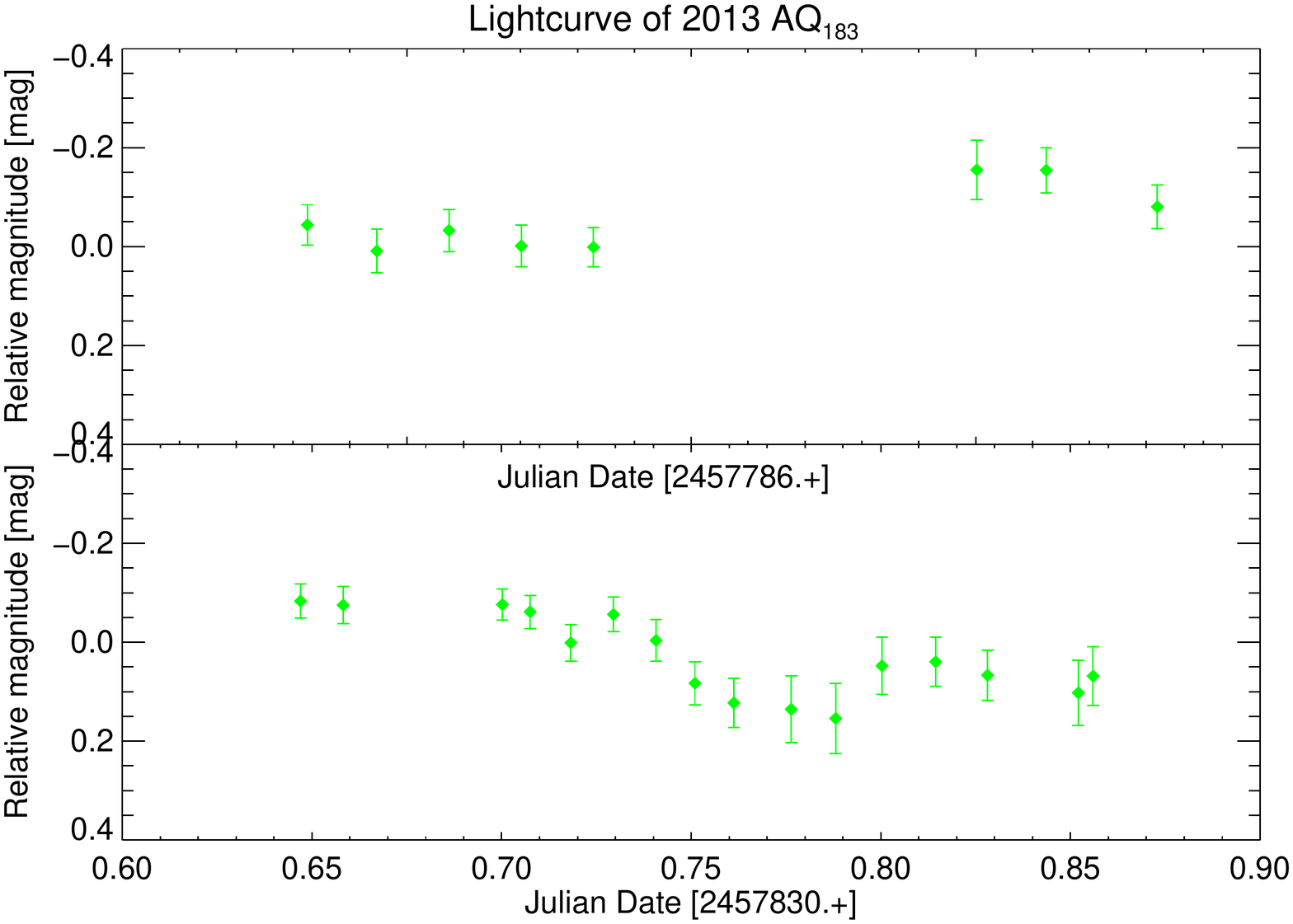}
\caption{Continued. }
\label{fig:LC}
\end{figure*}

 \begin{figure*}
  \includegraphics[width=9cm, angle=0]{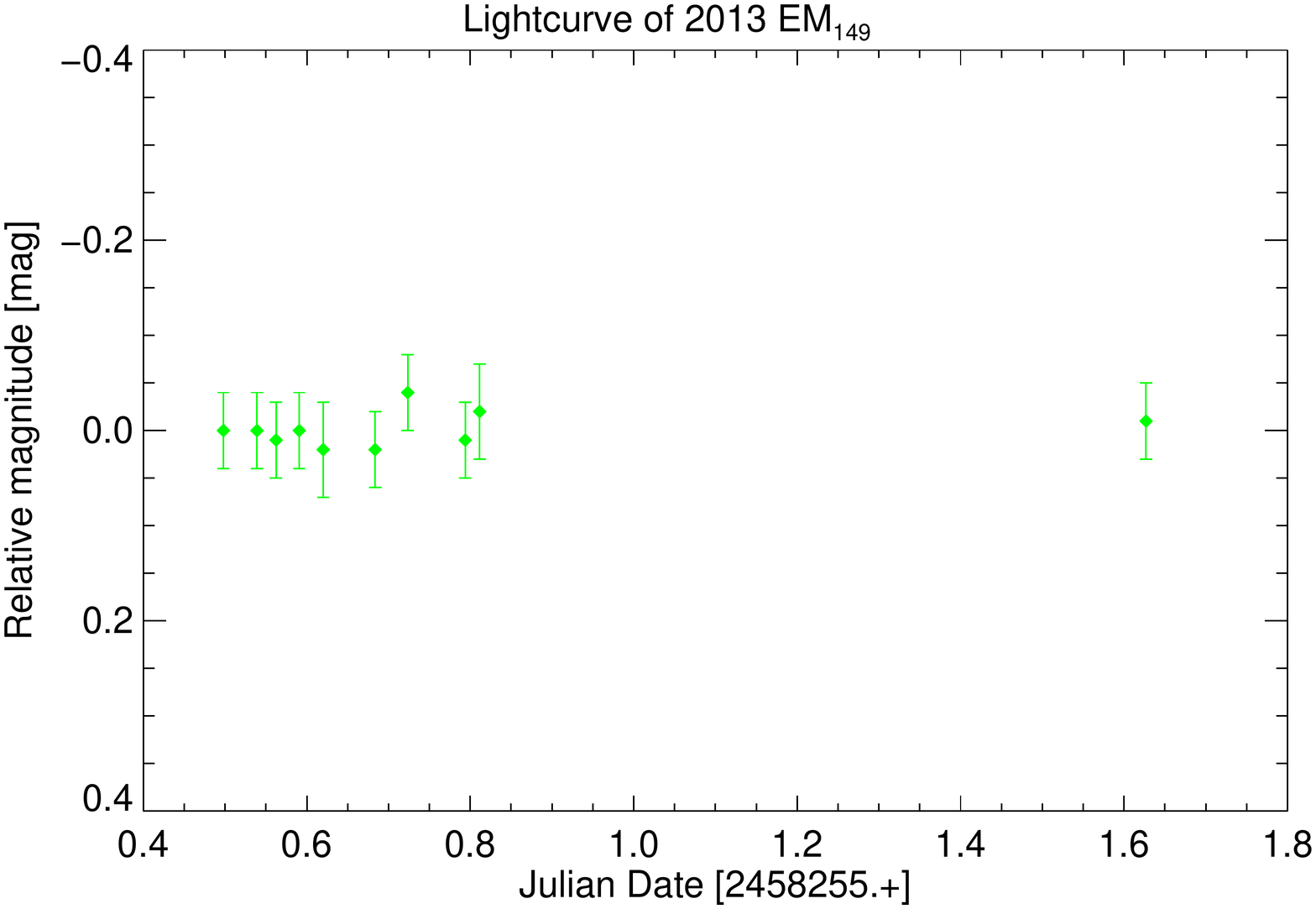}  
  \includegraphics[width=9cm, angle=0]{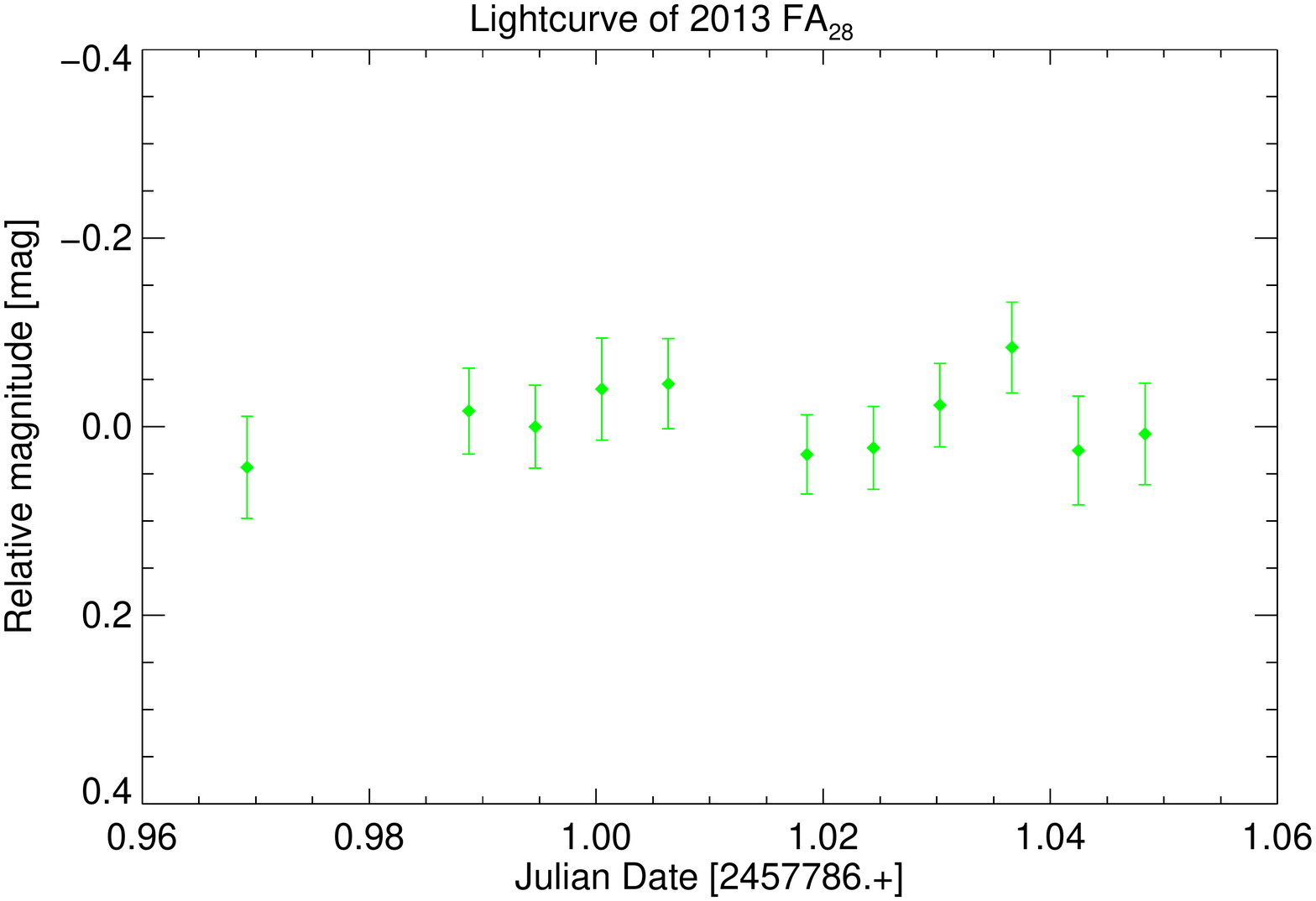}
    \includegraphics[width=9cm, angle=0]{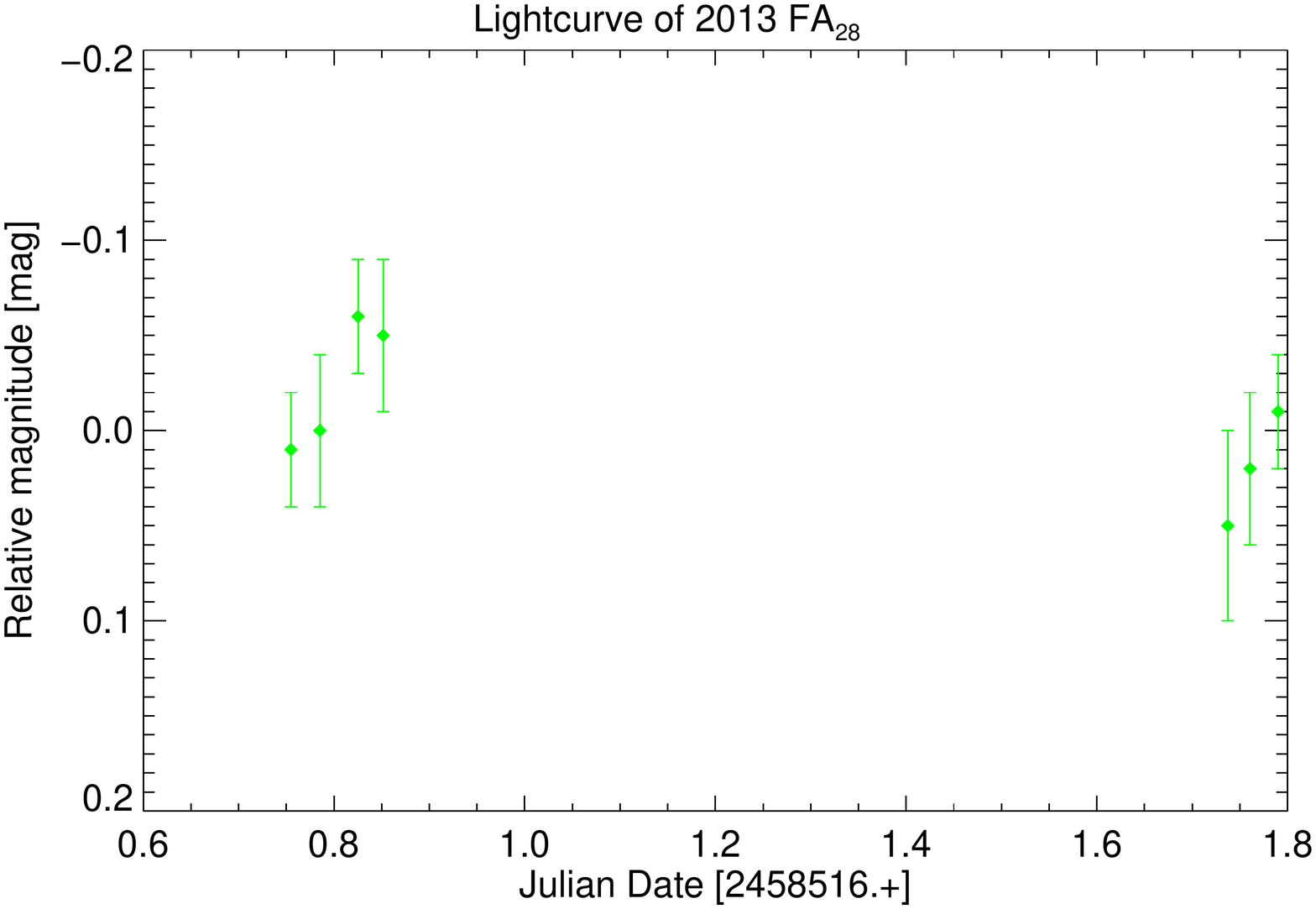}
        \includegraphics[width=9cm, angle=0]{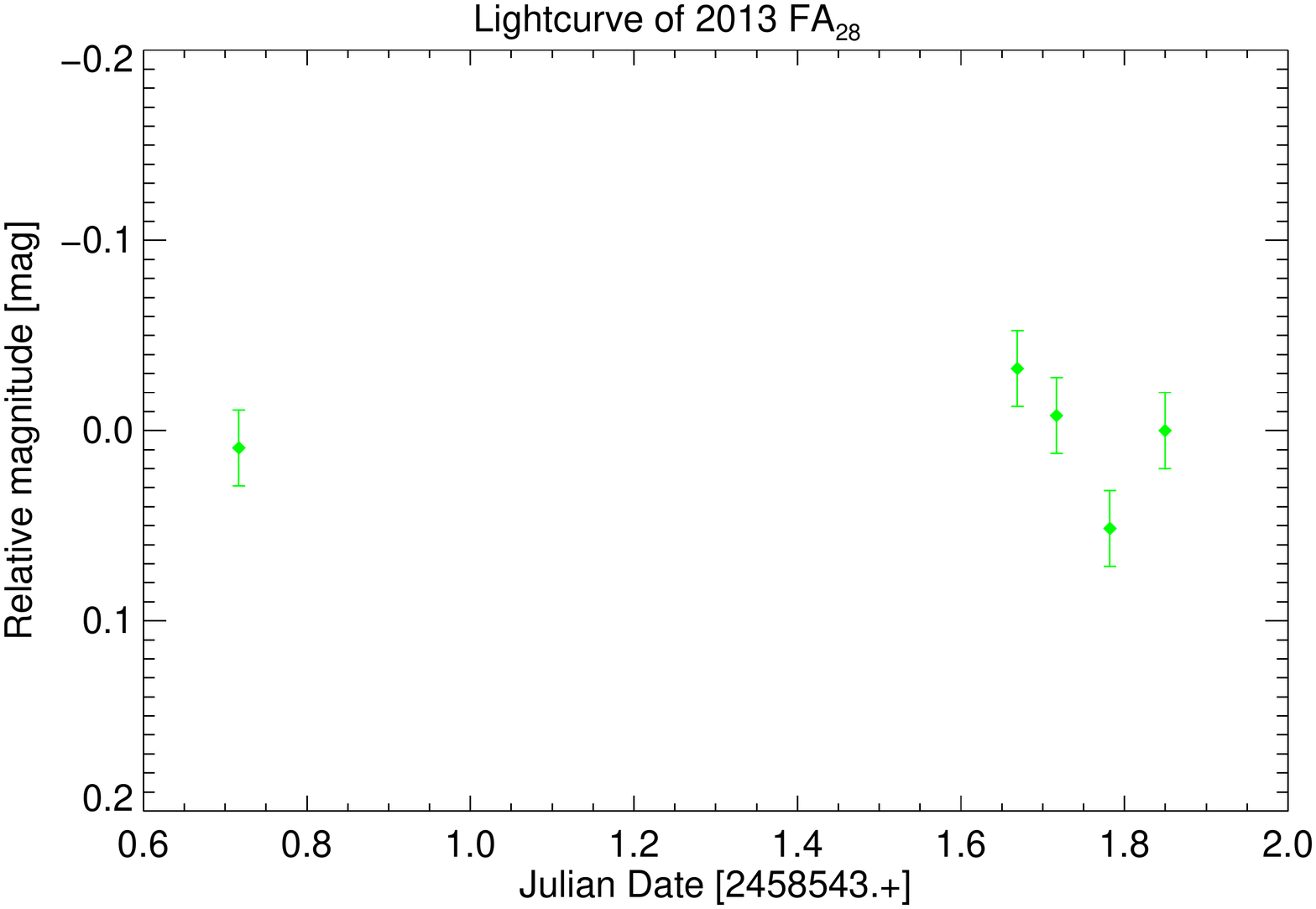}
   \includegraphics[width=9cm, angle=0]{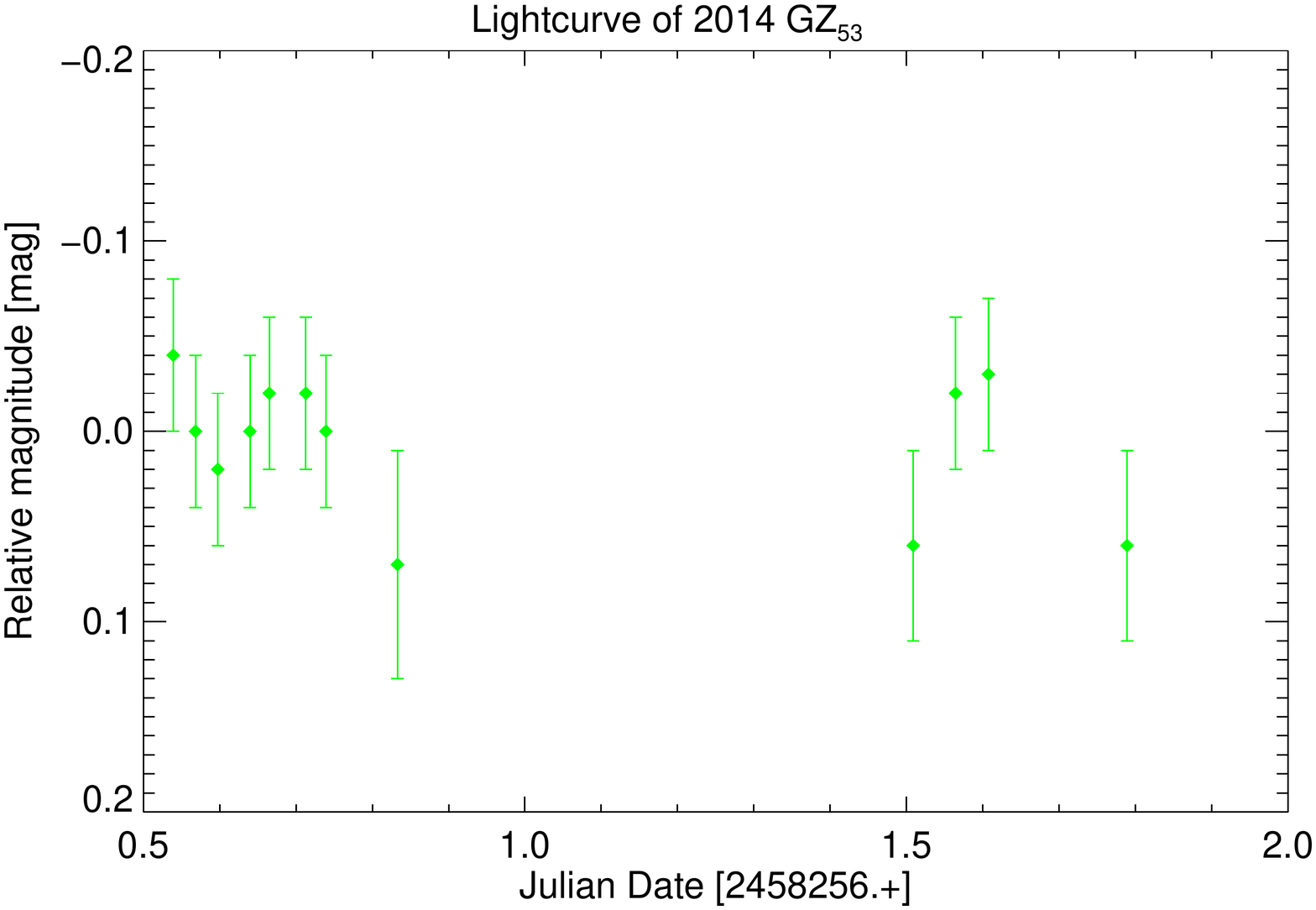}
   \includegraphics[width=9cm, angle=0]{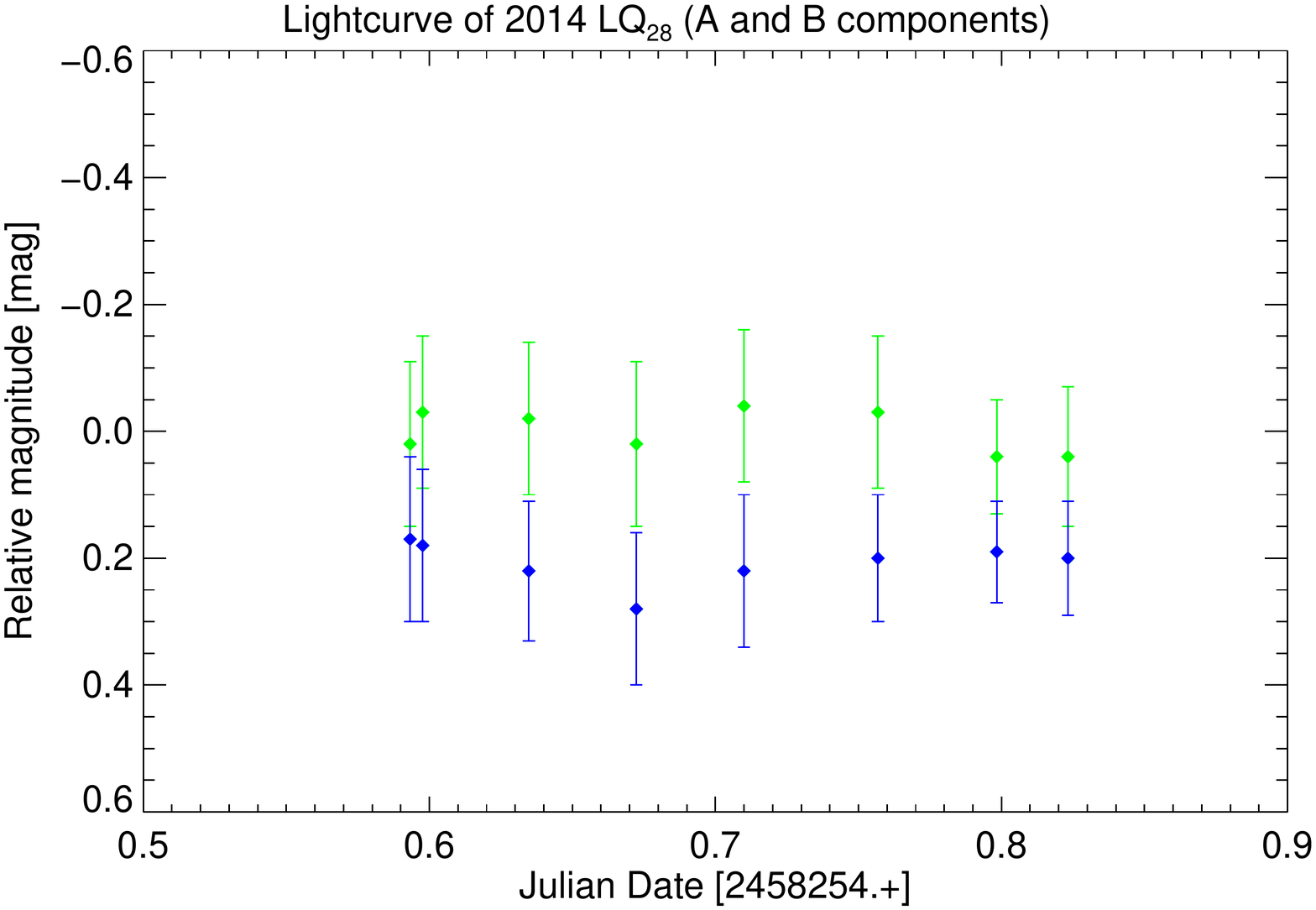}
\caption{ Continued. }
\label{fig:LC}
\end{figure*}

 \begin{figure*}
  \includegraphics[width=9cm, angle=0]{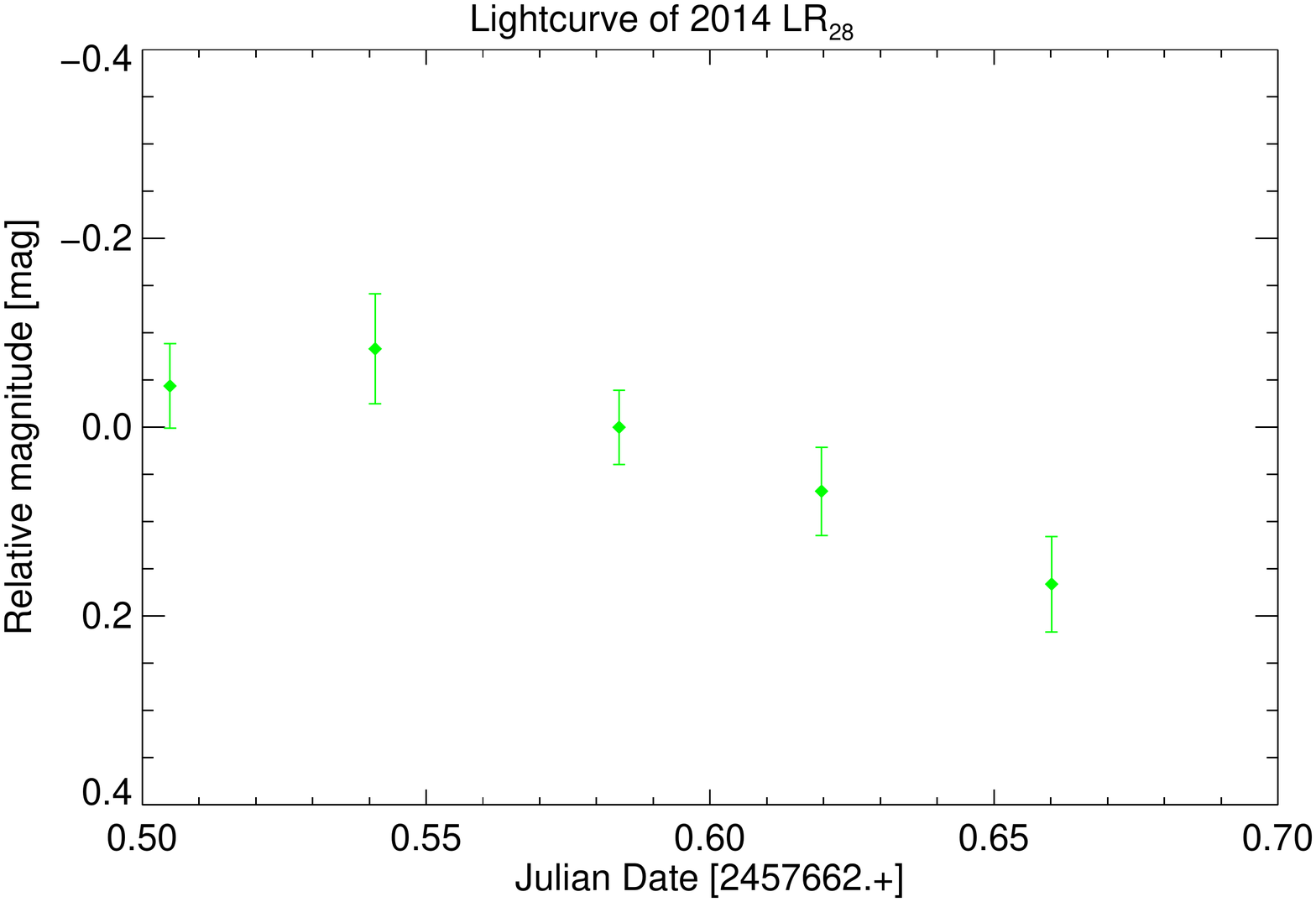}
    \includegraphics[width=9cm, angle=0]{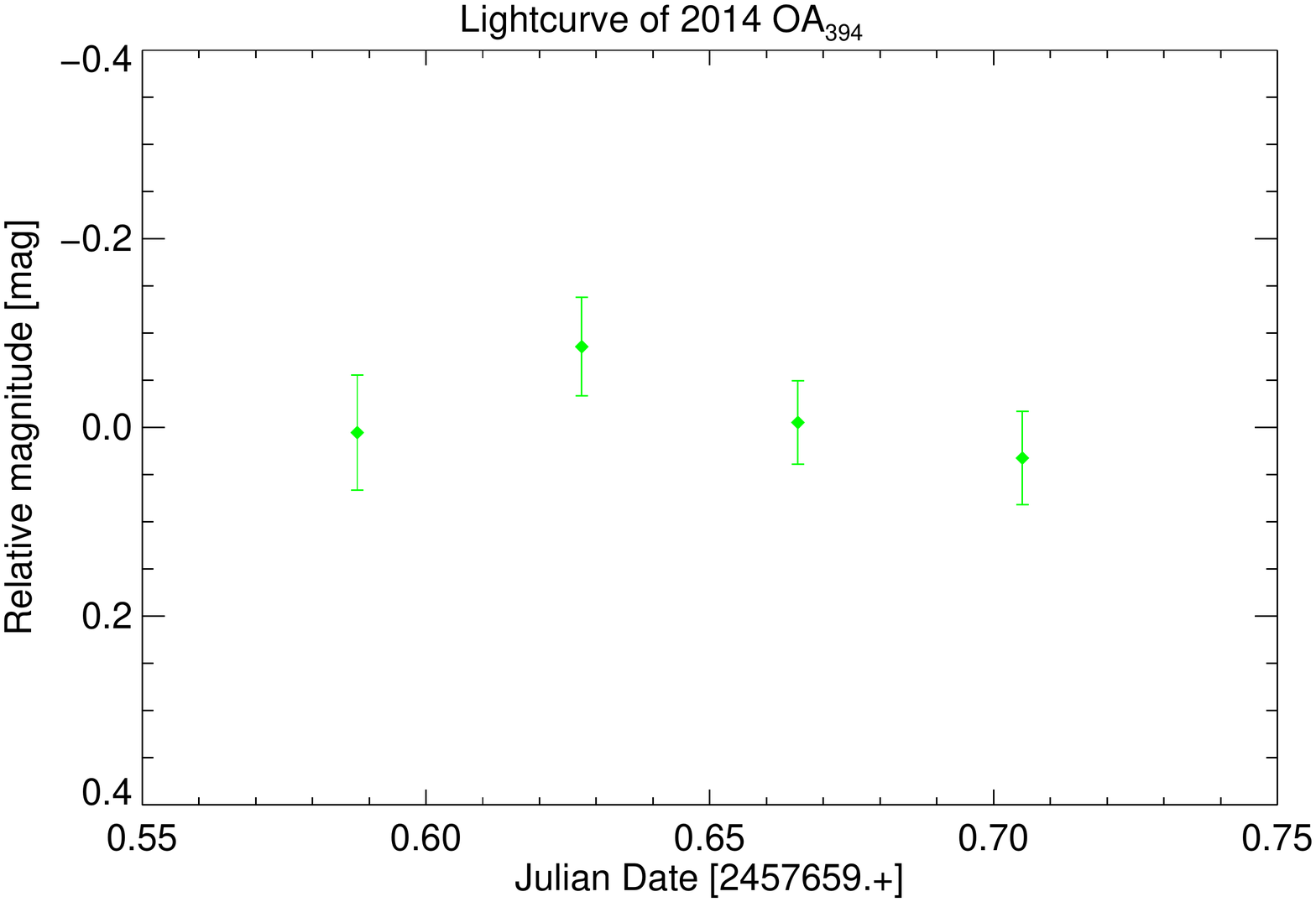}
     \includegraphics[width=9cm, angle=0]{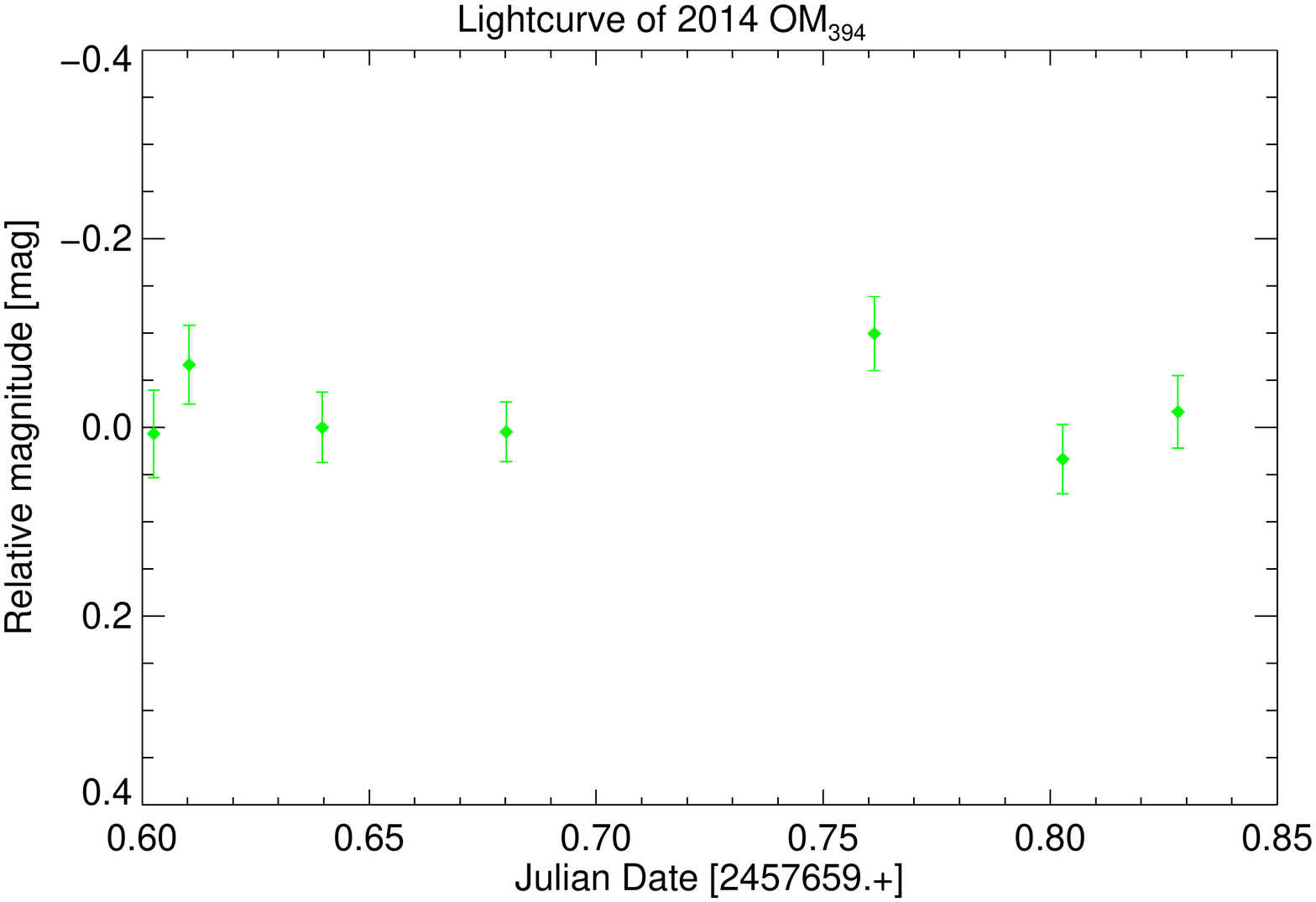}
\caption{ Continued. }
\label{fig:LC}
\end{figure*}

\clearpage
 \appendix
 \section*{Appendix B}
 \startlongtable
\begin{deluxetable}{lccc}
 \tablecaption{\label{Tab:Summary_photo2} Photometry used in this paper is available in the following table. Julian date is without light-time correction. Full table will be available in the published version.  }
\tablewidth{0pt}
\tablehead{
Object  & Julian Date & Relative magnitude &  Error  \\
        &          &   [mag]           &  [mag]\\
}
\startdata  
(58534) 1997~CQ$_{29}$ Logos-Zoe &   &   &   \\
 & 2457830.68213 & -0.14 & 0.07 \\ 
& 2457830.69215& 0.00 & 0.06 \\ 
& 2457830.73353 & 0.38 & 0.09 \\  
\hline
\enddata
\end{deluxetable}


\begin{thebibliography}{}
\bibitem[Alexandersen et al.(2018)]{Alexandersen2018} Alexandersen, M., Benecchi, S.~D., Chen, Y.-T., et al.\ 2018, arXiv:1812.04304 
\bibitem[Batygin et al.(2011)]{Batygin2011} Batygin, K., Brown, M.~E., \& Fraser, W.~C.\ 2011, \apj, 738, 13 
\bibitem[Benecchi et al.(2009)]{Benecchi2009} Benecchi, S.~D., Noll, K.~S., Grundy, W.~M., et al.\ 2009, \icarus, 200, 292 
\bibitem[Benecchi \& Sheppard(2013)]{Benecchi2013} Benecchi, S.~D., \& Sheppard, S.~S.\ 2013, \aj, 145, 124 
\bibitem[Benecchi et al.(2017)]{Benecchi2017} Benecchi, S.~D., Buie, M.~W., Porter, S.~B., et al.\ 2017, AAS/Division for Planetary Sciences Meeting Abstracts \#49, 49, 504.07 
\bibitem[Benecchi et al.(2018)]{Benecchi2018} Benecchi, S., Porter, S., Spencer, J., et al.\ 2018, arXiv:1812.04758 
\bibitem[Cellino et al.(1985)]{Cellino1985} Cellino, A., Pannunzio, R., Zappala, V., Farinella, P., \& Paolicchi, P.\ 1985, \aap, 144, 355 
\bibitem[Chandrasekhar(1987)]{Chandrasekhar1987} Chandrasekhar, S.\ 1987, 
New York : Dover, 1987. 
\bibitem[Collander-Brown et al.(1999)]{Collander1999} Collander-Brown, S.~J., Fitzsimmons, A., Fletcher, E., Irwin, M.~J., \& Williams, I.~P.\ 1999, \mnras, 308, 588 
\bibitem[Davis \& Farinella(1997)]{Davis1997} Davis, D.~R., \& Farinella, P.\ 1997, \icarus, 125, 50 
\bibitem[Duffard et al.(2009)]{Duffard2009} Duffard, R., Ortiz, J.~L., Thirouin, A., Santos-Sanz, P., \& Morales, N.\ 2009, \aap, 505, 1283 
\bibitem[Dunlap \& Gehrels(1969)]{Dunlap1969} Dunlap, J.~L., \& Gehrels, T.\ 1969, \aj, 74, 796 
\bibitem[Elliot et al.(2005)]{Elliot2005} Elliot, J.~L., Kern, S.~D., Clancy, K.~B., et al.\ 2005, \aj, 129, 1117
\bibitem[Feigelson \& Nelson(1985)]{Feigelson1985} Feigelson, E.~D., \& Nelson, P.~I.\ 1985, \apj, 293, 192 
\bibitem[Gladman et al.(2008)]{Gladman2008} Gladman, B., Marsden, B.~G., \& Vanlaerhoven, C.\ 2008, The Solar System Beyond Neptune, 43  
\bibitem[Grundy et al.(2011)]{Grundy2011} Grundy, W.~M., Noll, K.~S., Nimmo, F., et al.\ 2011, \icarus, 213, 678 
\bibitem[Grundy et al.(2012)]{Grundy2012} Grundy, W.~M., Benecchi, S.~D., Rabinowitz, D.~L., et al.\ 2012, \icarus, 220, 74 
\bibitem[Harris \& Warner(2018)]{Harris2018} Harris, A.~W., \& Warner, B.\ 2018, AAS/Division for Planetary Sciences Meeting Abstracts \#50, 50, 414.03  
\bibitem[Howell(1989)]{Howell1989} Howell, S.~B.\ 1989, \pasp, 101, 616 
\bibitem[Isobe et al.(1986)]{Isobe1986} Isobe, T., Feigelson, E.~D., \& Nelson, P.~I.\ 1986, \apj, 306, 490 
\bibitem[Isobe \& Feigelson(1990)]{Isobe1990} Isobe, T., \& Feigelson, E.~D.\ 1990, \baas, 22, 917 
\bibitem[Jeans(1919)]{Jeans1919} Jeans, J.~H.\ 1919, Cambridge, University press, 1919.
\bibitem[Kern \& Elliot(2006)]{Kern2006} Kern, S.~D., \& Elliot, J.~L.\ 2006, \icarus, 183, 179 
\bibitem[Kern(2006)]{Kern2006_phd} Kern, S.~D.\ 2006, Ph.D.~Thesis,  
\bibitem[Lacerda \& Luu(2006)]{Lacerda2006} Lacerda, P., \& Luu, J.\ 2006, \aj, 131, 2314 
\bibitem[Lacerda(2011)]{Lacerda2011} Lacerda, P.\ 2011, \aj, 142, 90 
\bibitem[Lacerda et al.(2014)]{Lacerda2014} Lacerda, P., McNeill, A., \& Peixinho, N.\ 2014, \mnras, 437, 3824 
\bibitem[Lavalley et al.(1990)]{LaValley1990} Lavalley, M.~P., Isobe, T., \& Feigelson, E.~D.\ 1992, \baas, 22, 917-918, 1990. 
\bibitem[Lavalley et al.(1992)]{LaValley1992} Lavalley, M.~P., Isobe, T., \& Feigelson, E.~D.\ 1992, \baas, 24, 839 
\bibitem[Leone et al.(1984)]{Leone1984} Leone, G., Paolicchi, P., Farinella, P., \& Zappala, V.\ 1984, \aap, 140, 265 
\bibitem[Levine et al.(2012)]{Levine2012} Levine, S.~E., Bida, T.~A., Chylek, T., et al.\ 2012, \procspie, 8444, 844419 
\bibitem[Lomb(1976)]{Lomb1976} Lomb, N.~R.\ 1976, \apss, 39, 447 
\bibitem[Moore et al.(2018)]{Moore2018} Moore, J.~M., McKinnon, W.~B., Cruikshank, D.~P., et al.\ 2018, arXiv:1808.02118 
\bibitem[Noll et al.(2002)]{Noll2002} Noll, K.~S., Stephens, D.~C., Grundy, W.~M., et al.\ 2002, \aj, 124, 3424 
\bibitem[Noll et al.(2004)]{Noll2004} Noll, K.~S., Stephens, D.~C., Grundy, W.~M., Osip, D.~J., \& Griffin, I.\ 2004, \aj, 128, 2547 
\bibitem[Noll et al.(2008a)]{Noll2008} Noll, K.~S., Grundy, W.~M., Chiang, E.~I., Margot, J.-L., \& Kern, S.~D.\ 2008, The Solar System Beyond Neptune, 345  
\bibitem[Noll et al.(2008b)]{Noll2008b} Noll, K.~S., Grundy, W.~M., Stephens, D.~C., Levison, H.~F., \& Kern, S.~D.\ 2008, \icarus, 194, 758 
\bibitem[Noll et al.(2014)]{Noll2014} Noll, K.~S., Parker, A.~H., \& Grundy, W.~M.\ 2014, AAS/Division for Planetary Sciences Meeting Abstracts, 46, 507.05 
\bibitem[Osip et al.(2003)]{Osip2003} Osip, D.~J., Kern, S.~D., \& Elliot, J.~L.\ 2003, Earth Moon and Planets, 92, 409 
\bibitem[P{\'a}l et al.(2015)]{Pal2015} P{\'a}l, A., Szab{\'o}, R., Szab{\'o}, G.~M., et al.\ 2015, \apjl, 804, L45 
\bibitem[Peixinho et al.(2008)]{Peixinho2008} Peixinho, N., Lacerda, P., \& Jewitt, D.\ 2008, \aj, 136, 1837 
\bibitem[Peixinho et al.(2015)]{Peixinho2015} Peixinho, N., Delsanti, A., \& Doressoundiram, A.\ 2015, \aap, 577, A35 
\bibitem[Penteado et al.(2016)]{Penteado2016} Penteado, P.~F., Trilling, D.~E., \& Grundy, W.\ 2016, AAS/Division for Planetary Sciences Meeting Abstracts \#48, 48, 120.20 
\bibitem[Pike et al.(2017)]{Pike2017} Pike, R.~E., Fraser, W.~C., Schwamb, M.~E., et al.\ 2017, \aj, 154, 101 
\bibitem[Rabinowitz et al.(2014)]{Rabinowitz2014} Rabinowitz, D.~L., Benecchi, S.~D., Grundy, W.~M., \& Verbiscer, A.~J.\ 2014, \icarus, 236, 72 
\bibitem[Santos-Sanz et al.(2009)]{SantosSanz2009} Santos-Sanz, P., Ortiz, J.~L., Barrera, L., \& Boehnhardt, H.\ 2009, \aap, 494, 693 
\bibitem[Romanishin \& Tegler(1999)]{Romanishin1999} Romanishin, W., \& Tegler, S.~C.\ 1999, \nat, 398, 129 
\bibitem[Sheppard \& Jewitt(2002)]{Sheppard2002} Sheppard, S.~S., \& Jewitt, D.~C.\ 2002, \aj, 124, 1757 
\bibitem[Sheppard \& Jewitt(2003)]{Sheppard2003} Sheppard, S.~S., \& Jewitt, D.~C.\ 2003, Earth Moon and Planets, 92, 207 
\bibitem[Sheppard \& Jewitt(2004)]{Sheppard2004} Sheppard, S.~S., \& Jewitt, D.\ 2004, \aj, 127, 3023 
\bibitem[Sheppard et al.(2008)]{Sheppard2008} Sheppard, S.~S., 
Lacerda, P., \& Ortiz, J.~L.\ 2008, The Solar System Beyond Neptune, 129 
\bibitem[Sheppard \& Thirouin(2018)]{Sheppard2018} Sheppard, S.~S., Thirouin, A.\ 2018, Central Bureau Electronic Telegrams, 4483. 
\bibitem[Showalter et al. (2019)]{Showalter2019} Showalter M. R., Buie M. W.,Grundy W. M.,   Hamilton D. P., Kaufmann D. E., et al.\ 2019, LPSC abstract 
\bibitem[Spearman (1904)]{Spearman1904} Spearman C. \ 1904, The American Journal of Psychology, 15, 72. 
\bibitem[Stellingwerf(1978)]{Stellingwerf1978} Stellingwerf, R.~F.\ 1978, \apj, 224, 953 
\bibitem[Stephens \& Noll(2006)]{Stephens2006} Stephens, D.~C., \& Noll, K.~S.\ 2006, \aj, 131, 1142 
\bibitem[Stern et al.(2019)]{Stern2019} Stern, S.~A., Spencer, J.~R., Weaver, H.~A., et al.\ 2019, arXiv:1901.02578 
\bibitem[Stetson(1987)]{Stetson1987} Stetson, P.~B.\ 1987, \pasp, 99, 191 
\bibitem[Thirouin et al.(2010)]{Thirouin2010} Thirouin, A., Ortiz, J.~L., Duffard, R., et al.\ 2010, \aap, 522, A93 
\bibitem[Thirouin et al.(2012)]{Thirouin2012} Thirouin, A., Ortiz, J.~L., Campo-Bagatin, A., et al.\ 2012, \mnras, 424, 3156 
\bibitem[Thirouin(2013)]{Thirouin2013} Thirouin, A. \ 2013, 
Ph.D.~Thesis, University of Granada.
\bibitem[Thirouin et al.(2014)]{Thirouin2014} Thirouin, A., Noll, K.~S., Ortiz, J.~L., \& Morales, N.\ 2014, \aap, 569, A3 
\bibitem[Thirouin et al.(2016)]{Thirouin2016} Thirouin, A., Sheppard, S.~S., Noll, K.~S., et al.\ 2016, \aj, 151, 148 
\bibitem[Thirouin et al.(2017)]{Thirouin2017} Thirouin, A., Sheppard, S.~S., \& Noll, K.~S.\ 2017, \apj, 844, 135 
\bibitem[Thirouin \& Sheppard(2017)]{ThirouinSheppard2017} Thirouin, A., \& Sheppard, S.~S.\ 2017, \aj, 154, 241 
\bibitem[Thirouin \& Sheppard(2018)]{Thirouin2018} Thirouin, A., \& Sheppard, S.~S.\ 2018, \aj, 155, 248 
\bibitem[Trilling \& Bernstein(2006)]{Trilling2006} Trilling, D.~E., \& Bernstein, G.~M.\ 2006, \aj, 131, 1149
\bibitem[Weidenschilling(1980)]{Weidenschilling1980} Weidenschilling, S.~J.\ 1980, \icarus, 44, 807 
\bibitem[Zangari et al. (2019)]{Zangari2019} Zangari A. M., Beddingfield C. B., Benecchi S. D.   Beyer R. A., Bierson C. J. et al.\ 2019, LPSC abstract 
\bibitem[Zappala(1980)]{Zappala1980} Zappala, V.\ 1980, Moon and Planets, 23, 345 
 
\end{thebibliography}
\end{document}